\documentclass[11pt]{article}

\usepackage{geometry}                
\usepackage{graphicx}
\usepackage{amssymb}
\usepackage{epstopdf}
\usepackage{color}
\usepackage{url}
\newcommand{\ifb}{fb$^{-1}$ }

\date{}                                           

\begin{document}

\begin{titlepage}

\begin{center}
\vspace*{-1cm}

\vskip 1.5in
{\LARGE \bf The State of Supersymmetry \\ \vspace{2mm} after Run I of the LHC} \\ \vspace{5mm}
{\it \normalsize Lectures at the training week of the GGI workshop \\
 ``Beyond the Standard Model after the first run of the LHC''} \\
\vspace{.15in}

\vskip 0.35in
{\large Nathaniel Craig}

\vskip 0.12in
{\em 
School of Natural Sciences
\\ Institute for Advanced Study \\
Princeton, NJ 08540 \\}

\vskip 0.12in
and
\vskip 0.12in

{\em 
Department of Physics \\
Rutgers University \\
Piscataway, NJ 08854}

\vskip 0.4in

\end{center}

\baselineskip=16pt

\begin{abstract}

\noindent

 In these lectures I survey the state of supersymmetric extensions of the Standard Model in light of data from the first run of the LHC. After assessing pre-LHC expectations based on principles of naturalness and parsimony, I review the landscape of direct and indirect search limits at the LHC, including the implications of the observed Higgs mass and couplings. Finally, I survey several broad classes of supersymmetric models that are consistent with current data and enumerate the most promising search strategies and model-building directions for the future.  
 
\end{abstract}

\end{titlepage}

\pagebreak

\tableofcontents

\pagebreak

\part{Preliminaries}

\section*{Preface}
\addcontentsline{toc}{section}{Preface}

These lecture notes are based on a series of three lectures delivered to an exceedingly patient audience during the training week of the GGI workshop ``Beyond the Standard Model after the first run of the LHC''. They're aimed at young physicists with a working grasp of supersymmetric phenomenology, intended primarily as an assessment of supersymmetry in light of LHC data rather than a review of fundamental principles. The original audience included both beginning graduate students and some of the pioneers of weak-scale supersymmetry, which made for an unusual balance. The first lecture (sections 1-4) is intended as a general, broad summary of the state of supersymmetry in light of LHC limits. The second lecture (sections 5 \& 6) is somewhat more technical, but remains focused on the direct implications of LHC limits for supersymmetry. The third lecture (sections 7 \& 8) ventures into model-building guided by LHC limits.  Needless to say, these lecture notes contain some additional content omitted during the actual lectures due to time constraints or simple absent-mindedness. Experimental figures have been updated to include results presented between delivery of the lectures and completion of these notes. The material covered -- and the tone of its presentation -- is necessarily idiosyncratic, and I apologize for any and all omissions; for an excellent alternative with somewhat different coverage of similar topics, I recommend Jonathan Feng's lecture notes \cite{Feng:2013pwa}. In any event, hopefully these notes  -- unlike the lectures themselves -- do not suffer from my general {\it tendenza a parlare troppo in fretta.}

\section{Introduction}

As of the time of these lectures -- June 2013 -- the LHC has completed a run of unprecedented success, accumulating $\sim 5$ fb$^{-1}$ of integrated luminosity at 7 TeV $pp$ collisions and $\sim 20$ fb$^{-1}$ of integrated luminosity  at 8 TeV $pp$ collisions in each detector. Not all of the data has been analyzed -- in part due to the deserved focus on analyzing the recently-discovered Higgs boson at 125 GeV, in part due to the countless channels available. So there remains considerable room for surprises in data that has already been written to tape; it is entirely possible that the tone of these lectures will be superseded by intriguing hints in Run I analyses that have yet to be completed or released.

But barring that cheerful eventuality, the general tone at the moment is fairly sombre for supersymmetry enthusiasts. Of course, the discovery of the Higgs boson is immensely exciting, and supersymmetry is one of few UV complete frameworks to predict a range for the Higgs mass that is compatible with observation. So that's good. But the various superpartners predicted by minimal realizations of supersymmetry have yet to appear, even as the mass reach of the LHC stretches out towards degrees of freedom at the TeV scale. To the extent that these superpartners are responsible for preserving the naturalness of the weak scale against radiative corrections, their apparent absence at the weak scale is cause for no small amount of unease. 

I should take great care to emphasize that this unease is by no means unique to supersymmetry. Effective field theory tightly ties the cutoff for radiative corrections to the Higgs mass to the mass scale of new physics; the non-observation of additional particles pushes that cutoff to higher and higher scales for any natural explanation of the weak scale. So essentially any model that explains the naturalness of the weak scale is under tension due to the onward march of null results. This is either a challenge to naturalness or a challenge to our ability to construct natural theories. 

My hope, in these lectures, is to survey in concrete terms the state of supersymmetry in light of Run I data -- to quantify the feeling of unease coming from null results, and to evaluate the directions for supersymmetry that seem most promising in light of data. The idea of giving a ``state of supersymmetry'' lecture brings to mind the State of the Union, an address that customarily begins with variations on the phrase ``the state of the Union is \dots''. In the modern era ``\dots'' is typically some moderately optimistic platitude like ``strong''. At the moment, I do not think that optimistic platitudes are a very helpful guide for our community. And so, as for the state of supersymmetry, I am inclined to paraphrase a president from a much earlier time:\footnote{From Andrew Johnson's 1867 address, ``Candor compels me to declare that at this time there is no union as our fathers understood the term, and as they meant it to be understood by us'.}

\begin{quote}
Candor compels me to declare that at this time there is no supersymmetry as our forebears understood the term, and as they meant it to be understood by us.
\end{quote}

I do not mean this to say that there is no supersymmetry in nature. Rather, I mean that the march of null results suggests that we were mostly wrong about precisely how supersymmetry would appear at the LHC. To me, this suggests immense opportunity to step back and re-evaluate the criteria that led us to this point. In particular, we built our expectations for supersymmetry at the LHC on the twin pillars of parsimony and naturalness. The null results at the LHC suggest that those two pillars were perhaps not the right foundations. Consequently, there is now tremendous opportunity to figure out the correct theory of the universe, with much room for it to be supersymmetric in some form.

A supersymmetry skeptic might understandably ask why we should attribute the lack of SUSY signals at the LHC to incorrect model-building criteria, rather than to the simple lack of SUSY itself. Indeed, it may well be that supersymmetry plays no role in stabilizing the weak scale. But there are many valid reasons to favor supersymmetry, some of which are only strengthened by what we've learned so far at the LHC:
\begin{itemize}
\item It robustly solves the hierarchy problem without UV sensitivity.
\item It predicts that electroweak symmetry is broken by an elementary scalar, in good agreement with observation.
\item It predicts the Higgs mass to lie below 135 GeV, in good agreement with observation.
\item It has excellent decoupling properties, ameliorating tension with limits on precision electroweak observables.
\item The degrees of freedom directly related to the naturalness of EWSB are not yet sharpy constrained by the LHC.
\end{itemize}

Against this we must measure the negative indication:
\begin{itemize}
\item The simplest versions of supersymmetry, with the simplest assumptions about the supersymmetric spectrum, are under stress.
\end{itemize}

In the balance, it seems to me to be worthwhile to pursue supersymmetric explanations of the hierarchy problem, at least until we've fully explored the range of supersymmetric models available. Indeed, it's a splendid opportunity for young physicists to inject new ideas and take SUSY phenomenology in new directions. The motivation is still strong, but data has told us that our pre-LHC criteria were perhaps not correct, and new criteria (such as the Higgs mass and couplings) provide suggestive avenues for further development. Thankfully, this is a more or less falsifiable proposition -- if we turn on the LHC at 13 TeV and have still discovered no indication of new physics within a few years of starting Run II, even convoluted models of weak-scale supersymmetry will be strongly disfavored. Of course, this says nothing about models of supersymmetry above the weak scale, perhaps motivated by gauge coupling unification -- but that's a matter for another day\dots

\section{What did we expect?}

So far I have made much of the idea that our pre-LHC expectations regarding supersymmetry were perhaps incorrect, without breaking down precisely what those expectations were. In my mind, the two principle guideposts for model-building in the pre-LHC era were the clear criteria for physics motivated by the hierarchy problem: naturalness and parsimony. The general moral of these lectures will be that SUSY is in fine shape if we relax one criterion or the other. But first, let's examine where those criteria led us.

\subsection{Naturalness}

Naturalness is the deepest underlying motivation for weak-scale SUSY, so let's briefly examine the motivation and precisely how it shaped thinking in the run up to the LHC.

We start in our field with Dirac, asking why the proton is much lighter than the Planck scale. And we understand this through asymptotic freedom of QCD, via 
$$ \frac{m_p}{M_P} \sim \exp \left[-c/g_3^2 \right] \ll 1$$

We see the same problem recapitulated in the electroweak sector, namely by the observation $m_W, m_Z, m_h \ll M_P$. However, these physical states are not obviously composites of some asymptotically free dynamics, and so it is less clear that the same reasoning applies. 

And the problem is actually sharper. The Higgs, whose vacuum expectation value sets $m_W$ and $m_Z$ -- if a fundamental scalar -- is generally quadratically sensitive to the scale of physics to which it couples. Generic new physics induces contributions to the Higgs mass on order of the new physics scale times a loop factor, and gravity attests to the existence of such scales in the most troubling way. 

This was captured clearly by Wilson, as conveyed by Susskind \cite{Susskind:1978ms}. In their assessment, satisfactory explanations of the weak scale should require more or less that ``observable properties of a system should not depend sensitively on variations in the fundamental parameters.'' In subsequent years we have identified two broad classes of solutions with some promise of satisfying this criterion: (1) introduce a symmetry to control radiative corrections to elementary scalar masses to arbitrarily high scales, or (2) lower the cutoff of the effective theory containing the elementary scalar (either by lowering the cutoff entirely, as in large extra dimensions, or introducing a scale above which there is no elementary scalar, as in compositeness and warped extra dimensions). Supersymmetry is an exemplar of the first class, introduced to control the large contribution to the Higgs mass from states at the highest cutoffs of the theory by enhancing the symmetry of the theory. Supersymmetric extensions of the Standard Model trade quadratic dependence on high scales for logarithmic dependence on the cutoff and quadratic dependence on smaller sparticle masses. Of course, just eliminating quadratic sensitivity to the UV cutoff is not enough for naturalness; one must subsequently ensure that the logarithmic dependence on the cutoff and quadratic sensitivity to sparticle masses does not make the theory unnatural. So much effort over the last thirty years has been devoted to taking the idea of ``insensitivity'' quantitatively.

We seek to elevate this into a quantitative principle that we can use to evaluate models. This leads to the naturalness prescription followed by our field. The typical prescription is to choose a framework and fundamental parameters (call them $a_i$ -- what precisely qualifies among the $a_i$ depends on the context), then compute the sensitivity parameters $\Delta$, defined by \cite{Barbieri:1987fn}

\begin{equation} \label{eq:delta}
\Delta[a_i] = \frac{\partial \ln m_Z^2}{\partial \ln a_i^2}, \Delta \equiv \max_i \Delta[a_i]
\end{equation}
and impose $\Delta < \Delta_{max}$ with social choice of $\Delta_{max}$. (Alternately, some prescriptions vary $m_Z$ with respect to $a$, not $a^2$.) Crucially, the acceptable value of $\Delta$ has drifted up over the years, starting around $\sim 10$ and floating towards 100. Many would now call 1000 a reasonable value. Yet there is no intrinsic measure; this is simply a prescription. 

In what follows, I will use the historical $\Delta \sim 10$ benchmark to set LHC expectations for the connection between naturalness and the spectrum of superpartners. But before proceeding further, it is helpful to critique the role of the sensitivity parameter in shaping model-building. There are many questions left unanswered by the naturalness prescription embodied by (\ref{eq:delta}).

\begin{itemize}
\item First, how to combine sensitivity parameters $\Delta$? Take the maximum value? Add in quadrature? Multiply together? Arguments can be made for each, and various practitioners have made various choices, but it's not at all clear which is appropriate. Needless to say, models do not behave uniformly under changes in this prescription.

\item What fundamental parameters $a_i$ are included? The SUSY-breaking parameters alone?  Also Standard Model dimensionless parameters? A/the cutoff? Which cutoff? We often try to factorize SUSY-breaking from the supersymmetric structure of SM couplings, but more ambitious models frequently connect them; these may or may not be tied in some deep way. It's not obvious that fixing the SM dimensionless parameters and exploiting them to reduce the apparent tuning in SUSY-breaking parameters is necessarily natural.

\item What about the parameterization of the fundamental parameters? For example, consider sensitivity of the $Z$ boson mass to the top quark Yukawa in the MSSM. The top pole mass and soft terms of $H_u, \tilde t_L, \tilde t_R$ are weak functions of $\lambda_t(M_{GUT})$ due to an approximate IR fixed point of the top Yukawa. If our tuning measure chooses $M_t$ as a free parameter, then tuning can be large, but if it chooses $\lambda_t(M_{GUT})$, the tuning is small. So the tuning depends on the parameterization. 

\item More distressingly, this sensitivity measure also suggests dynamical solutions could be unnatural. For example, for the QCD scale,
$\Delta[g_3] \sim \ln(M_P^2 / \Lambda_{QCD}^2) \sim 90$ is large. Similarly, if the electroweak scale is set by some dynamical process, as in dynamical SSB, we would have $\Delta[g_H] \sim 80$. Does that mean we should take $\Delta \sim 100$ as natural? Or is that natural for theories with cutoff $M_P$, and SUSY theories with lower cutoff should have appropriately smaller benchmark sensitivity?

\end{itemize}

So it is clear that measures of tuning have no intrinsic meaning. They may have some comparative value in terms of contrasting models, but even this is not absolute. Different models perform differently under different measures, and the ordering of naturalness may change. One frequently comes across models that are constructed using a legalistic interpretation of naturalness that fails an intuitive sniff test.

I would prefer we exercise our physical judgment when weighing naturalness. Perhaps a reasonable criterion in the context of model-building is to ask if the IR theory is a generic function of the UV parameters, but not commit overly much to specific measures. Another very nice guide is the naturalness criterion of Veltman \cite{Veltman:1980mj}, which posits that {\it radiative effects not exceed tree-level effects in size}. I will return to this criterion a fair bit. A natural theory should rest comfortably among these various requirements.

In any event, the idea of naturalness provided several useful guideposts for what IR physics might entail. For most of the last few decades we developed expectations of the scale based on the usual $\Delta \sim 10$ limit on tuning. SUSY regulates the hierarchy problem, but in terms of the sensitivity of the electroweak scale to the theory, not all sparticles are equally important. Thus a supersymmetric theory that naturally explains the weak scale does not necessarily have all sparticles clustered around the same scale. There are two direct sources of concern, tree-level contributions and loop-level contributions. Both play a role primarily through the relation between the weak scale and soft parameters, viz.
\begin{equation} \label{eq:ewsb}
m_Z^2 = -2 (m_{H_u}^2 + |\mu|^2) + \dots
\end{equation}
This relation arises from minimizing the Higgs potential in the MSSM at large $\tan \beta$, where the second doublet $H_d$ is essentially a spectator. By inspection, this suggests that $m_{H_u}^2$ and $\mu$ are the most important parameters for tuning. This is, however, somewhat misleading. Indeed, one often hears incorrect interpretations of where (\ref{eq:ewsb}) comes from and what it means, so let's first examine it a bit more carefully. Really, (\ref{eq:ewsb}) is just coming from the usual relation between the mass, quartic, and vev of a single Higgs doublet $H$, 

\begin{equation}
v^2 = - \frac{m_H^2}{2 \lambda}
\end{equation}
where in the MSSM, $H_u \sim H$ at large $\tan \beta$, and the quartic is fixed by $D$-terms as $\lambda = \frac{1}{8} (g^2 + g'^2)$. The physical mass of the doublet $H_u$ is $m_H^2 = m_{H_u}^2 + |\mu|^2$, where the first term comes directly from the soft Lagrangian and the second term comes from integrating out auxiliary components of the $H_u$ chiral multiplet in minimal theories. This makes clear several salient points:

\begin{itemize}
\item It's not always appropriate to think of the $|\mu|^2$ term separately. Really what appears in  (\ref{eq:ewsb}) is the physical mass of the Higgs doublet. There are theories where non-trivial dynamics in the hidden sector drives the physical mass of the Higgs doublet $H_u$ to zero \cite{Murayama:2007ge, Roy:2007nz, Craig:2013wga}. This manifests itself dynamically as $m_{H_u}^2 \approx - |\mu|^2$, so that there can be a natural cancellation between ostensibly large parameters. 

\item The MSSM is disadvantaged from the perspective of tuning since the quartic is so small; $m_Z$ only appears in (\ref{eq:ewsb}) because the MSSM quartic is fixed in terms of the gauge couplings. More generally, though the LHS of  (\ref{eq:ewsb}) should just be read as $4 \lambda v^2$. Extensions of the MSSM with new contributions to the quartic therefore improve the tuning of the theory.
\end{itemize}

\begin{figure}[htbp] 
   \centering
   \includegraphics[width=6in]{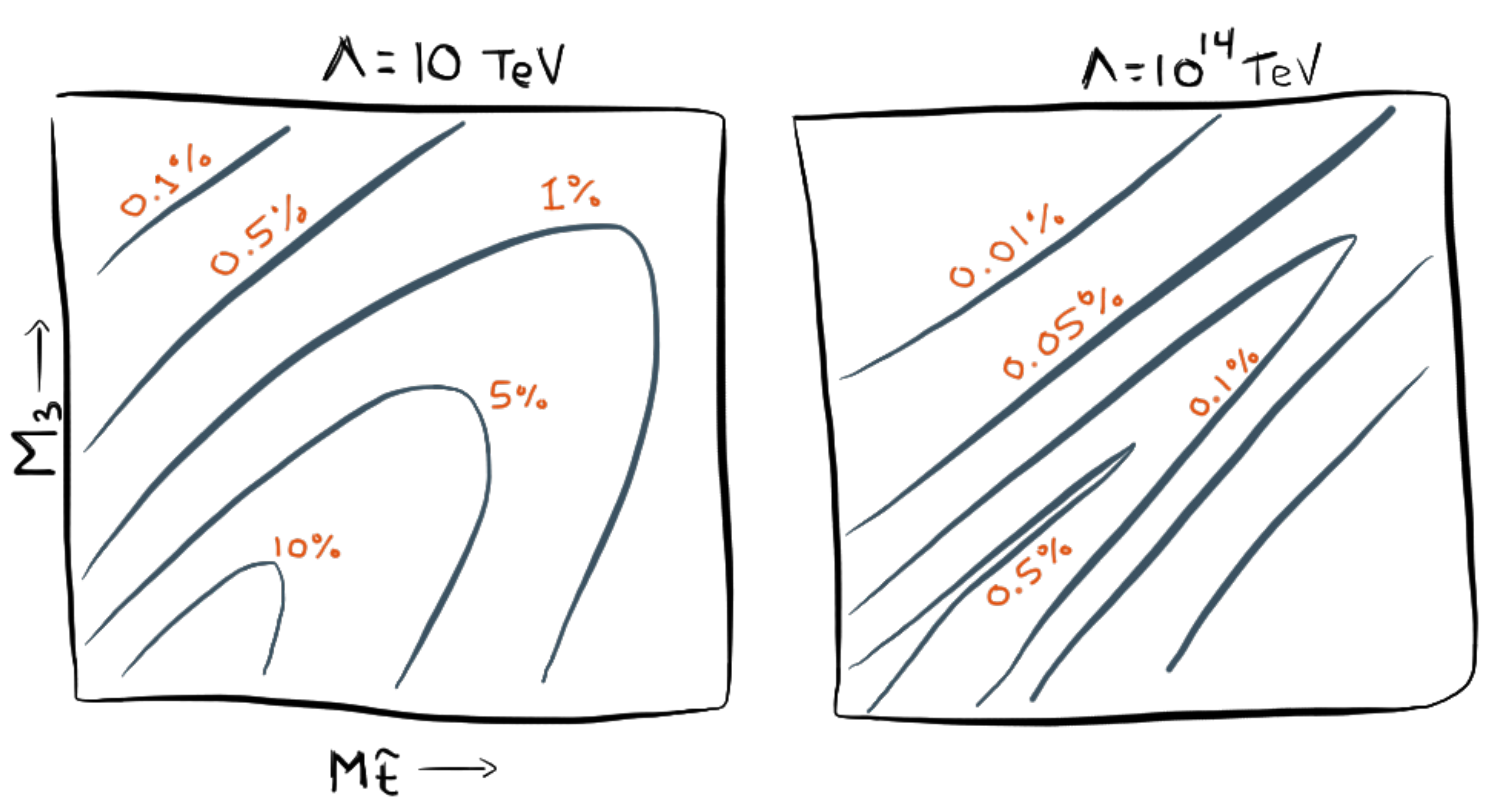} 
   \caption{Illustration of the correlative effects of RG evolution on the stop mass and gluino mass given even a small amount of RG evolution from the UV scale $\Lambda$. The contours denote fine-tuning as a function of the stop mass and gluino mass for $\Lambda = 10$ TeV (left) and $\Lambda = 10^{14}$ TeV (right). As $\Lambda$ is increased, RG evolution ties the gluino and stop masses tightly together, so that separating the mass scales from their RG relation amounts to a fine-tuning of the UV parameters. Adapted from a figure in \cite{Arvanitaki:2012ps}.}
   \label{fig:rg}
\end{figure}

With those caveats in mind, let's return to the simplest case of the MSSM with no interesting dynamics at low scales. We can take  (\ref{eq:ewsb}) at face value, and so naturalness morally crops up in three places:
\begin{itemize} 
\item The first is the tree-level potential, which involves certain combinations of soft masses that set the weak scale vev. At tree-level the naturalness of the weak scale implies something about the soft parameters $m_{H_u}^2$ and $\mu$, which itself controls the higgsino masses. This is usually construed as implying that the $\mu$ parameter is small, and hence higgsinos are light. The sensitivity associated with $\mu$ is $\Delta[\mu]\sim 2 \mu^2/m_Z^2$, so naturalness suggests $\mu \lesssim 200$ GeV and correspondingly light Higgsinos.\footnote{However, as mentioned above, one could also imagine a theory where $m_{H_u}^2 + |\mu|^2 \approx 0$ is natural despite both parameters being large. The simple fact is that in this limit $m_{H_u}^2 + |\mu|^2$ is the physical mass parameter of the doublet $H_u$ in the symmetric phase. So if some dynamics drives its mass to zero, an apparent cancellation could be affected. One could alternately introduce effective hard SUSY-breaking terms that separate the physical Higgsino mass from the $\mu$ parameter, in which case small $\mu$ would not imply light Higgsinos. So treat any claim that naturalness demands Higgsinos around 200 GeV with extreme caution!}
\item The second is immediate loop-level corrections. The soft mass parameter $m_{H_u}^2$ accumulates one-loop corrections from other soft parameters. A simple heuristic is that the sparticles are as important as their partner particles for this contribution. By far the largest is due to the stops, since the top chiral superfields couple most strongly to $H_u$, with correction of order \cite{Brust:2011tb}
\begin{equation}
\delta m_{H_u}^2 = - \frac{3 y_t^2}{4 \pi^2} m_{\tilde t}^2 \ln \left(\Lambda / m_{\tilde t} \right)
\end{equation}
 Naturalness required stops $\sim 400$ GeV with a cutoff $\Lambda \sim 10$ TeV. Other particles are also tied to naturalness, though less directly. After the SM top loop, the gauge and Higgs loops drive the mass corrections, so unsurprisingly the wino and higgsino corrections play a role, with \cite{Brust:2011tb}
 \begin{equation}
\delta m_{H_u}^2 = - \frac{3 g^2}{8 \pi^2} (m_{\tilde W}^2 + m_{\tilde h}^2) \ln \left(\Lambda / m_{\tilde W} \right)
\end{equation}
Having already bounded Higgsinos, for winos this translates to $m_{\tilde W} \lesssim$ TeV. Note that sbottoms need not be directly connected to naturalness, but since the left-handed sbottom gauge eigenstate transforms in the same $SU(2)$ multiplet as the left-handed stop gauge eigenstate, at least one sbottom is typically found in the same mass range as the stops.
\item The third is two-loop corrections, due to the naturalness of other sparticles. Of course, the degrees of freedom that control the Higgs mass must themselves have protected masses, so there is a naturalness problem for the remaining states. In certain cases this is quite stringent, specifically for states with color charge due to the large couplings and Casimirs involved.  The stop mass is corrected by the gluino mass due to the size of $g_3$, so it is hard to separate the gluino substantially from the stops, with e.g. \cite{Brust:2011tb}
\begin{equation}
\delta m_{\tilde t}^2 = \frac{2 g_s^2}{3 \pi^2} m_{\tilde g}^2 \ln \left(\Lambda / m_{\tilde g} \right)
\end{equation}
which ties $m_{\tilde g} \lesssim 2 m_{\tilde t}$. Indeed, these corrections typically tie the masses of the gluino and all squark flavors quite tightly given even a modest amount of running. This is illustrated in Fig.~\ref{fig:rg}, which sketches the fine-tuning associated with separating the stop mass from the gluino mass as a function of the cutoff $\Lambda$. In this respect certain two-loop corrections are more important for naturalness than one-loop.\footnote{Here I am being a little quick. There are two-loop contributions from the gluino directly to the $H_u$ soft mass that are important for naturalness but should not be interpreted as one-loop renormalization of the stop mass. However, they are of the same order, and -- unlike the above -- they do not link the mass scale of the gluino and stop as tightly.}
 Of course, this also rests on certain assumptions of minimality, about which more in a moment. 
\end{itemize}

\begin{figure}[htbp] 
   \centering
   \includegraphics[width=4in]{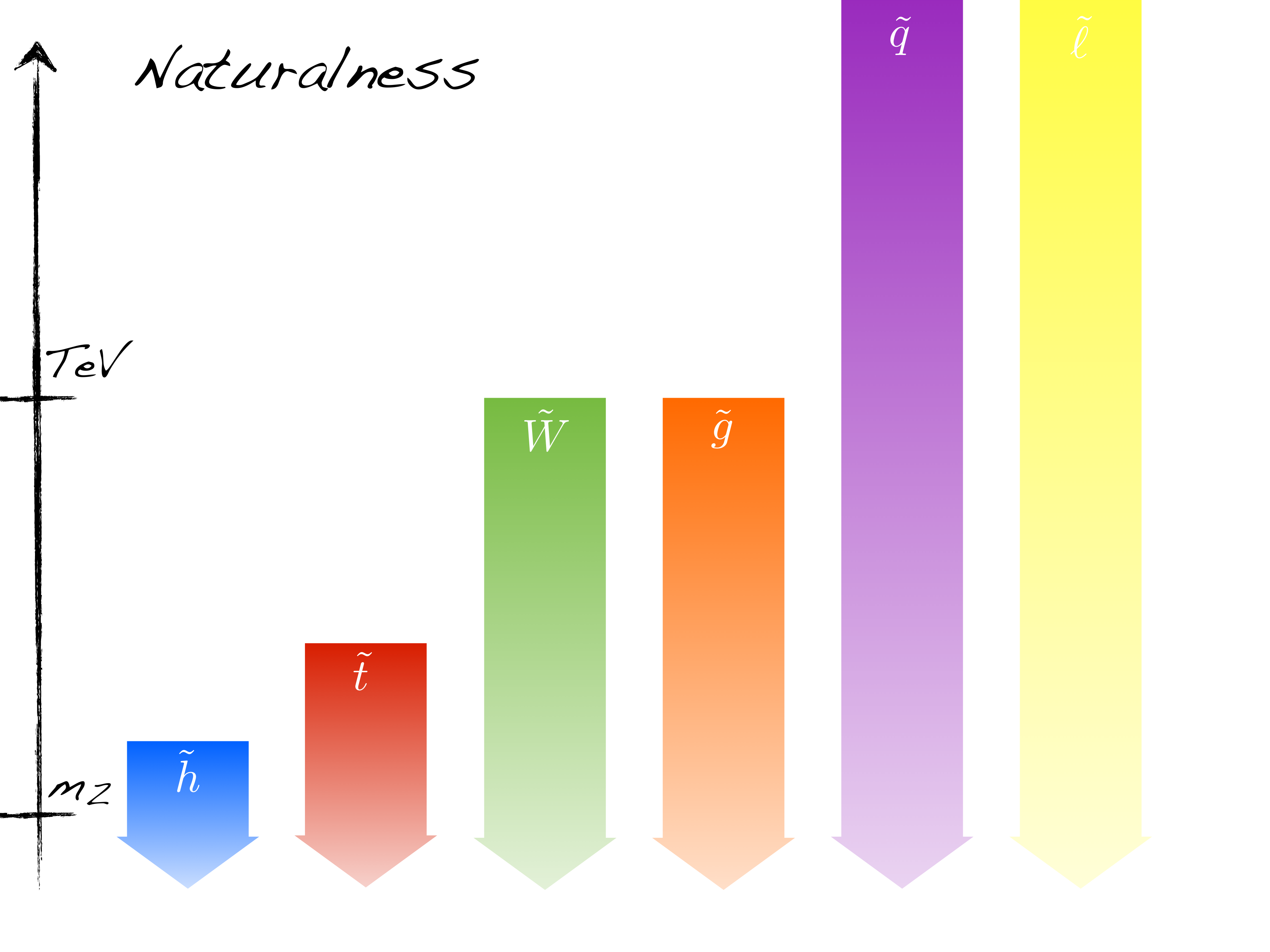} 
   \caption{Cartoon illustration of the mass scales for various sparticles dictated solely by electroweak naturalness with sensitivity parameter $\Delta \lesssim 10$. }
   \label{fig:nat}
\end{figure}

Taken together, the implications of naturalness lead to LHC expectations illustrated in Figure \ref{fig:nat}.

\pagebreak

\subsection{Parsimony}

\begin{quote}
{\it ``Numquam ponenda est pluralitas sine necessitate.''}\\
-William of Ockham\\
{\it ``Patients can have as many diseases as they damn well please.''} \\
-Hickam's Dictum\footnote{Thanks to Eva Silverstein for bringing this lovely bon mot to my attention.}
\end{quote}

Although not a quantitative principle, parsimony as a qualitative principle has played a key role in shaping model-building. The MSSM is, after all, the {\it minimal} extension of the Standard Model consistent with supersymmetry; in addition to the extension of all known particles into their corresponding supermultiplets, it comes with the minimal extension of the Higgs sector consistent with holomorphy \cite{Dimopoulos:1981zb}. This ties back to another definition of naturalness in the literature that predates radiative naturalness of the electroweak scale -- namely, that the number of fundamental parameters should be less than the number of physical parameters, leading to predictive relations among the physical parameters rather than ad hoc values \cite{Georgi:1974yw}. This earlier idea of naturalness arose as a philosophy tied to spontaneous symmetry breaking, where indeed the many parameters in the broken phase enjoyed predictive relations arising from the symmetry of the unbroken phase. This is certainly a well-motivated philosophy, especially in lieu of data. However, there is nothing intrinsic about the philosophy within the context of a given physical theory. If the combination of philosophy plus theory is incompatible with observation, we should consider the consequences of changing philosophies before we completely discard the theory itself.

The application of parsimony as a governing principle has led to a number of heuristic rules with respect to SUSY: 

\begin{itemize}
\item No fields are added beyond those required to promote the Standard Model to an anomaly-free $\mathcal{N} = 1$ supersymmetric gauge theory with chiral matter (hence the ``M'' in MSSM). 
\item The scale of SUSY breaking is high, perhaps tied to physics at the GUT scale or above. This implies many decades of RG running, so that RG correlations between parameters are strong.
\item The number of fundamental SUSY breaking parameters is small. So there are typically correlations between parameters even before RG effects are taken into account.
\item Colored sparticles are typically heavier than uncolored sparticles due to both of the above effects. 
\item The generations of the Standard Model are not differentiated by SUSY breaking, since any flavor structure in SUSY breaking requires considerable additional structure in order to remain consistent with limits on FCNC. The simplest models are universal, and have no bearing on flavor; the only flavor violation arises proportional to Standard Model yukawa couplings. (This is commonly called the  ``minimal flavor violation'' (MFV) ansatz \cite{D'Ambrosio:2002ex}.)
\end{itemize}

Taken together, this emphasis on parsimony leads to expectations that 
\begin{enumerate}
\item Colored sparticles lie at the top of the spectrum
\item The gluino and squarks are always approximately coincidental in mass
\item There is no substantial splitting between sfermion generations, apart from splittings generated radiatively proportional to the observed Standard Model flavor hierarchy.
\end{enumerate}

\begin{figure}[htbp] 
   \centering
   \includegraphics[width=4in]{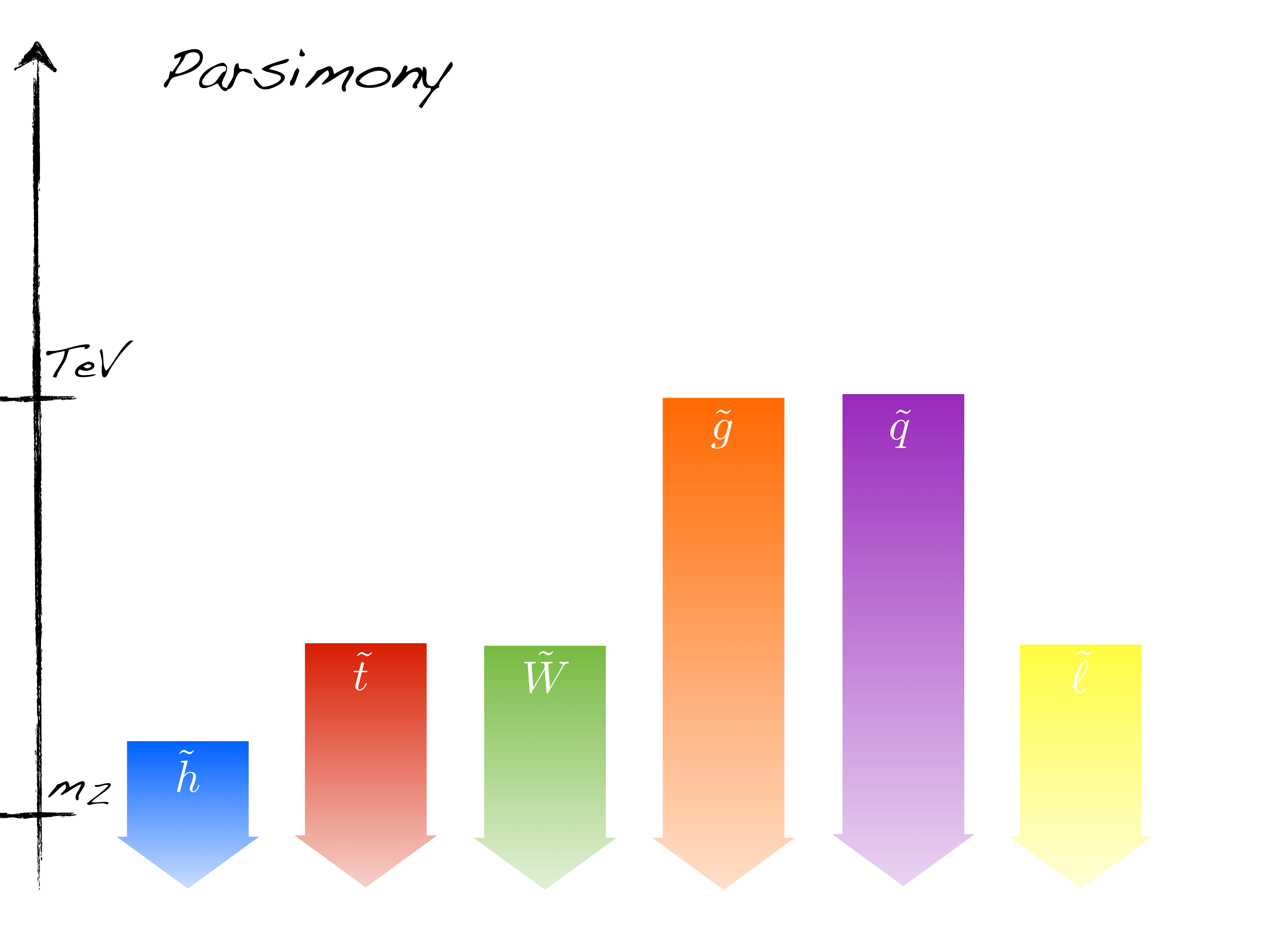} 
   \caption{Cartoon illustration of the mass scales for various sparticles dictated by the combination of parsimony and electroweak naturalness with sensitivity parameter $\Delta \lesssim 10$.}
   \label{fig:par}
\end{figure}

If you then take the input from parsimony, the expectations for SUSY at the LHC change substantially from the expectations driven strictly by naturalness, as illustrated in Figure \ref{fig:par}. 

Parsimony also favored four interconnected observations/assumptions:

\begin{enumerate}
\item The added matter content (in particular, the added higgsinos) in the MSSM improves the prediction for gauge coupling unification relative to that of the Standard Model. Ultimately this is not a terribly precise guide -- sparticles at $\sim 10$ TeV serve quite well for unification, and indeed do somewhat better than sparticles at $\sim 1$ TeV.
\item The added matter content allows new baryon- and lepton-number violating operators, leading to prohibitive rates for proton decay. The most parsimonious way of dealing with these operators up to dimension 5 is to impose $R$-parity. We'll discuss this in more detail later in these lectures.
\item Given the imposition of $R$-parity, the MSSM could provide a natural dark matter candidate in the form of the lightest $R$-odd particle, provided it's uncharged. As with unification, this was not a sharp guide to scales, as the mass scale favored by standard thermal abundance could range as high as several TeV depending on the DM candidate and the rest of the spectrum.
\item The stability of the lightest $R$-odd particle guarantees that SUSY decay chains end with pairs of the LSP which, if neutral, translates into significant missing energy signatures.
\end{enumerate}

These additional observations have done much to shape our expectations for both spectra and signals at the LHC.

\subsection{Summary}

So broadly speaking, before the LHC turned on, the above principles and observations implied
\begin{itemize}
\item Light sparticles, as close to the weak scale as possible. If not all sparticles, at the very least stops around 400 GeV, higgsinos around 200 GeV, and probably gluinos beneath a TeV. The presence of colored states guarantees considerable rates of production.
\item From parsimony, we typically expect all other colored scalars around the mass of the stops, perhaps slightly heavier due to RG effects or finite threshold effects.
\item Generically lighter uncolored sparticles, especially sleptons, which can be produced in cascade decays or directly.
\item A stable neutral LSP, potentially a good dark matter candidate. Implies SUSY processes end in the LSP and are therefore distinguished by significant missing energy. 
\end{itemize}

Taken together, these ideas -- substantially naturalness and parsimony -- told us to expect a wealth of new particles accessible at even the 7 and 8 TeV collision energies of Run 1. Some were directly related to naturalness, others tied by implications of parsimony.

In what follows, I will largely address how these expectations have fared in light of LHC and ancillary data. Subsequently, we'll turn to ways in which these principles can be systematically upended. The moral will be that, barring missing subtleties, the LHC has largely falsified supersymmetric models governed by the twin pillars of parsimony and naturalness. However, discarding either principle opens a panoply of interesting possibilities consistent with data.

\pagebreak

\part{What do we know?} 

\section{Direct Limits}

\begin{figure}[htbp] 
   \centering
   \includegraphics[width=5in]{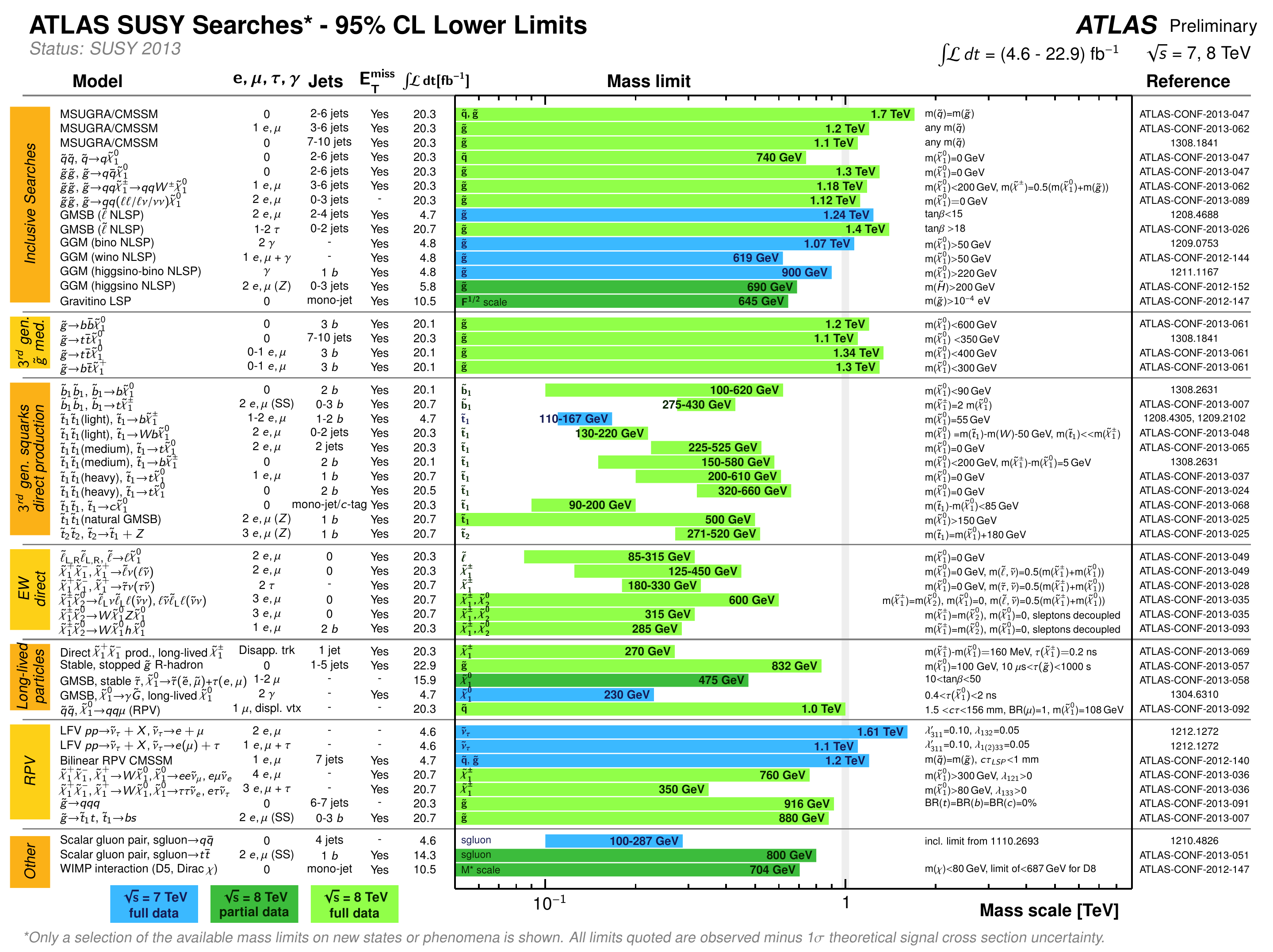}
   \caption{ATLAS summary of SUSY limits as of SUSY 2013 \cite{ATLAS_SUSY}.}
   \label{fig:atlas}
\end{figure}

\begin{figure}[htbp] 
   \centering
   \includegraphics[width=5in]{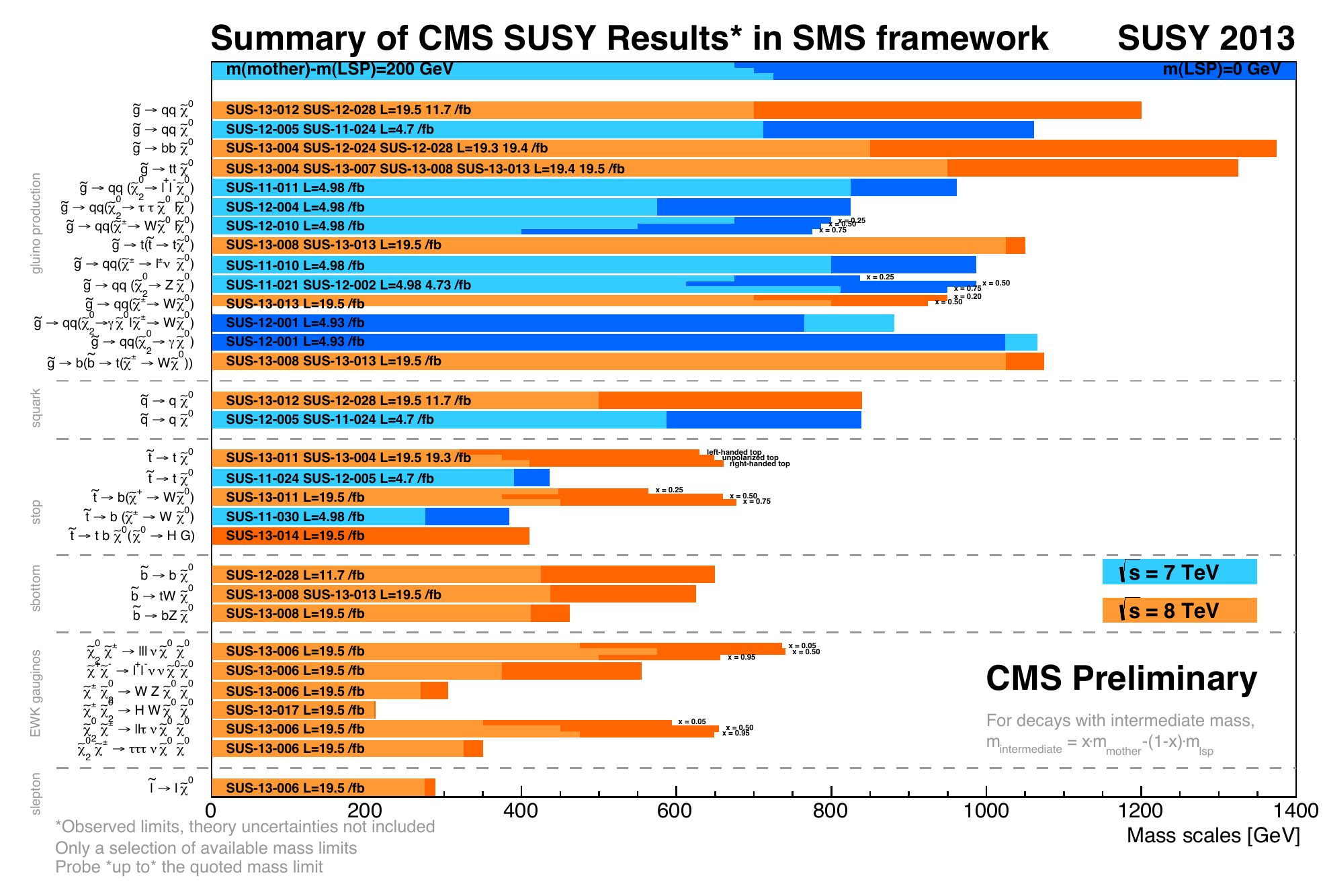}  
   \caption{CMS summary of SUSY limits as of SUSY 2013 \cite{CMS_SUSY}.}
   \label{fig:cms}
\end{figure}

So we turned on the LHC with these expectations in mind, and proceeded to look in all of the channels that seemed motivated.

Historically, SUSY limits were presented in a model-dependent framework. This maximized sensitivity by leveraging all available signal modes, but made reinterpretation challenging. One of the great developments at the Tevatron and LHC has been the presentation of SUSY limits in terms of simplified models, which focus on a single production mode and a one-step (or occasionally two-step) decay chain to the relevant final states. Of course, simplified models can't capture all details, and can't always be mapped onto the important features of a given model. But they provide an extremely useful way of setting limits, and can often be easily re-interpreted. So for clarity, to understand the state of play in these lectures, I'll focus on individual topologies in the context of simplified models. We can of course improve limits by taking combinations of particles, but individual topologies gives a nice sense of the irreducible limits on the parameter space.

ATLAS and CMS both provide helpful summary plots that qualitatively represent the mass reach of limits in various channels, subject to simplifying assumptions about the masses and branching ratios involved, shown in Figures \ref{fig:atlas} and \ref{fig:cms} \cite{ATLAS_SUSY, CMS_SUSY}.

\begin{figure}[htbp] 
   \centering
   \includegraphics[width=4in]{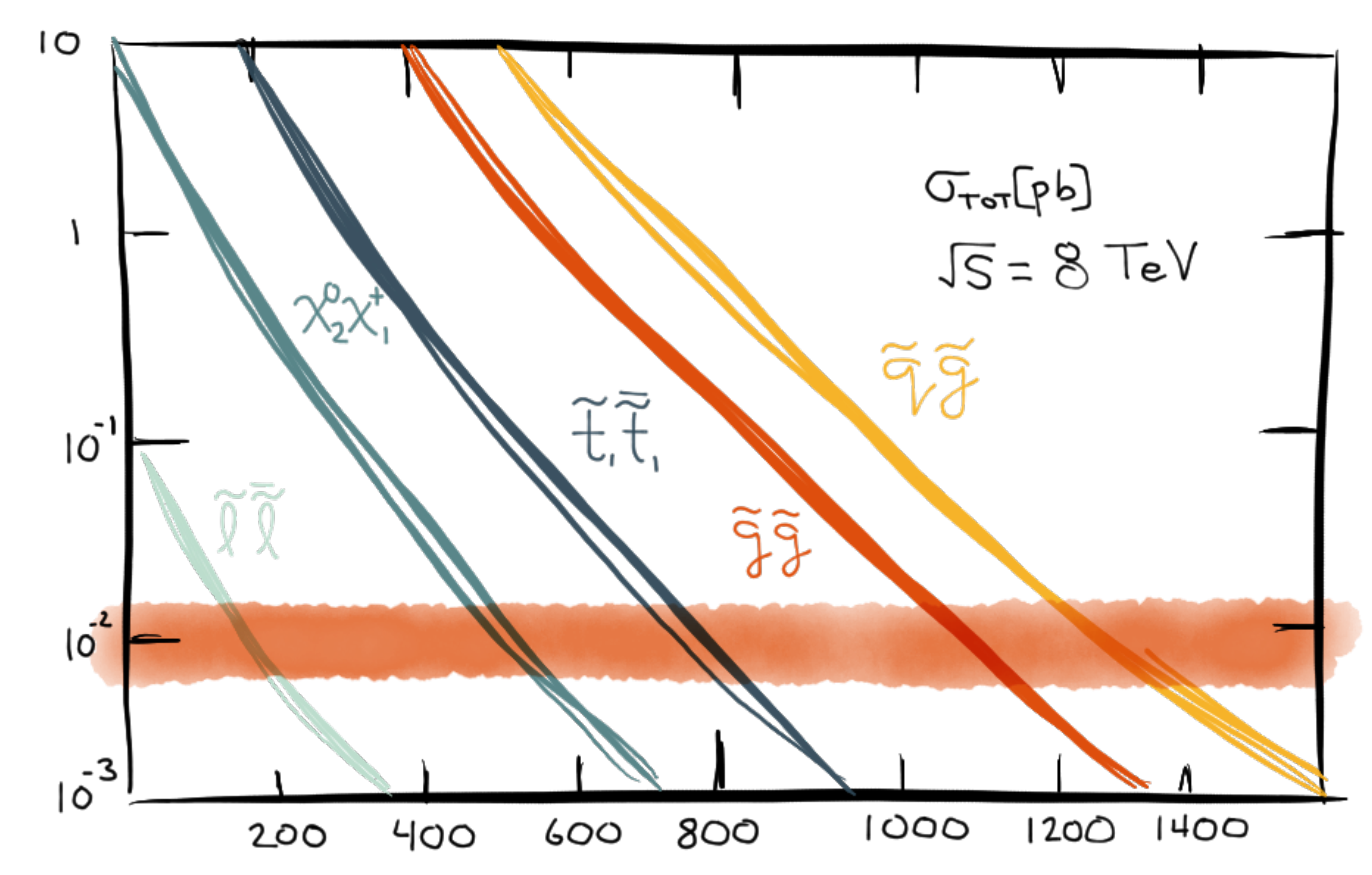}  
   \caption{Heuristic cross section exclusion. The diagonal lines correspond to the production cross sections for various SUSY processes, while the horizontal red band corresponds to 10 fb$^{-1} \pm$ few. For each process, the intersection of the production cross section and the sensitivity band tells you the current mass scale probed at the LHC. Provided the spectrum is not compressed, the current sensitivity is $\sigma \cdot {\rm Br} \sim 10$ fb across the board with improvements in distinctive final states, while with compressed MET it's $\sim$ 1 pb. Cartoon inspired by the Prospino2 propaganda plot \cite{prospino}.}
   \label{fig:heuristic}
\end{figure}

However, mass is not really the most useful variable; there is more or less an even sensitivity to cross section, with some modulation depending on the final state. Sensitivity to colored sparticles is on the order of $\sigma \cdot {\rm Br} \sim 10$ fb. This corresponds to event counts on the order of 100-200 events so far. Sensitivity to sleptons is slightly better, $\sim$ few fb, due to the favorable kinematics of two-body decays to lepton plus MET. Sensitivity to electroweakinos is of the order $\sigma \cdot {\rm Br} \sim 100$ fb for decays to $W,Z$, but back to $\sim 10$ fb if you are given at least two leptons from the gauge bosons or if intermediate sleptons are involved. This sensitivity degrades in compressed corners of parameter space to $\sigma \cdot {\rm Br} \sim 1$ pb. So instead of trying to keep all these mass numbers in your head, you can understand limits heuristically as the intersection of cross sections with this approximate ballpark sensitivity. I've tried to illustrate this in Fig.~\ref{fig:heuristic}.

Let's discuss various states and their relevant search modes in turn.

\subsection{Stops}

The stop is the essential particle for supersymmetric naturalness at the LHC, and so much effort has focused on constraining stops directly. Their production rates are highly suppressed relative to other colored sparticles at the LHC, and their final states are often challenging to distinguish from $t \bar t$ backgrounds. This poses a two-fold challenge for searches at the LHC.

\begin{figure}[htbp] 
   \centering
   \includegraphics[width=5in]{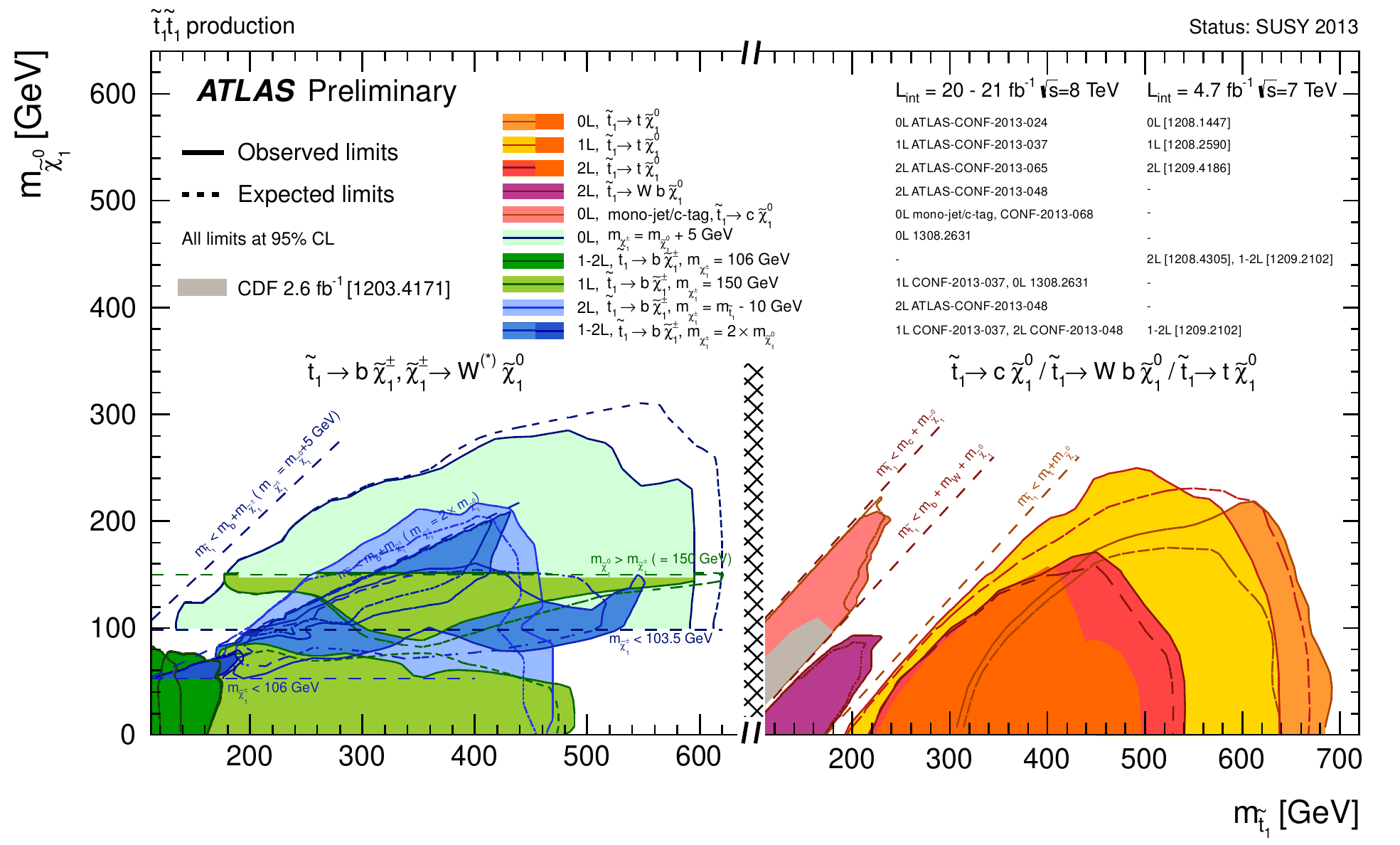} 
   \includegraphics[width=3in]{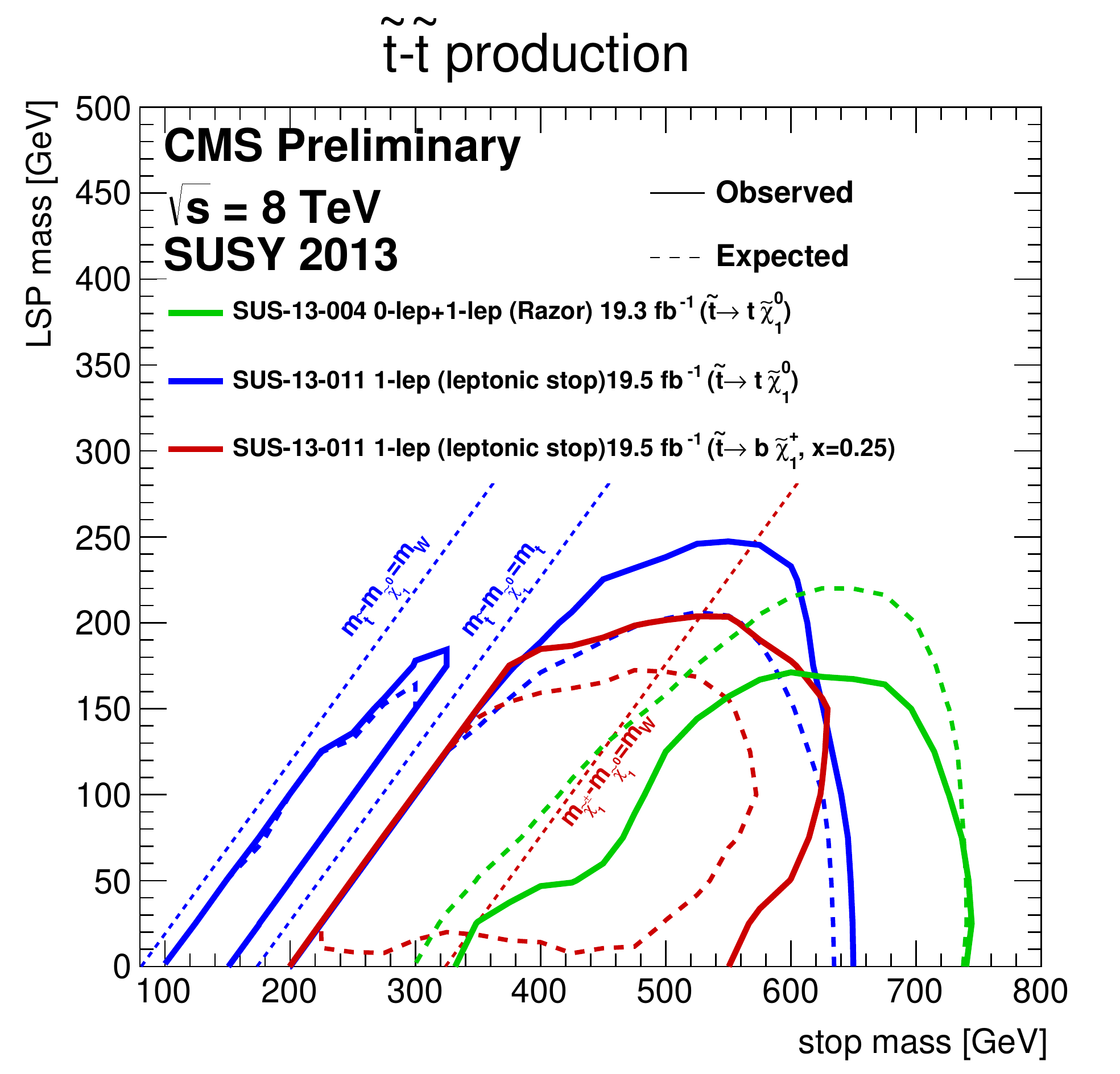} 
   \caption{Current stop limits from ATLAS (top) \cite{ATLAS_STOP} and CMS (bottom) \cite{CMS_STOP}.}
   \label{fig:stops}
\end{figure}

Direct production of stops proceeds primarily through s-channel gluon and $t$-channel stop exchange. Stop production through $q \bar q$ initial states is $p$-wave suppressed since the final state needs to carry angular momentum but the stops have no spin. This leads to a rate $\propto \beta^3$ near threshold. The direct production cross section for stops ranges from 10 pb down to 1 fb  at 8 TeV as the stop mass is varied between 200-900 GeV.  

On the decay side, there are two primary modes assuming that other colored states are kinematically decoupled: $\tilde t \to t \chi_1^0$ and $\tilde t \to b \chi^\pm \to b W^\pm \chi_0^1$. LHC searches typically focus on the semileptonic decay mode, with one $W$ going hadronically and the other leptonically, plus tagging on the two $b$ jets. This final state is $4j + \ell + MET$, with various search regions involving 0, 1, 2 $b$-tags. However, other decay products of the $W$'s provide complementary kinematic coverage, so various final states of the gauge bosons are probed. The current limits from ATLAS and CMS are shown in Fig.~\ref{fig:stops}, including results from \cite{Aad:2012ywa, Aad:2012xqa, Aad:2012uu, ATLAS-CONF-2013-037, ATLAS-CONF-2013-024,  ATLAS-CONF-2013-065,ATLAS-CONF-2013-068, ATLAS-CONF-2013-048, Aad:2012tx, Aad:2012yr, ATLAS-CONF-2013-053}  and \cite{CMS-PAS-SUS-13-004, Chatrchyan:2013xna}.

Different kinematic search regions are defined by the splittings available. In the case of $\tilde t \to \chi_1^0$, one can look for final states for both on-shell and off-shell $t$. For $\tilde t \to b \chi^\pm$, LHC searches currently look for final states with on-shell $W$. 
Note that the detailed sensitivity in the final state depends somewhat on the polarization of the decay modes, which in turn depends on the properties of the stops (i.e., the admixture of RH and LH states) as well as the composition of the electroweakinos involved in the decay modes. This changes the angular distributions of decay products and therefore the efficiency of signal discrimination.  This typically leads to variations on the order of tens of GeV in the limit setting.

The current reach is out to 650 GeV, corresponding to cross sections on the order of 10fb. The generic tuning associated with this bound is about $\Delta \sim 20$.

The most challenging region is the one where $\tilde t \to t \chi_1^0$ dominates but $m_{\tilde t} \sim m_t + m_{\chi_1^0}$, in which case the signal is essentially degenerate with $t \bar t$ and very difficult to distinguish from background. Similar challenges arise for the other topology when $m_{\tilde t} \sim m_b + m_{\chi_1^\pm}$ and $m_{\chi^\pm} \sim m_{\chi_1}^0$, leaving the event with very little MET and soft leptons. 

Kinematic reach will simply improve with increased center of mass energy. As for the squeezed regions, there are a variety of sensitive techniques developed by theorists to probe the kinematically squeezed regimes and unfavorable combinations of polarizations in the stops and their decay modes. I won't discuss those in detail here.

\subsection{Sbottoms}

Although the sbottom does not necessarily play a strong role in naturalness, the mass of $\tilde b_L$ is typically close to that of $\tilde t_L$ since the two transform as an electroweak doublet and must acquire the same soft mass. This does not necessarily imply that sbottoms are in the same mass region as stops, but in many models they are correlated.

Sbottom searches are essentially the complement of stop searches. The production modes and rates are similar, with slight relative enhancement due to electroweak corrections. The decay modes are the natural complement, e.g., the primary mode is $\tilde b \to b \chi_1^0$, as well as $\tilde b \to t \chi^\pm \to t W^\pm \chi_1^0$. One also can look for the process $\tilde b \to b \chi_2^0 \to b Z \chi_1^0$. This topology requires an additional neutralino.

\begin{figure}[htbp] 
   \centering
   \includegraphics[width=2.8in]{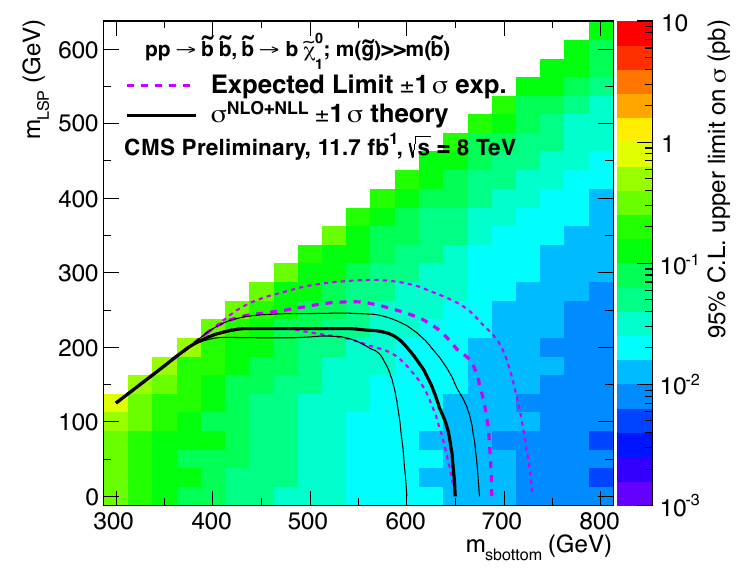} 
   \includegraphics[width=2.8in]{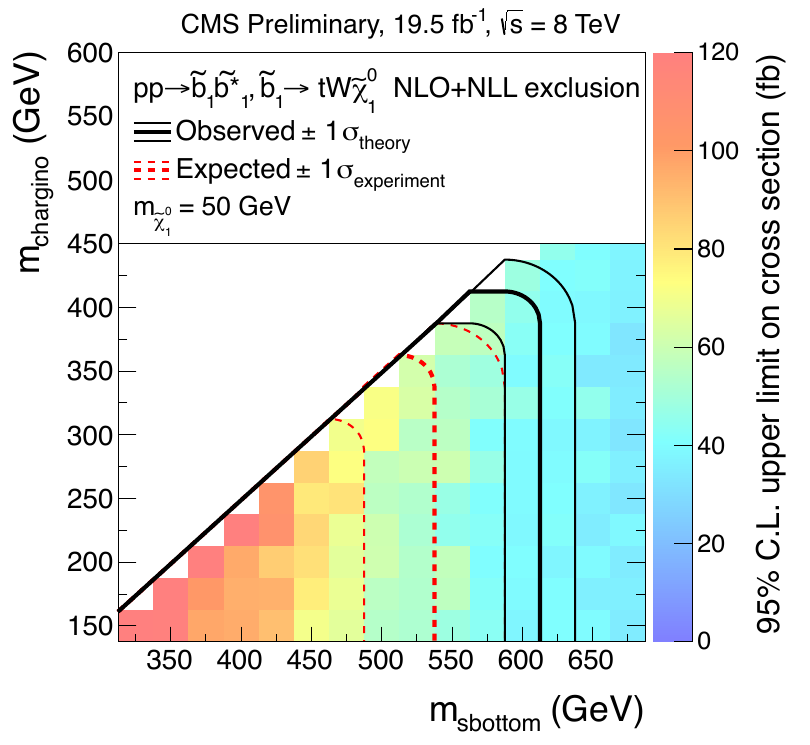} 
   \caption{Current sbottom limits from CMS \cite{Chatrchyan:1527115, CMS-PAS-SUS-13-008}; ATLAS limits are similar.}
   \label{fig:sbottoms}
\end{figure}

The first process $\tilde b \to b \chi_1^0$ is looked for in purely hadronic states using 1-2 $b$ tags plus missing energy.
The other processes can be efficiently probed using trileptons plus one or more b-tagged jets, given the high multiplicity of $W$ and $Z$ bosons in the final state. The primary decay mode has four $W$ bosons, while the alternate decay mode has two $Z$ bosons, and in conjunction with $b$ tags this provides considerable sensitivity. Current CMS limits from \cite{Chatrchyan:1527115, CMS-PAS-SUS-13-008} are shown in Fig.~\ref{fig:sbottoms}; ATLAS limits are similar.

Ultimately, the mass reach in these various channels is comparable to that of stops. This sensitivity corresponds to cross sections on the order of 10fb. There is no direct tuning associated with this, though one expects $\tilde b_L \sim \tilde t_L$.

\subsection{Gluinos}
Gluinos are one of the driving forces of supersymmetric signals at the LHC, given their considerable production cross section and radiative connection to squark masses. The pair production cross section is approximately two orders of magnitude larger than that of stops, ranging from 10pb -1fb at 8 TeV for gluinos between 400 and 1300 GeV.

\begin{figure}[htbp] 
   \centering
   \includegraphics[width=3in]{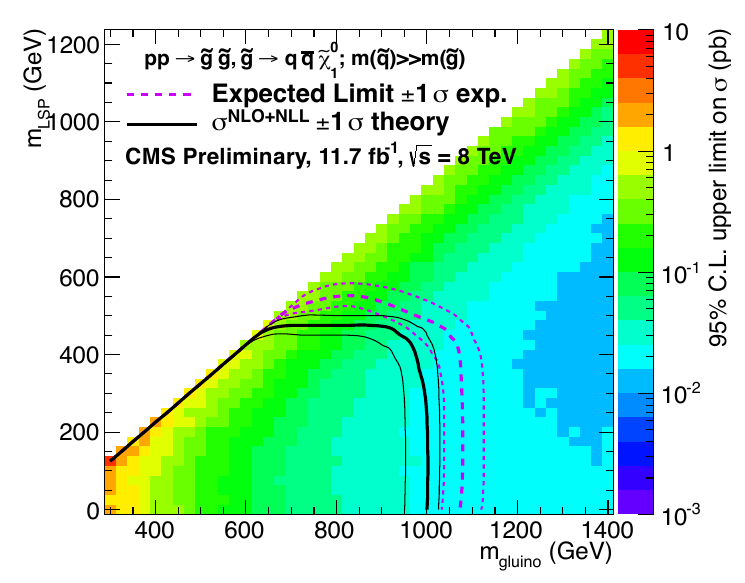} 
   \includegraphics[width=2.5in]{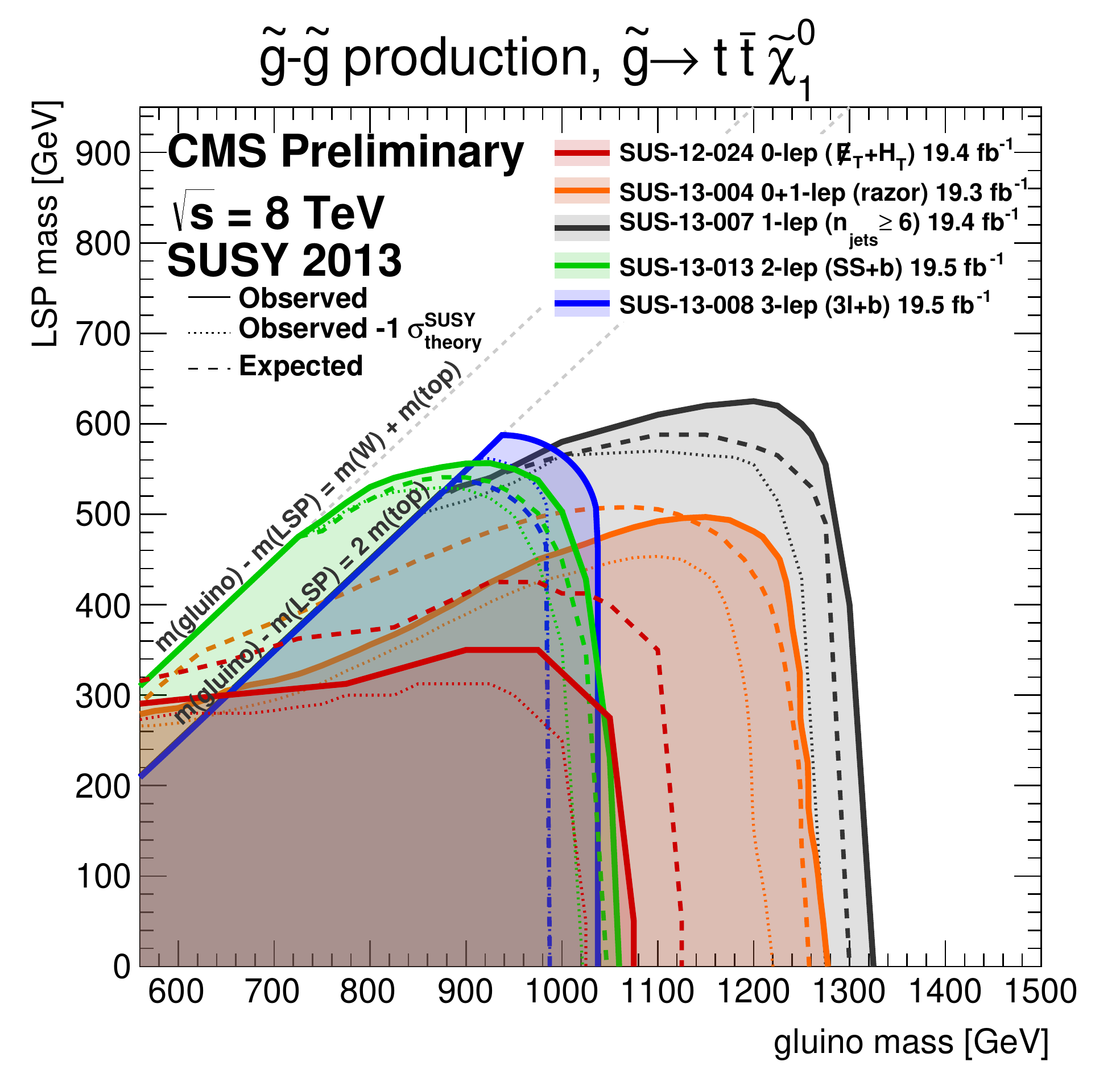} 
   \caption{Current CMS limits on gluinos \cite{CMS_GLUINO, Chatrchyan:2013lya}; ATLAS limits are similar.}
   \label{fig:gluinos}
\end{figure}

First, we can consider ``pure'' gluino limits, under the assumption that squarks are significantly heavier. In this case, the gluino decay occurs primarily into three-body final states involving off-shell intermediate squarks, $\tilde g \to q \bar q \chi_1^0$. If the lightest squark is third-generation, then the quarks are predominantly third-generation, $\tilde g \to t \bar t \chi_1^0$ or $\tilde g \to b \bar b \chi_1^0$. Representative limits from CMS are shown in Fig.~\ref{fig:gluinos}  \cite{CMS_GLUINO, Chatrchyan:2013lya}; ATLAS limits are similar, with somewhat greater mass reach due to differences in the search procedure and background characterization. Note that these limits assume the decays of the gluino are prompt. If the intermediate squarks are sufficiently heavy, the gluino may become long-lived on collider timescales. In this case it forms a quasi-stable bound state, called an $R$-hadron, with correspondingly spectacular signatures that are probed in different ways.

In the case of decays proceeding through light quarks, the final states are distinguished by high jet multiplicity, $\gtrsim 4j$, and in the case of decays through heavy flavor, this jet multiplicity is supplemented by a large number of $b$-tags. Searches are simply designed for missing energy and high jet multiplicity. At low mass, the cross section is sufficiently large that even reduced amounts of MET provide sensitivity, but at higher masses this sensitivity plateaus. For the generic light-flavor case, limits extend out to $\sim 1$ TeV. Limits are similar for the heavy flavor case with tops, albeit with reduced sensitivity in kinematically squeezed regimes, but improvements when searching for leptons. Leptonic final states have reach out to $\sim 1.2$ TeV. The limits for heavy flavor with bottoms extend out to $\sim 1.2$ TeV, since the kinematics are more open and the $b$-tags improve efficiency.  This sensitivity corresponds to cross sections on the order of 10 fb, and somewhat smaller for heavy flavor.

\subsection{Squarks}

First- and second-generation squarks are a different matter entirely compared to stops and bottoms, since their production cross sections benefit from direct $q \bar q$ contributions. The squark pair production cross section is comparable to the gluino pair production cross section. 

\begin{figure}[htbp] 
   \centering
   \includegraphics[width=4in]{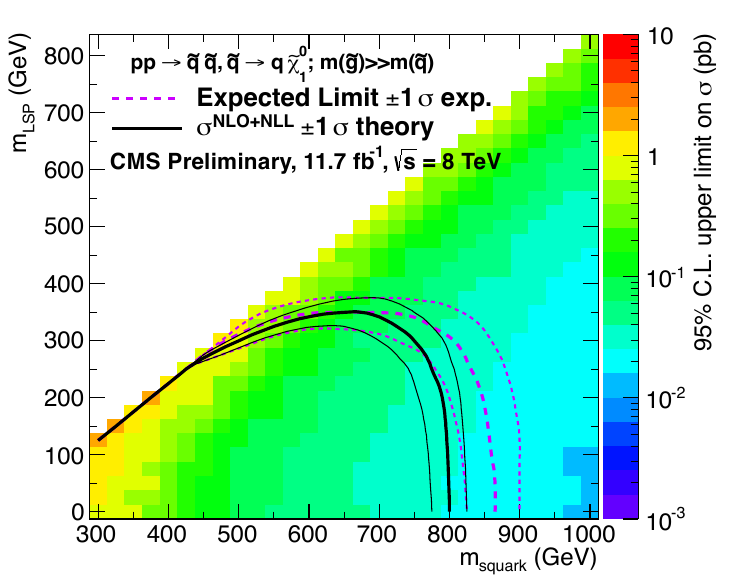}
     \caption{Current CMS squark limits \cite{ Chatrchyan:2013lya}; ATLAS limits are similar.}
   \label{fig:squarks}
\end{figure}

The primary mode here when other states are decoupled is $\tilde q \tilde q \to q \chi_1^0 q \chi_1^0$, and so efficiently searched for in final states with $\geq 2$ jets and MET. Direct limits, shown in Fig.~\ref{fig:squarks} \cite{ Chatrchyan:2013lya}, lie around 800 GeV at both ATLAS and CMS, a bit weaker than 10fb due to the low-multiplicity final state. There is no direct tuning associated with this limit. However, if all squarks were around the same mass, it would imply a tuning on the order of $\Delta \sim 30$.

\begin{figure}[htbp] 
   \centering
   \includegraphics[width=4in]{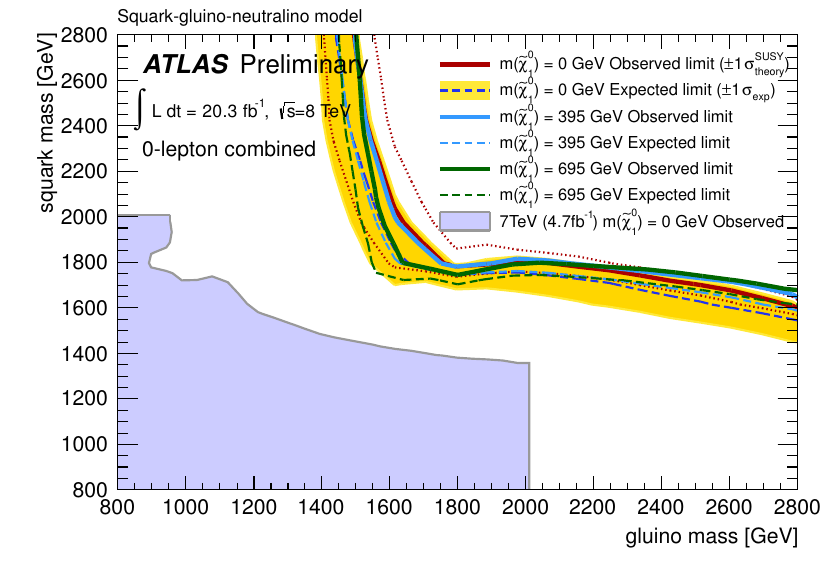}
     \caption{Current ATLAS limits on a squark-gluino simplified model \cite{ATLAS-CONF-2013-047}. CMS limits are similar.}
   \label{fig:squarkgluino}
\end{figure}

Also of interest are scenarios where squarks and gluinos are of comparable mass, as is typically the case in mSUGRA-inspired scenarios. In this case, squark-gluino associated production is available, with a cross section nearly an order of magnitude larger than gluino pair production. This added source of cross section, combined with production in various modes and the multiplicity of available final states, leads to sensitivity substantially above separate squark or gluino scenarios. Current ATLAS limits on a squark-gluino simplified model are shown in Fig.~\ref{fig:squarkgluino} \cite{ATLAS-CONF-2013-047}.

If squarks, including stops, were all of the same mass and similar in mass to the gluino, this would imply a tuning of $\Delta \sim 100$. Thus we begin to see clearly how spectra dictated by parsimony are under increasing tension from the LHC.

\subsection{Electroweakinos}

\begin{figure}[htbp] 
   \centering
   \includegraphics[width=3in]{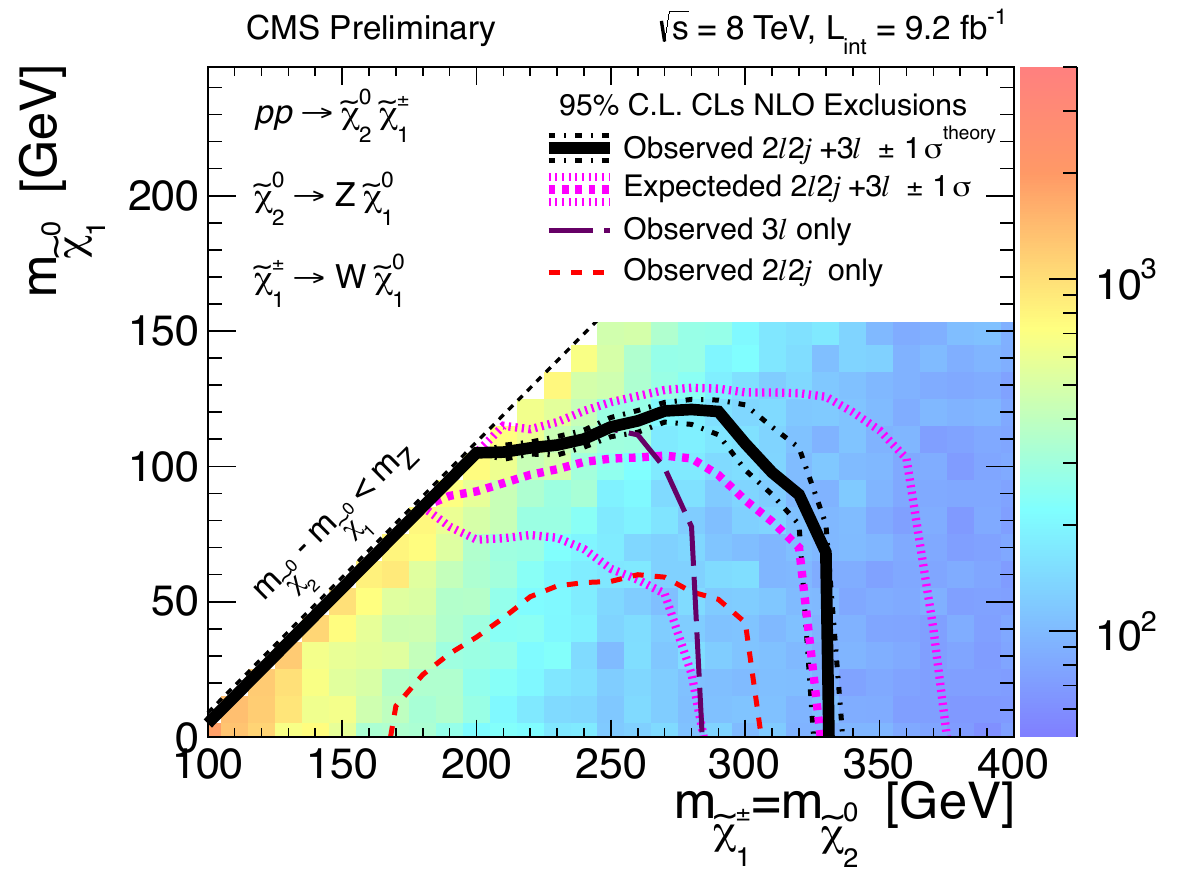}
     \caption{Current electroweakino limits from CMS \cite{CMS-PAS-SUS-13-006}; ATLAS results are similar.}
   \label{fig:ewk}
\end{figure}

One of the most impressive and exciting developments (to me, at least) has been the improving sensitivity to pure electroweak production of electroweakinos, with processes such as $\chi_2^0 \chi^\pm \to Z \chi_1^0 W \chi_1^0$ beginning to exceed LEP limits in mass reach.  These searches are challenging due to the substantial irreducible background from SM diboson production, but can be effectively searched for using leptonic final states and refinements involving the flavor and charge properties of the leptons. The most effective channels are trileptons including one opposite-sign, same-flavor pair reconstructing a $Z$ boson, as well as $2 \ell 2 j$ final states where the leptons again reconstruct a $Z$, with limits shown in Fig.~\ref{fig:ewk} \cite{CMS-PAS-SUS-13-006}.

Sensitivity to $WZ$ + MET is of order 100 fb, but including leptonic BR, sensitivity is of order 10 fb on par with other searches. As illustrated in Fig.~\ref{fig:ewk}, this pushes out to $\sim 325$ GeV, which entails a tuning of $\Delta \sim 25$.

\begin{figure}[htbp] 
   \centering
   \includegraphics[width=6in]{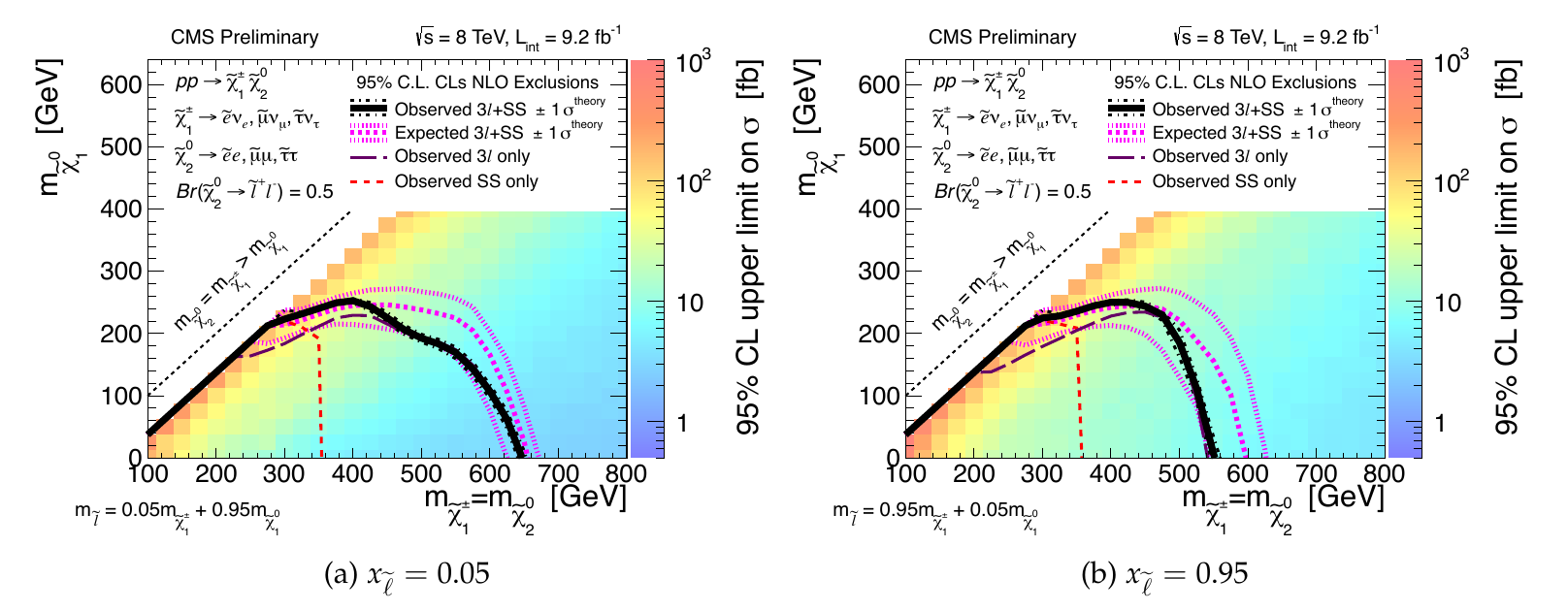}
     \caption{Current CMS electroweakino limits for spectra with light sleptons \cite{CMS-PAS-SUS-13-006}; ATLAS results are similar.}
   \label{fig:ewkslepton}
\end{figure}

One can also search for more optimistic scenarios where the electroweakinos decay through sleptons, yielding additional leptons in combinations that populate channels with smaller SM backgrounds. In general, trilepton final states without a $Z$ boson are relatively rare in the SM, so sensitivity to these scenarios is good and the cross section reach is at the $\sim$ few fb level. This is illustrated cleanly in Fig.~\ref{fig:ewkslepton} \cite{CMS-PAS-SUS-13-006}.

The considerable hole in current searches at the LHC is to the pair production of charginos \cite{Curtin:2012nn}. Production and decay of $\chi^+ \chi^- \to W^+ W^- +\chi_1^0 \chi_1^0$ is extremely challenging to search for given the large irreducible $WW$ background at the LHC. At present the only genuine direct limit is set by ATLAS, which is $2-3$ times the theory cross section below 200 GeV and then worsens to 5 times the theory cross section by 250 GeV due to the falling rate \cite{ATLAS-CONF-2013-049}. This is the one final state for which limits have not improved relative to LEP.

\subsection{Sleptons}

\begin{figure}[htbp] 
   \centering
   \includegraphics[width=3in]{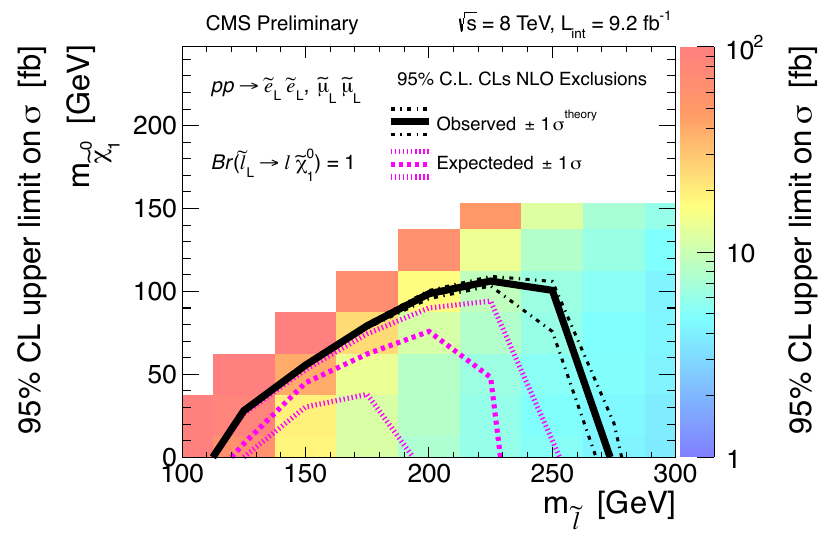}
     \includegraphics[width=2.5in]{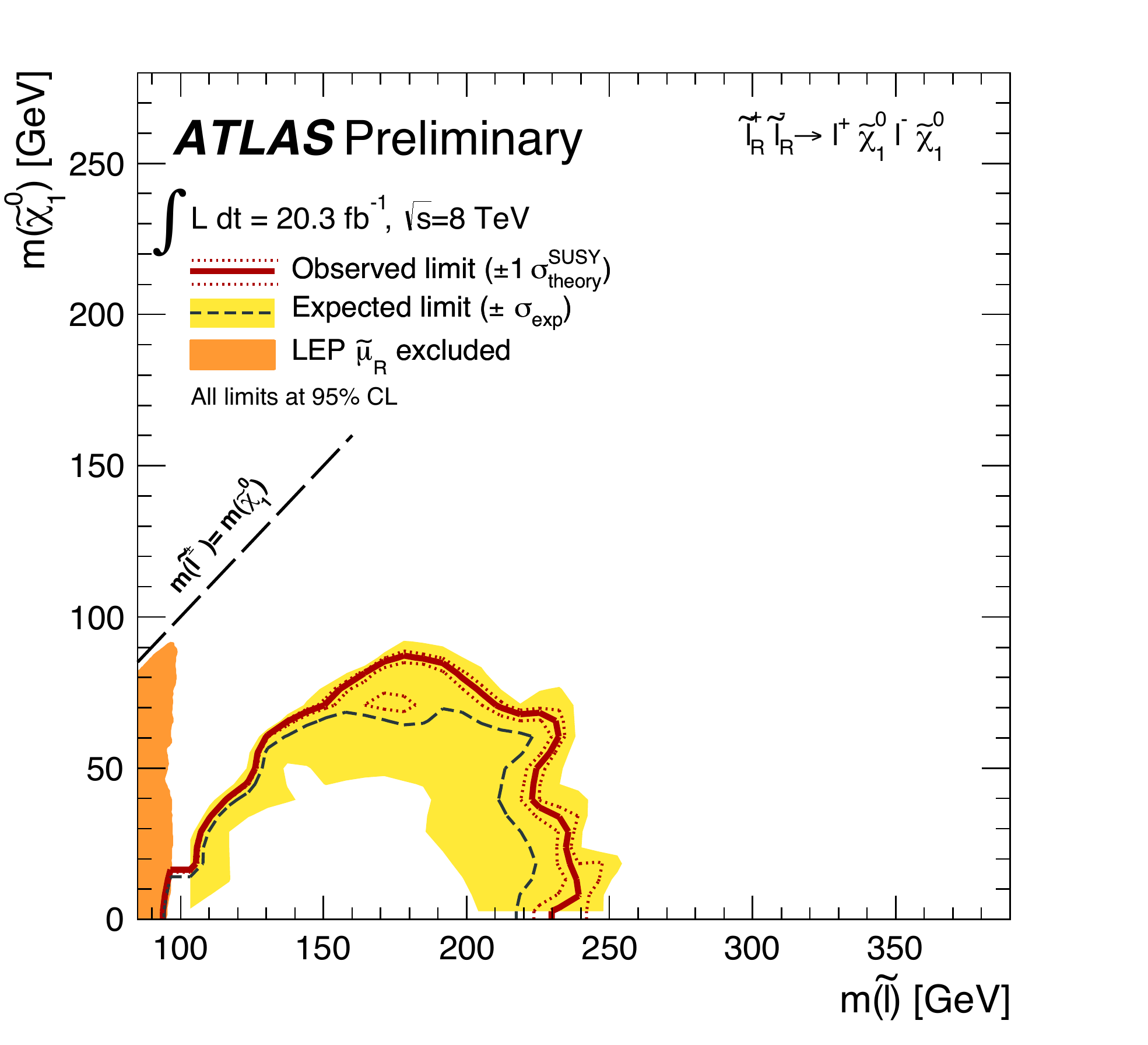}
     \caption{Left: Current CMS limits on pure left-handed slepton pair production \cite{CMS-PAS-SUS-13-006}. Right: Current ATLAS limits on pure right-handed slepton pair production \cite{ATLAS-CONF-2013-049}.}
   \label{fig:sleptons}
\end{figure}

Amazingly, we now have direct limits on the pair production of sleptons purely through electroweak processes, now considerably exceeding the LEP direct limit. As you can see in Fig.~\ref{fig:sleptons} \cite{CMS-PAS-SUS-13-006}, the latest CMS results can exclude left-handed first- and second-generation sleptons out to 275 GeV, where the pair production cross section is of order $\sim$ few fb. There is no tuning associated with this. Note that these exclusions are only for left-handed sleptons; right-handed sleptons possess much smaller production cross sections and are correspondingly more challenging to probe at this stage. However, impressively, ATLAS has recently set limits on the right-handed case \cite{ATLAS-CONF-2013-049} by maximally leveraging the stransverse mass variable $m_{T2}$. This demonstrates the considerable power of discriminating kinematic variables for such clean final states.

You might think that you could leverage the slepton final state (OSSF lepton pair plus MET) to set competitive limits on chargino pair production, but that is far from true. For instance, in the CMS slepton search, sensitivity to slepton pair production is obtained by using a variable called $M_{CT \perp}$, a variable that peaks around the mass difference between a visible parent and invisible LSP in pairwise two-body decays. This variable allows discrimination from the $WW$ background. However, if the signal is $W$-like, as in the case of charginos, then $M_{CT \perp}$ is not a useful variable and sensitivity vanishes. 

\subsection{Additional Higgs states}

The MSSM is a well-known example of a 2HDM with couplings fixed by holomorphy and gauge invariance to be Type 2. This suggests that the light CP-even state we've discovered is not necessarily aligned exactly with the EWSB vacuum condensate. I'll discuss the implications of Higgs couplings later, but for now we can discuss the implications of direct searches for the extra Higgs states. There is the CP even neutral Higgs $H$, the pseudoscalar $A$, and the charged Higgs $H^\pm$; these typically have comparable masses assuming some parametric separation from the state at 125 GeV. 

\begin{figure}[htbp] 
   \centering
   \includegraphics[width=3in]{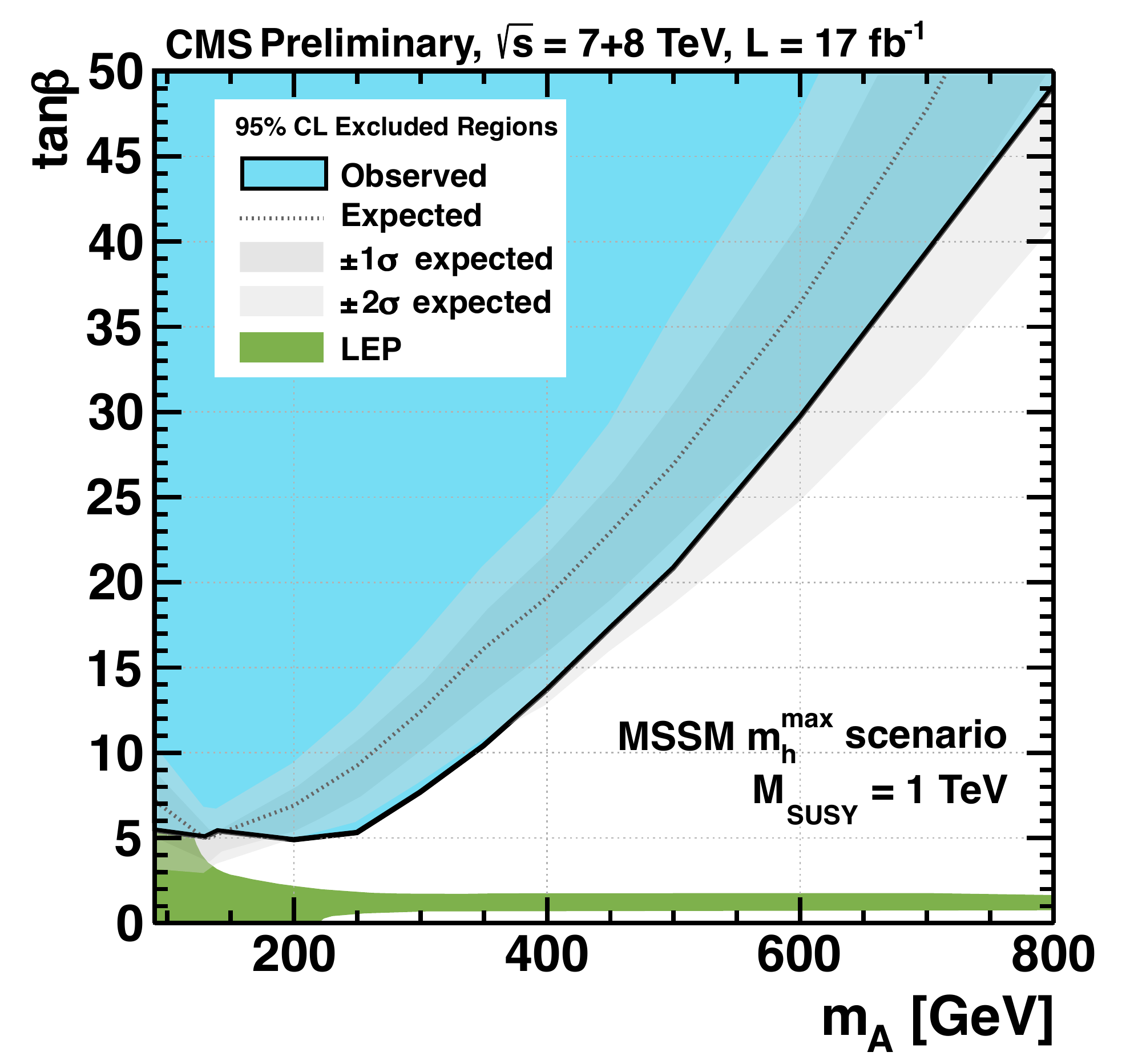}
     \caption{Current MSSM Higgs limits from CMS \cite{CMS-PAS-HIG-12-050}; ATLAS limits are similar. The ``LEP'' exclusion is an artifact of the MSSM prediction for the Higgs mass, rather than a direct search limit, and shouldn't be taken too seriously.}
   \label{fig:higgs}
\end{figure}

The strongest limits by far come from decays of all Higgs states to $\tau \tau$. Current limits are shown in Fig.~\ref{fig:higgs} \cite{CMS-PAS-HIG-12-050}. The search combines ditau signals from $h, H, A$ and gains sensitivity at large $\tan \beta$ where the production modes are enhanced both by $bb$ contributions to gluon fusion and also $bb \Phi$ associated production. There is not much advantage garnered from the actual enhancement of the $\Phi \tau \tau$ coupling, since the $\Phi bb$ coupling is similarly enhanced and therefore the total width grows in proportion to the partial width of interest. Thus the high $\tan \beta$ sensitivity comes predominantly from enhancement of the production mode.

\begin{figure}[htbp] 
   \centering
   \includegraphics[width=4in]{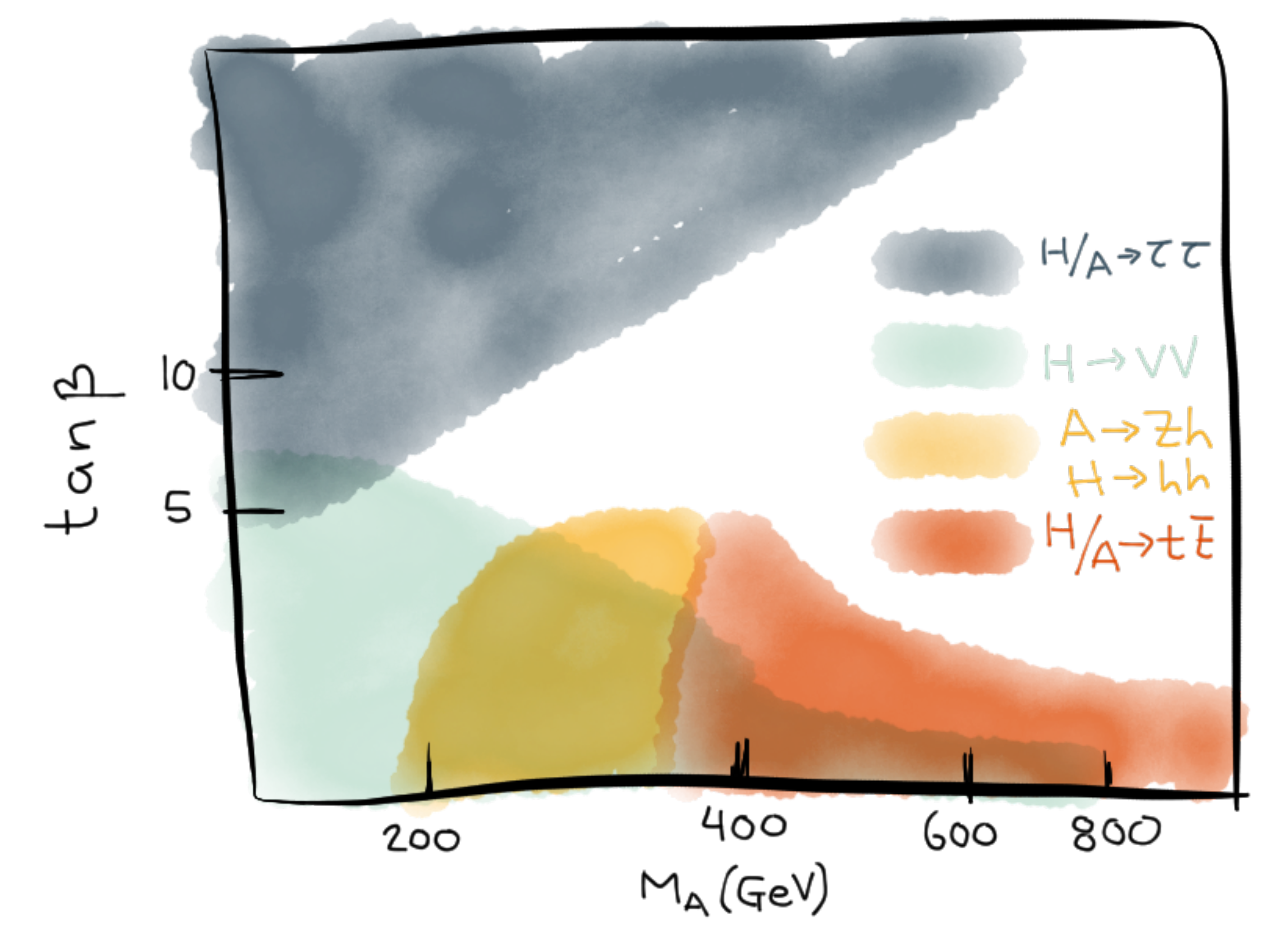}
     \caption{Projected MSSM Higgs limits using the $VV, Zh, hh,$ and $t \bar t$ final states with current data. Cartoon of a figure appearing in \cite{Djouadi:2013vqa}.}
   \label{fig:lowtb}
\end{figure}

Historical prejudice disfavored low $\tan \beta$ due to the LEP limit on the Higgs mass. But, as we'll discuss later, the observed Higgs mass already favors additional mechanisms to enhance the Higgs mass. In my mind, this means we shouldn't take the ``LEP exclusion'' region of this plot too seriously, and should be open to signals at low $\tan \beta$ as well. (To be clear, we should take the LEP exclusion on physical Higgs states seriously, but should remain open to probing the phenomenology of scenarios whose signals populate the low $\tan \beta$ region and require physics beyond the MSSM to explain the observed Higgs mass.) At low $\tan \beta$ the decays of heavy scalars to $VV$ and $hh, Zh$ become important. While the decay of a heavy Higgs into vectors has long been studied, quite recently CMS presented the first limits on $hh$ and $Zh$ in a general 2HDM framework using multi-lepton final states \cite{CMS:2013eua}. A cartoon projection of sensitivity in these additional channels in the MSSM parameter space using the current data set is shown in Fig.~\ref{fig:lowtb}, adapted from \cite{Djouadi:2013vqa}.

\subsection{GMSB}

So far our discussion has focused on simplified models with some SUSY production mode followed by prompt decay to a neutralino and SM states. Of course, there are broad classes of models where the final decay occurs not to a neutralino, but to the goldstino/gravitino. Although I don't have time to do the subject justice, it's worth briefly mentioning the state of gauge-mediated models, which are distinguished by having a gravitino LSP. This raises several prospects of interest. One is that the NLSP, if a neutralino, can decay to the LSP by emitting a photon, opening a final state not frequently exploited in other searches. Another is that the NLSP can be long-lived, giving rise to displaced vertices or charged tracks. At present many of these searches present only 7 TeV data. Limits are consistent with searches in other channels, provided one accounts for the reduced integrated luminosity-- namely gluino mass limits between 600-1000 GeV depending on the final state. Similarly, long-lived stau limits are out to 400 GeV, and long-lived neutralino limits out to 230 GeV. It will be interesting to see how these limits progress as GMSB searches are updated with the full data set. Of course, in many GMSB scenarios the decay to the gravitino is prompt, and most search limits discussed above can be directly applied.

\subsection{Summary}

Across the board, both ATLAS and CMS have sensitivity to sparticle production at the level of $\sigma \cdot {\rm Br} \sim$ few-ten fb at 8 TeV. This translates into $\sim$ TeV limits on light squark flavors and gluinos, $\sim 650$ GeV limits on third-generation squarks, $\sim 350$ GeV limits on electroweakinos without sleptons (as much as $500$ GeV limits including light sleptons), and $\sim 250$ GeV limits on sleptons. If for some reason stops are distinguished from other squarks, the irreducible tuning is $\Delta \sim 40$ and barely consistent with $m_{\tilde g} \sim 2 m_{\tilde t}$.  In scenarios with light-flavor squarks and gluinos of comparable mass, the reach approaches nearly 2 TeV. If all squarks, including stops, are of similar mass, the tuning worsens by an order of magnitude. The only marked hole is in chargino pair production, for which there is no limit improving upon LEP. 

All of these limits assume kinematically available decays to a $\chi_1^0$ whose mass is well-separated from the mass of the decaying particle. In compressed regions, the sensitivity worsens to cross sections on the order of $\sim pb$, and tighter relations erode this sensitivity entirely. However, there is generally no symmetry reason for such exact degeneracy in all appreciably-produced states. We'll discuss means of exploiting these caveats later. 

If we have a reason for stops to be parametrically light, then we are not in terrible shape in terms of naturalness. However, this requires decoupling the other squarks. To determine whether this is cause for discomfort, it's important to build models. Models provide a measure for whether it is generic to populate such parameter space.

\section{Indirect Limits}

Of course, there are various limits on the sparticle spectrum coming from indirect limits as well. Some of these limits have been improved directly by LHC measurements that have surpassed the sensitivity of previous indirect measurements, but many have remained essentially unchanged during the LHC era. Since these lectures are focused on the consequences of the LHC for supersymmetry, I'll focus strictly on new results arising directly from the LHC or limits that have been substantially improved by other experiments in the LHC era -- namely, measurements of $B_s \to X_s \gamma$ and $B_s \to \mu^+ \mu^-$. I will assume that the general state of limits on flavor-violating soft terms from measurements of FCNC is already familiar. Crudely speaking, in the presence of flavor violation in the soft mass spectrum involving the first and second generation, stringent limits on flavor-changing neutral currents (particularly $K^0 - \bar K^0$ mixing) constrain new scalars to live above $\sim$ few tens of TeV in the absence of CP-violating phases, or above $\sim$ few hundreds of TeV in the presence of such phases. If flavor violation is concentrated in heavy flavor, such that all flavor-violating processes go through the third generation, the limits are ameliorated by roughly an order of magnitude. In any event, these precise limits have remained unaltered in recent years, and are discussed in depth elsewhere.

\begin{figure}[htbp] 
   \centering
   \includegraphics[width=5in]{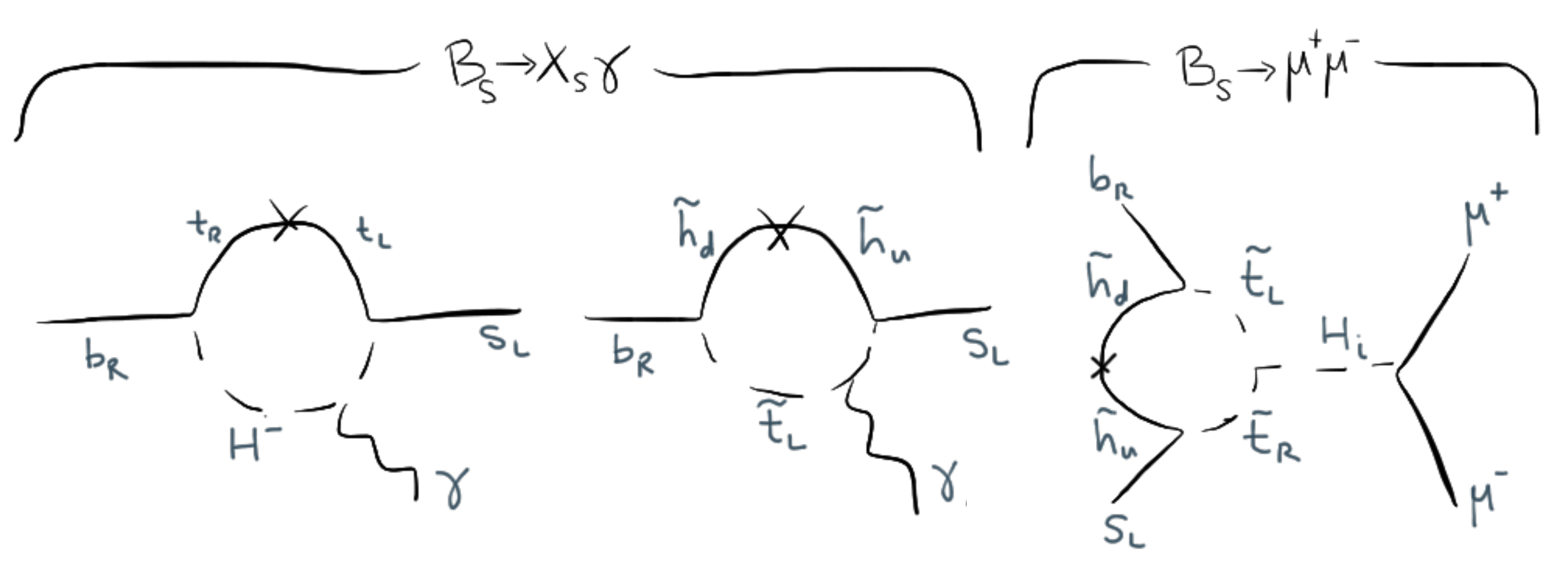}
   \caption{Sample diagrams for SUSY processes contributing to $B_s \to X_s \gamma, \mu^+ \mu^-$.}
   \label{fig:bsdiag}
\end{figure}

\begin{figure}[htbp] 
   \centering
   \includegraphics[width=2.5in]{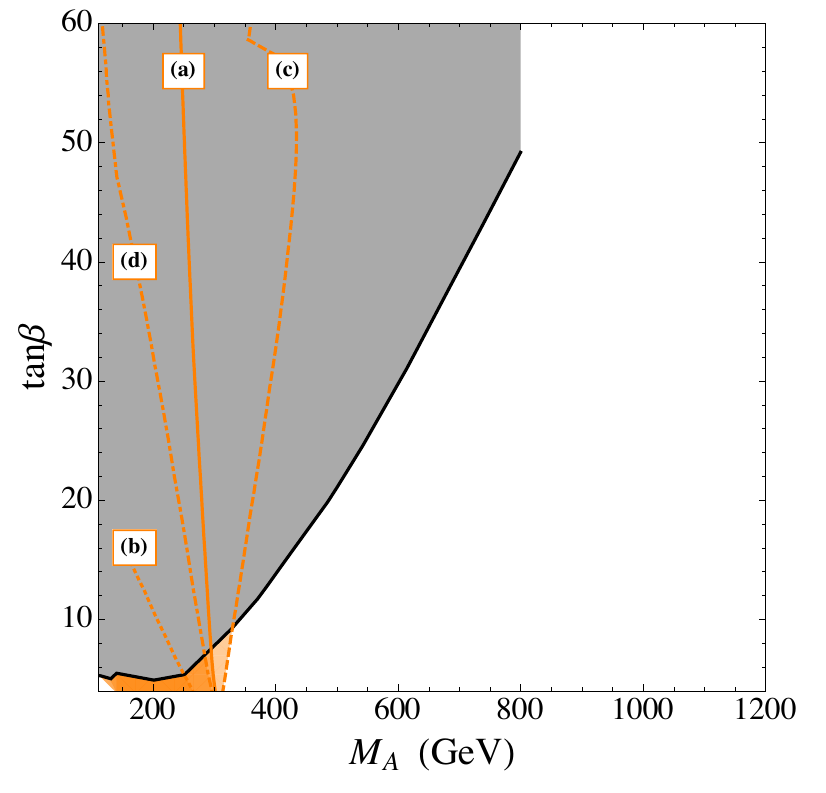}
     \includegraphics[width=2.5in]{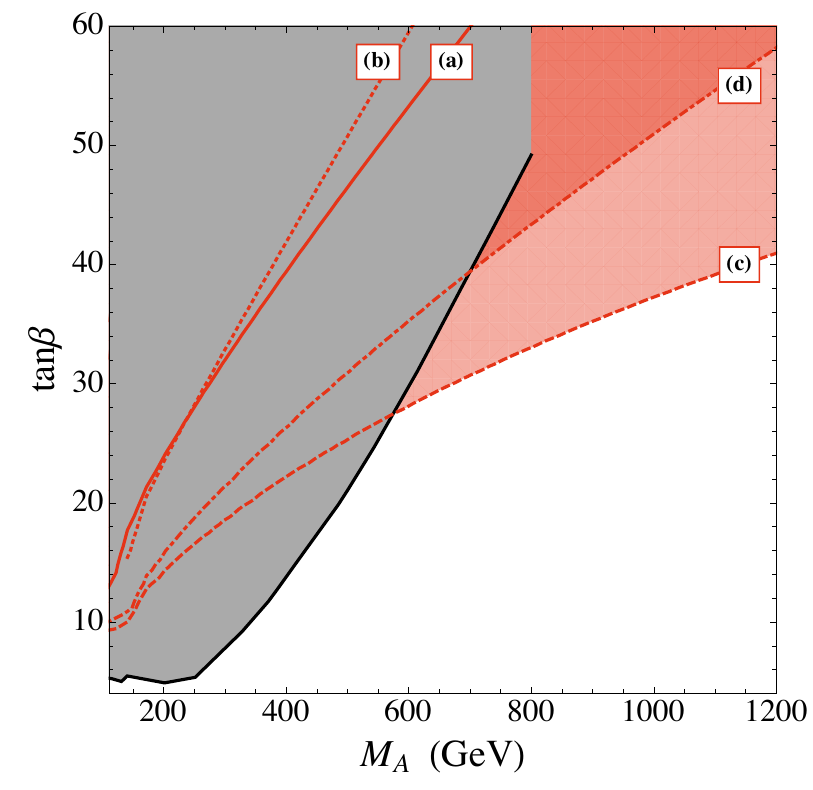}
   \caption{Left: Limits from $b \to s \gamma$ on charged Higgs mass, assuming no light stop cancellation. Right: Limits from $B_s \to \mu^+ \mu^-$. The various scenarios correspond to different choices of $\mu$ with $A_t$ fixed to explain the Higgs mass, and all sfermions at 2 TeV. The most aggressive limits come when $\mu$ and $A_t$ have relative opposite sign due to constructive or destructive contributions from various MSSM processes. Figures from \cite{Altmannshofer:2012ks}.}
   \label{fig:bslims}
\end{figure}

Sample diagrams of SUSY processes contributing to $B_s \to X_s \gamma, \mu^+ \mu^-$ are shown in Fig.~\ref{fig:bsdiag}, while limits for various benchmark spectra are shown in Fig.~\ref{fig:bslims}. The measurement of $ B_s \to  X_s \gamma$ places an interesting constraint on various SUSY scenarios that grows more constraining as time goes on. The experimental measurement and SM NNLO prediction yield \cite{Altmannshofer:2012ks}
\begin{eqnarray}
{\rm Br}(B \to X_s \gamma)_{exp} = (3.43 \pm 0.22) \times 10^{-4} \\
{\rm Br}(B \to X_s \gamma)_{SM} = (3.15 \pm 0.23) \times 10^{-4}
\end{eqnarray}
and the room allowed in the ratio is of order $0.18 \pm 0.13$, limiting new physics to approximately 30\% of the SM contribution. This constraint has tightened in recent years due to improvements in both the measurement and the SM theory prediction. Since the SM contribution is already a one-loop process, this translates into a stringent constraint on new flavor violation -- if additional physics contributes starting at one loop, there must be small couplings, mass decoupling, or accidental cancellations to ensure that it remains a fraction of the SM contribution.

In SUSY, it's well-known that there is a contribution to $B \to X_s \gamma$ from light mixed stops and light higgsinos that is $\tan \beta$ enhanced, and constrains $A_t \mu \tan \beta / m_{\tilde t}^2 < {\rm few}$. This is particularly interesting given the emphasis on the lightness of both stops and higgsinos from naturalness considerations. However, the onward march of SUSY limits alleviates any possible tension; for stop masses consistent with current limits, this does not place a qualitatively new constraint, though it is a limit for scenarios that squash direct limits using kinematics.

Of course, there are also familiar contributions to $B \to X_s \gamma$ coming from the 2HDM sector of the MSSM. In particular, there is a contribution from loops involving the charged Higgs and top quark; these additional contributions can interfere destructively with the stop-higgsino contribution. Absent cancellations, the 2HDM contributions on their own require $m_{H^\pm} \gtrsim 300$ GeV, which helps to squash some of the low-$\tan \beta$ window that would otherwise still be allowed by direct 2HDM searches. 

In addition to $B \to X_s \gamma$, there are also limits coming from two-sided measurement of $B_s \to \mu^+ \mu^-$, which has improved substantially during the LHC era. There is a leading SUSY contribution to this process from penguin diagrams involving the exchange of heavy $H$ and pseudoscalar $A$, using their one-loop flavor-changing $b \to s$ couplings. These contributions decouple with $m_A$ but scale as $\tan^3 \beta$, placing an interesting constraint on large $\tan \beta$. However, the sensitivity of these contributions to the details of the 2HDM sector and SUSY parameters -- and the freedom to decouple the extra Higgs degrees of freedom without substantially impairing naturalness -- means that there is no irreducible limit on SUSY models coming from $B_s \to \mu^+ \mu^-$.

Finally, a word on flavor violation in the top sector. Historically, flavor violation involving the top quark is poorly constrained, given the fact that the top decays before hadronizing (so that detailed studies of meson oscillation that set such stringent FCNC limits involving lighter flavors are not possible, though certain combinations of flavor-violating couplings involving the top quark can be probed through $D$ meson oscillations) and the relatively small number of top quarks produced prior to turning on the LHC. However, the LHC is a top factory, so one might hope that limits on top FCNC coming from the LHC might provide a qualitatively new probe of SUSY FCNC. However, current sensitivity is at the level of $\sim 0.1 \%$, with anticipated sensitivity to the level of $\sim 0.01\%$, with SUSY signals starting several orders of magnitude below this level. At this stage, LHC probes of top flavor violation say nothing interesting about SUSY parameter space.

\subsection{Summary}

The state of indirect limits has not changed radically in the LHC era.  Arbitrary SUSY flavor violation requires sparticles above 10 TeV without CPV phases, and above 100 TeV with CPV phases. This suggests either that flavor violation is small or the mass scale is in the 10-100 TeV range. 

$B$-physics constraints limit certain scenarios with light stops, providing a potentially useful handle if light stops are not kinematically obvious at LHC. At present, however, the new measurement of $B_s \to \mu^+ \mu^-$ does not place a qualitatively better limit than direct searches for additional Higgs states at LHC. New measurements of top flavor violation at the LHC do not yet play a useful role.

\section{Implications of the Higgs}

Considering that the Higgs was the whole reason for the hierarchy problem -- and the first triumphant discovery at the LHC -- it's crucial to evaluate the implications of the Higgs discovery for SUSY models. Generally, it seems the state we've found is an elementary scalar, which is favorable for SUSY -- it's not transparently a composite of some strong dynamics. Thankfully, we can learn a great deal about possible UV physics from both the mass and couplings.

\subsection{Mass}
To a certain extent, the mass of the Higgs is the most important variable with respect to SUSY parameter space. 
At one loop in the decoupling limit, the mass of the lightest CP-even Higgs is
\begin{equation}
m_h^2 = m_Z^2 \cos^2(2 \beta) + \frac{3 m_t^4}{4 \pi^2 v^2} \left[ \log  \left( \frac{m_S^2}{m_t^2} \right) + \frac{X_t^2}{m_S^2} \left( 1 - \frac{X_t^2}{12 m_S^2} \right) \right]
\end{equation}

where $X_t \equiv A_t -\mu \cot \beta$ and $m_S^2 = m_{\tilde t_1} m_{\tilde t_2}$. The first loop correction should be thought of as the usual logarithmic correction, and the second as finite threshold corrections. Both are corrections to the quartic coupling, since I have fixed the Higgs mass in terms of the quartic and the vev.

If radiative effects are unimportant, then SUSY predicts $m_h \leq m_Z$, in tension with observation. In fact, observationally we have $m_h \sim \sqrt{2} m_Z$, or $m_h^2 \sim 2m_Z^2$. This implies the corrections to the tree-level expression must be as important as the tree-level value itself. If the corrections are radiative, it certainly runs up against the Veltman definition of naturalness.

\begin{figure}[htbp] 
   \centering
   \includegraphics[width=3in]{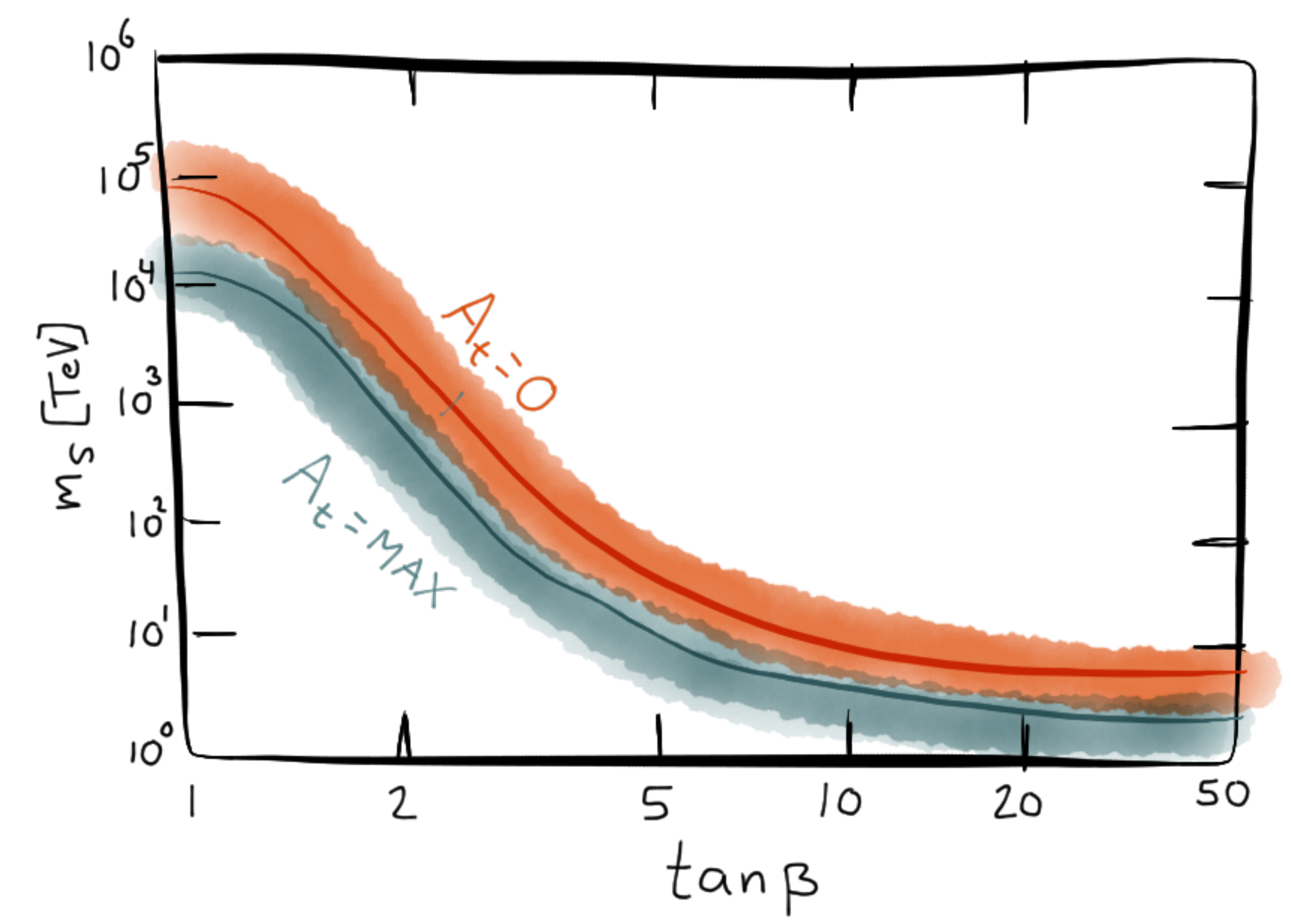} 
   \caption{The region accommodating the observed Higgs mass in the MSSM as function of stop mass scale and $\tan \beta$. The red band corresponds to zero stop mixing, while the blue band corresponds to maximal stop mixing. Adapted from a figure in \cite{Arvanitaki:2012ps}. }
   \label{fig:hmass}
\end{figure}

There are essentially four ways to accommodate the observed Higgs mass:
\begin{enumerate}
\item {\it Leverage the logarithm by making $m_S$ large.}  The region accommodating the observed Higgs mass as a function of the stop mass scale is shown in Fig.~\ref{fig:hmass}. Without substantial stop mixing, reproducing this requires $m_S \gtrsim 4$ TeV and points to a mini-split spectrum, about which more later. If this is the reason for the Higgs at 125 GeV, naturalness is imperiled, with a tuning on the order of $\Delta \gtrsim 1000$.

\item {\it Leverage the threshold corrections by living close to maximal mixing, $X_t \approx \pm \sqrt{6} m_S$.} The consequences of stop mixing for the Higgs mass are illustrated in Fig.~\ref{fig:hmix}, adapted from a figure in \cite{Draper:2011aa}. This allows $m_S \sim$ TeV, but requires a close tuning of $X_t$ relative to $m_S$. Moreover, in calculable models such as gauge mediation, generating such large $A$-terms without increasing the fine-tuning of the theory is prohibitive.

\begin{figure}[htbp] 
   \centering
   \includegraphics[width=4in]{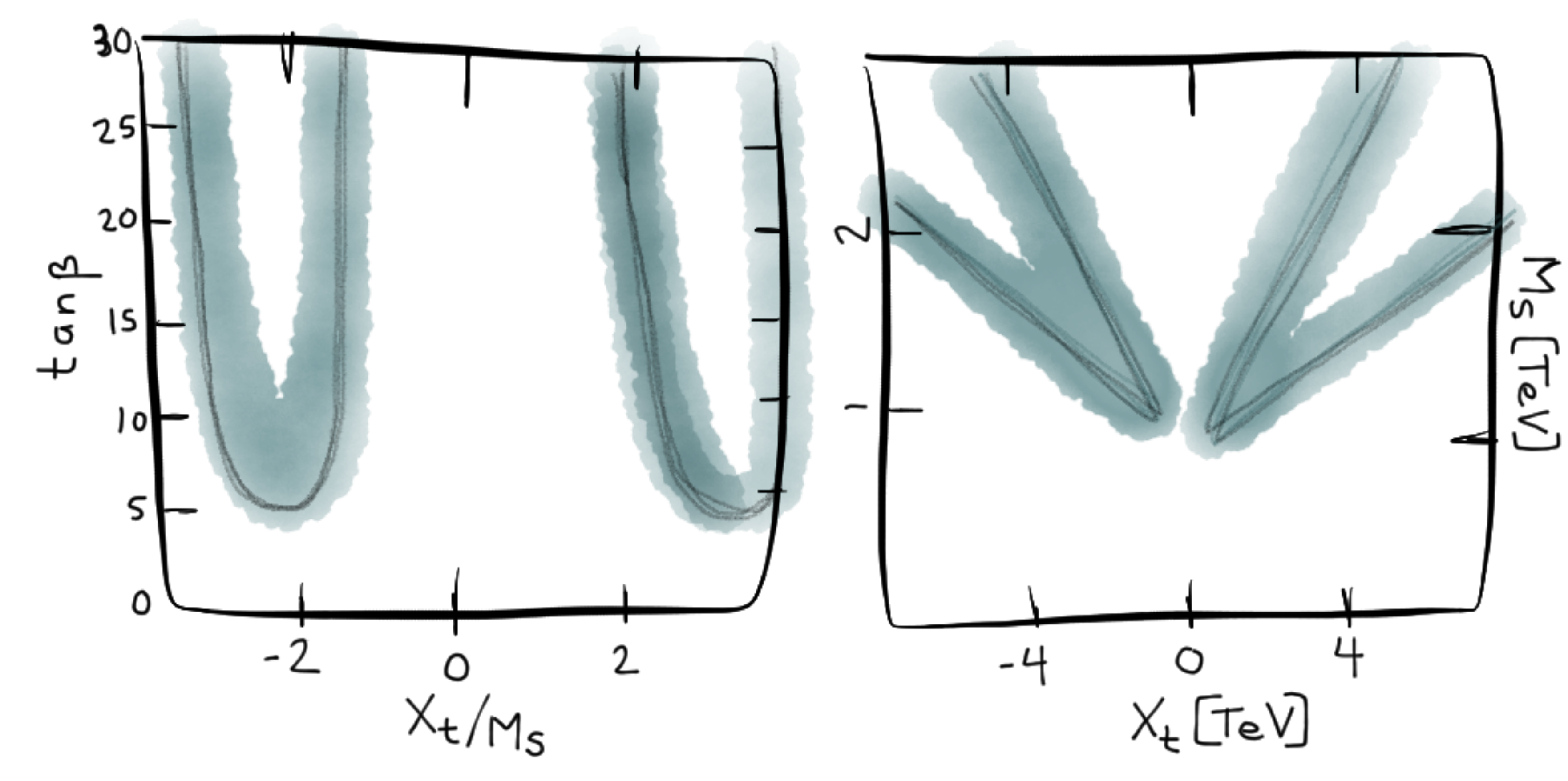} 
   \caption{The region accommodating the observed Higgs mass in the MSSM as a function of stop mixing, the average stop mass scale, and $\tan \beta$. Adapted from a figure in \cite{Draper:2011aa}.}
   \label{fig:hmix}
\end{figure}

To be more specific, in gauge mediation the $A$-terms are zero at one loop at the messenger scale, which is naturally problematic for maximal mixing. One option is to generate sufficiently large $A$-terms by substantial running, in which case they are driven to be large and negative primarily via the gluino mass \cite{Draper:2011aa}:

\begin{equation}
\frac{d A_t}{dt} \sim y_t^2 A_t + g_3^2 M_3
\end{equation}
 The disadvantage here is that the gluinos must be quite heavy -- at least 2.5 TeV for GUT-scale messengers -- and the messenger scale generally quite high.

 Alternately, nonzero messenger-scale masses can be generated at one loop by coupling MSSM superfields directly to messengers. In order to avoid potentially problematic flavor violation, the MFV way to do this is by coupling the Higgs doublets to messengers via couplings of the form $\lambda_u H_u \Phi_i \bar \Phi_j$. The problem is that while this generates one-loop $A$-terms of order $(\alpha_u /4 \pi) F/M$ (where $\sqrt{F}$ is the scale of SUSY breaking and $M$ is the messenger mass scale), it also generates one- and two-loop contributions to $m_{H_u}^2$. The one-loop contribution may be suppressed if the messengers are of the minimal gauge mediation form (i.e., their mass comes entirely from the lowest component of a single SUSY-breaking spurion), but the two-loop contribution is of order $A_t^2$. This is necessarily large, since the $A$-terms need to be large, implying $\mu$ is large, and the tree-level tuning of EWSB is more or less as bad as simply taking heavy, unmixed stops \cite{Craig:2012xp}.

\item {\it Enhance the tree-level contribution to the Higgs mass.} Since the size of the tree-level contribution is set by the Higgs quartics, this suggests new contributions to the quartic couplings. These effects are in some sense the most natural, since they do not require radiative corrections to exceed tree-level contributions. Either $F$- or $D$-terms will suffice. New $F$-term contributions imply additional degrees of freedom coupled to the Higgs, such as the NMSSM and variants, while new $D$-term contributions imply new gauge groups broken at a low scale.

For the $F$-term, imagine introducing a singlet $S$ with marginal superpotential interaction
\begin{equation}
W = \lambda S H_u H_d + \dots
\end{equation}
for which
\begin{equation}
\delta m_h^2 =  \lambda^2 v^2 \sin^2 2 \beta
\end{equation}
The problem is that this is largest at $\tan \beta =1$, where the standard MSSM contribution to the Higgs mass vanishes. Also $\lambda$ runs up at higher scales, since there is no gauge contribution to regulate the self-interaction contribution to the beta function. The perturbative max (i.e., the value of the weak-scale coupling required to ensure no Landau pole below the GUT scale) lies around $\lambda \sim 0.7$. The largest tree-level value with this bound is 122 GeV, and in general stops around 400 GeV or heavier are required \cite{Ellwanger:2009dp, Hall:2011aa}.  So this scenario is constrained, but there are few deadly drawbacks. The generic problem is that symmetries are necessary to protect $S$ from Planck-scale tadpoles.

For the $D$-term, imagine extending the theory to, e.g., $SU(2)_A \times SU(2)_B \to SU(2)_L$ at a scale $f$, with $H_u$ charged under $SU(2)_A$ in the UV. This adjusts the SM quartics by an amount $g^2 \to g^2(1+\Delta)$, where \cite{Batra:2003nj}

\begin{equation}
\Delta \equiv \frac{g_A^2}{g_B^2} \frac{m_S^2}{m_V^2+m_S^2} 
\end{equation}
where $m_V^2 \sim (g_A^2 + g_B^2) f^2$ is the supersymmetric vector mass, and $m_S$ is the soft mass of the fields that Higgs the group to the diagonal.
 
 Then the correction to the Higgs mass is
 \begin{equation}
 \delta m_h^2 = \frac{g^2 \Delta}{2} \cos^2(2 \beta)
 \end{equation}
 
 Now there is also a shift in soft masses that bounds $m_S$, namely \cite{Batra:2003nj}
 \begin{equation}
 \delta m_{H_u}^2 \sim \frac{3}{4} \frac{g_A^2}{g_B^2} \frac{g^2}{16 \pi^2} m_S^2
 \end{equation}
 which requires $m_S \lesssim 10$ TeV to avoid worsening the tuning of EWSB, and then preserving a large effect makes $m_V$ low enough to raise tension with precision electroweak.

\item {\it Enhance the loop-level contribution to the Higgs mass with additional matter.} We know that chiral multiplets with large couplings to the Higgs enhance the radiative contribution to the physical Higgs mass, so we can imagine adding new fields to assist the top multiplet. These new fields should be vector-like, in order to avoid stringent limits on a chiral fourth generation. This is only a partial solution, since it requires the log still be large, although now distributed among many fields. Imagine adding doublets $\Phi, \bar \Phi$ and singlets $\phi, \bar \phi$ with couplings
\begin{equation}
W = M_\Phi \Phi \bar \Phi + M_\phi \phi \bar \phi + k H_u \Phi \bar \phi + \dots + {\rm soft \; terms}
\end{equation}
This is like adding extra top multiplets, and \cite{Martin:2009bg, Graham:2009gy}
\begin{equation}
\delta m_h^2 = \frac{N_c}{4 \pi^2} k^4 v^2 \sin^4 \beta \log\left( m_S^2 / m_F^2 \right) + \dots
\end{equation}
(where here $m_S$ and $m_F$ are the scalar and fermion masses of the new multiplet, respectively) and one can explain the Higgs mass.

However, there are similarly radiative contributions to the Higgs soft parameters of order \cite{Martin:2009bg}
\begin{equation}
\delta m_H^2 \sim - \frac{N_c}{4 \pi^2} k^2 m_{S}^2 \ln(\Lambda/m_S)
\end{equation}
and so it is far from obvious that tuning improves substantially. One can live along certain directions of parameter space where the physical mass correction is large but the radiative correction to the soft mass isn't maximal, but these are not necessarily generic. On the whole, adding new vector-like states to raise the Higgs mass more or less just distributes tuning associated with explaining the Higgs mass among added degrees of freedom.

\end{enumerate}

Unsurprisingly, the Higgs mass is a very important target! The observed mass is close to SUSY expectations, but implies that a correction of the same size as the tree-level MSSM piece is necessary. If we retain the minimal framework, we are forced into a position of either tolerating large loops -- in violation of Veltman's naturalness criterion -- or attempting to induce large threshold corrections, which have tuning problems of their own in calculable models. Alternately, we can introduce new tree-level contributions by extending the Higgs sector or add new degrees of freedom to run in loops. Both of these possibilities imply new physics that is sensitive to SUSY breaking and whose mass scale is generically  bounded from above. Such extensions typically also give rise to corrections to Higgs couplings and other precision observables \cite{Arvanitaki:2011ck}

\subsection{Couplings}
We also learn substantially from the couplings of the observed state. The Higgs sector of the MSSM is a Type 2 2HDM, which comes with certain tree-level predictions for changes in the couplings of the Higgs. In general, the CP-even excitations of the Higgs are not aligned with the vacuum condensate, and so one can expect tree-level deviations in the Higgs couplings. Radiative corrections from new colored and charged states can also play an important role. To the extent that we've observed a largely SM-like scalar, this leads to additional constraints on the parameter space.

\subsubsection{Tree level}

As a Type 2 2HDM, there can be tree-level changes in the Higgs couplings due to the 2HDM scalar potential, corresponding to rotations of $h$ away from the vacuum condensate. This can be simply parameterized in terms of $\tan \beta$ and $\alpha$, the rotation between CP-even mass eigenstates and the original doublets. 

We have for $r_i \equiv \frac{g_{hii}}{g_{hii}^{SM}}$ \cite{Blum:2012ii},
\begin{equation}
r_b = - \frac{\sin \alpha}{\cos \beta} \hspace{1cm} r_t = \frac{\cos\alpha}{\sin \beta} \hspace{1cm} r_V = \sin (\beta - \alpha)
\end{equation}
In general $\alpha, \beta$ are free parameters, but in the MSSM the predictive quartic couplings allow us to exchange $\alpha$ for the physical mass scale $m_A$, via
\begin{equation}
\cot \alpha = - \tan \beta - \frac{2 m_Z^2}{m_A^2} \tan \beta \cos (2 \beta) + \mathcal{O}(m_Z^4/m_A^4)
\end{equation}
Then at large $\tan \beta$,
\begin{equation}
r_b \to 1 + 2 \frac{m_Z^2}{m_A^2} \hspace{1cm} r_t \to 1 - \mathcal{O}(m_Z^4/m_A^4) \hspace{1cm} r_V \to 1 - \mathcal{O}(m_Z^4/m_A^4)
\end{equation}
So the dominant effects are in the bottom quark coupling at large $\tan \beta$. This leads to limits on $m_A$ primarily from adjusting the total width of the Higgs, rather than any deviations in the couplings of well-measured channels.

Current coupling measurements are at the 25\% level. So taken alone, at large $\tan \beta$ this implies the lower bound on $m_A$ is $\gtrsim 250$ GeV, which is not a tight constraint. Of course, this is purely tree-level, and loop-level effects can be very important, especially if new charged and colored states are light.

\subsubsection{Loop level}

Two of the most important Higgs couplings arise in the SM at one loop, namely the coupling to gluons and photons. So both of these are easily altered by loops of new colored and charged particles, which SUSY provides in abundance.

The contributions from new physics are fairly simple to compute; the Higgs low-energy theorems tell us that the correction to the Higgs-gluon-gluon coupling due to new physics goes as 
\begin{equation}
\mathcal{L}_{hgg} = \frac{\alpha_s}{12 \pi} \frac{h}{v} \left( 2 \sum_F t_F \frac{\partial \log \det m_F(v)}{\partial \log v} + \frac{1}{2} \sum_S t_S \frac{\partial \log \det m_S(v)}{\partial \log v} \right) G_{\mu \nu}^a G^{a \mu \nu}
\end{equation}
where $t$ is the Dynkin index. 

First, let's work out the implications for stops:
\begin{itemize}
\item Applying this to the top-stop system,  the stop contribution to the gluon effective coupling $hGG$ is \cite{Blum:2012ii}
\begin{equation}
r_G -1 \approx \frac{1}{4} \frac{m_t^2}{m_S^2} \left(\frac{m_{\tilde t_1}^2 + m_{\tilde t_2}^2 - X_t^2}{m_S^2} \right)
\end{equation}
which includes only the loop effects, i.e., neglects tree-level corrections to the top coupling discussed earlier. This enhances the $gg$ coupling for unmixed stops, but can suppress  the $gg$ for mixed stops; this latter effect is hard to realize given vacuum stability. However, it's clear that the interplay of mixing and masses can be used to cancel or reduce the contribution to the effective coupling.

\item To get a ballpark sense for the effects under consideration, we have $m_{\tilde t_1} = m_{\tilde t_2} = 250$ GeV $\to r_G = 1.24$, so a $50\%$ enhancement of gluon fusion rate. This falls to $r_G = 1.06$ by 500 GeV, so generic bounds ensure small contributions consistent with current measurements even without mixing. Note that if stops were light, unmixed, and nearly degenerate with the top, the effects could be considerable, as much as 150\% of the SM coupling. This provides another handle on light stops.

\item As for photons, the contribution to the photon effective coupling is computed in a similar way, with 

\begin{equation}
r_\gamma = \frac{A_W^\gamma + A_t^\gamma + A_{\tilde t}^\gamma}{(A_W^\gamma + A_t^\gamma)^{SM}} \approx 1.28 r_V - 0.28 r_G
\end{equation}
The dominance of the $W$ loop mean that the stop contribution enters here with a flipped sign relative to the gluon coupling. So one expects enhanced gluon couplings to be correlated with suppressed photon couplings, etc.
\end{itemize}

Of course, the stop is not the only degree of freedom that can run in loops. Other contributions include:
\begin{itemize}
\item Sbottom contributions are typically not large due to the suppressed coupling. Stau effects are possible but typically not large except in somewhat small corners of parameter space. 

\item Charginos may have an important effect. Useful since we expect higgsinos $\sim 200$ GeV. The coupling has no simple form, but parametrically, the chargino correction is \cite{Blum:2012ii} 
\begin{equation}
r_\gamma^{\chi^\pm} \lesssim 1 + \frac{2}{5} \frac{m_W^2}{m_{\chi^\pm}^2} \left( 1 + \frac{2 m_W}{m_{\chi^\pm}} \right)^{-1}
\end{equation}
which leads to $r_\gamma^{\chi^\pm} -1 \lesssim 10\%$ given bounds from LEP. However, this is also quite sensitive to the wino and higgsino content of the chargino. So in general we expect these corrections to be fairly small and not to give irreducible limits on the parameter space.
\end{itemize}

\begin{figure}[htbp] 
   \centering
   \includegraphics[width=3in]{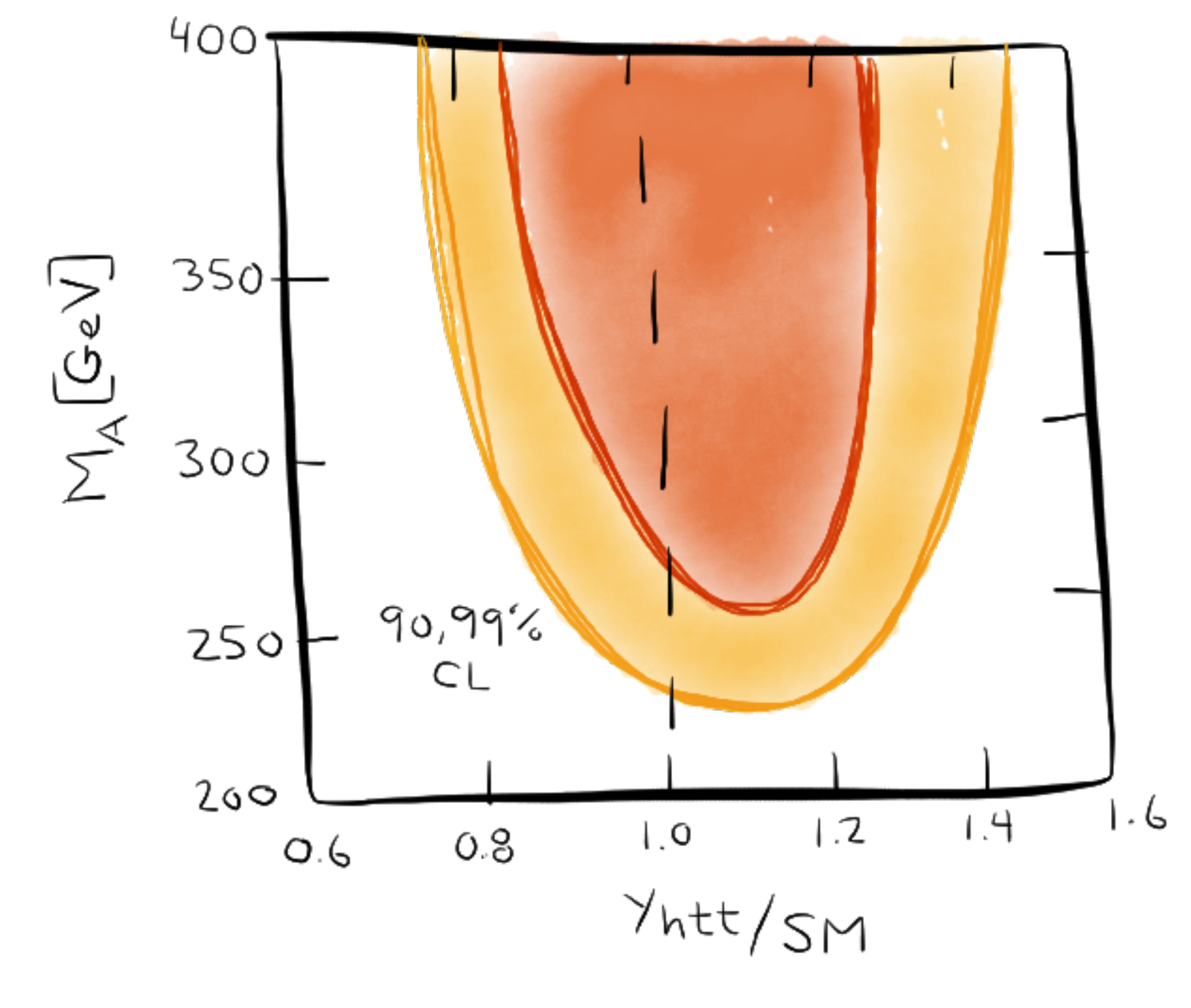} 
   \caption{Combined coupling fit in the MSSM as a function of $m_A$ and the effective top-higgs coupling appearing in the gluon and photon couplings (sensitive to loop effects from stops) at large $\tan \beta$. Cartoon of a figure appearing in \cite{Giardino:2013bma}.}
   \label{fig:strumia}
\end{figure}

Ultimately, we can combine the tree-level and loop-level effects in a given coupling fit to Higgs measurements, as illustrated in Fig.~\ref{fig:strumia}. The fit is insensitive to $\tan \beta$ above $\tan \beta \approx 10$, where bounds are $m_A \gtrsim 250$ GeV \cite{Giardino:2013bma}. At low $\tan \beta$, the tree-level alterations of couplings shrink, and fit allows lower values of $m_A$.

So at present coupling fits to $H$ do not provide qualitatively new data regarding the extended Higgs sector, but this could change with improved precision.

\subsection{Summary}

The Higgs mass and couplings both provide useful new input to the structure of SUSY theories. 
\begin{itemize}
\item The Higgs mass is somewhat tense with respect to minimal predictions, requiring a correction that is the same size as the minimal tree-level contribution. While this may be alleviated in various ways, most mechanisms require additional degrees of freedom bounded in scale. 
\item Explaining the Higgs mass simply in the context of the MSSM involves some degree of tuning from the log-enhanced and finite corrections. This points to high scales or non-minimal scenarios with new degrees of freedom.
\item Coupling limits are interesting, but they lead to no strong limit on scales for either loop-level or tree-level corrections. Varying mixing in new degrees of freedom can tune corrections away, so that corrections at a given mass scale can be substantially reduced relative to naive expectations.
\item In general, coupling measurements at present only place useful constraints if the relevant new particles are very light.
\end{itemize}\pagebreak

\part{Where do we go?}

In light of the natural expectations for supersymmetric signals and null results at the LHC, there are two logical directions to pursue: ``breaking the signal'', i.e., finding ways to undermine the conventional signals of supersymmetric models and thereby erode the sensitivity of LHC searches; and ``breaking the spectrum'', i.e., constructing novel models that comfortably sit on the periphery of LHC sensitivity and are perhaps under less tension than  their conventional counterparts.

In the former case, the presumption is that supersymmetry is natural and more or less consistent with our pre-LHC expectations, with a spectrum of sparticles beneath a TeV that we have yet to discover simply because kinematics or other features make them difficult to distinguish from backgrounds. In the latter case, the presumption is that supersymmetry provides a correct description of the universe at high energies but takes a form radically different from minimal realizations.

\section{Breaking the signal}

To figure out how to break the signal, we should first assess what the signal is -- i.e., what distinctive properties of SUSY theories provide handles for discrimination relative to Standard Model backgrounds. SUSY manifests itself above SM backgrounds in two decisive ways: through large missing transverse energy associated with decays of heavy states to a stable LSP, and through large activity (typically hadronic) associated with heavier mass scales.  If either one of these is substantially undermined, it can erode the signal space and leave the spectrum natural. Of course, these are not the only signals; if SUSY events throw off other tagging objects (such as photons or leptons) in conjunction with a large rate or energy scale, then even the erosion of both missing transverse energy and large hadronic activity can be overcome.

There are three principal ways to break these signals: compressing SUSY, stealthifying SUSY, or breaking R-parity. Each comes with its own advantages or disadvantages. In each case, one can think of these avenues as model-building modules -- they can be attached to a conventional SUSY model that on its own can explain the Higgs mass and electroweak naturalness.

\subsection{Compressed SUSY}

We typically assume that all SUSY production processes end in a collider-stable LSP that is substantially lighter than other sparticles. This intuition is an assumption born of parsimony -- for example, models based on mSUGRA give large splitting between the bino and gluino due to long RG running from the fundamental scale of SUSY breaking. Thus typically the LSP is a bino or higgsino with substantial separation from colored sparticles. This in turn gives maximal sensitivity at the LHC, since cascade decays to the LSP throw off impressive amounts of event activity in the form of energetic jets and leptons. It also leads to a substantial missing energy signal, since the LSP is pair-produced in essentially uncorrelated directions, guaranteeing large total MET in the event.

However, this separation is not an intrinsic feature of SUSY. If the LSP is closer in mass to the scale of colored sparticles, the spectrum becomes compressed and limits reduce substantially \cite{LeCompte:2011cn}. This reduction is due to two effects:

\begin{enumerate}
\item The hadronic activity in the event decreases since less energy is available to visible particles appearing in the cascade decay from heavy initial sparticle to the LSP.
\item The MET in each event decreases since the LSPs are pair produced in more of a back-to-back manner, so that there is cancellation in the total missing energy even though two invisible particles are carrying off a substantial amount of energy.
\end{enumerate}

\begin{figure}[htbp] 
   \centering
   \includegraphics[width=4in]{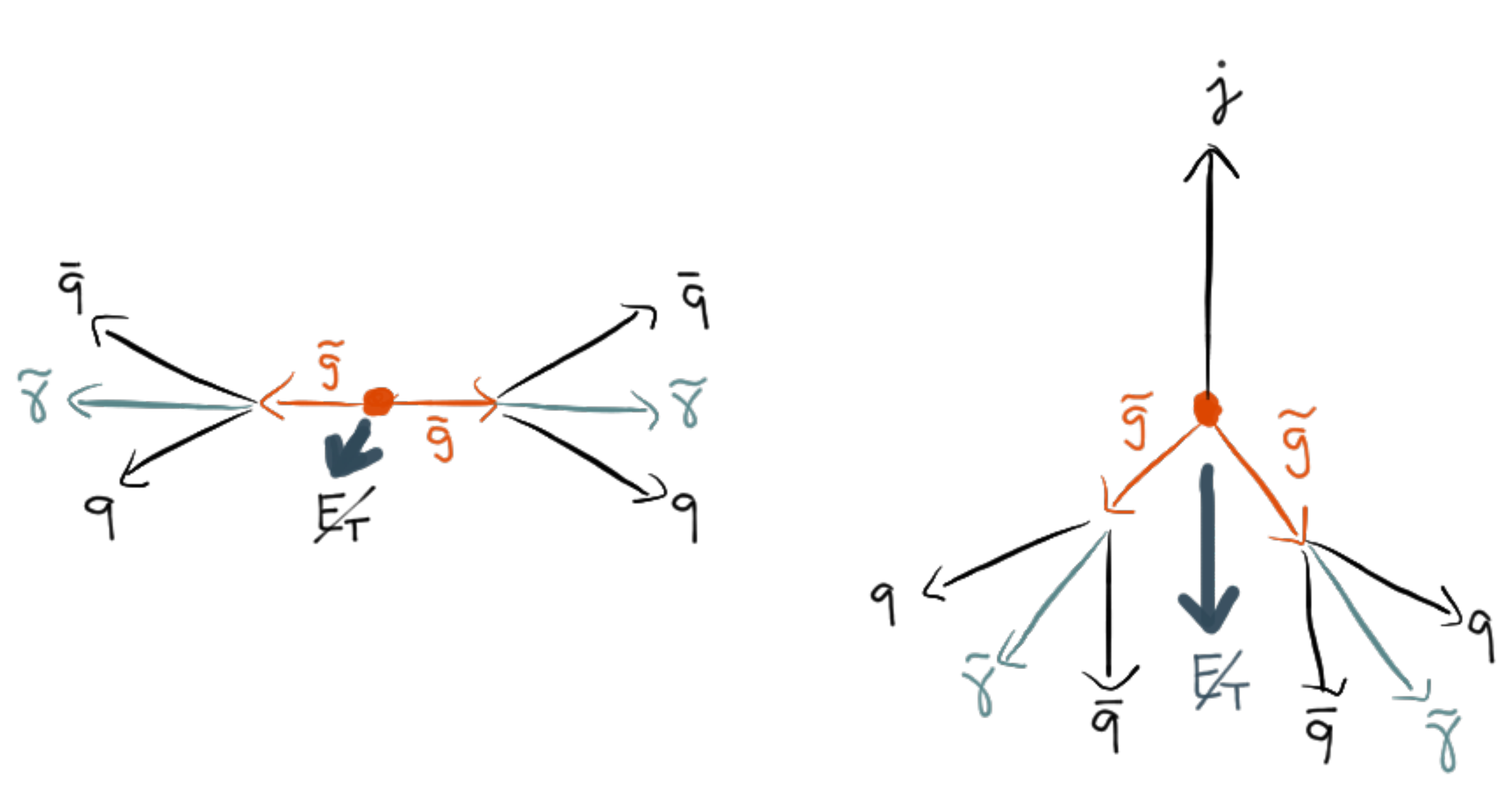} 
   \caption{Kinematics of compression. MET is accidentally reduced in pair-production, but the presence of hard ISR provides enough recoil to restore the intrinsic MET signal. Adapted from a figure in \cite{Alwall:2008ve}. }
   \label{fig:compdiagram}
\end{figure}

The kinematics of MET reduction are illustrated in Fig.~\ref{fig:compdiagram}. We see the effects of compression in every SUSY limit plot by looking close to the diagonal where the mass of the LSP grows close to the mass of the sparticle under study; limits are always eroded as we near the compressed region.
However, although these effects buy some reduction in the sensitivity of standard search strategies, they can be easily outmaneuvered by slight improvements in the search. In particular, only the reduction in hadronic activity is robust. The reduction in MET is not robust for events with initial state radiation (ISR). ISR jets provide something for the SUSY process to recoil against, so that the LSP pairs are not strictly back-to-back and there is a net missing transverse energy in the event. Thus compressed signals can be efficiently constrained by lowering $m_{eff}$ requirements and increasing jet multiplicity. By focusing more on MET, rather than combinations of MET and $m_{eff}$, the compressed signal can be dug out. On the flip side, this may become more challenging at Run II of the LHC as pile-up contributes more to backgrounds for MET-based searches.

At the moment, given the kinematic reach of the LHC at 7 and 8 TeV, if the LSP is heavier than 500 GeV then there is no real limit in any simplified model. But this will be eroded with more data, dedicated searches, and higher center-of-mass energy. Compression is therefore an annoyance, but not a panacea for eroding search limits.

However, to the extent that compression does reduce limits, it's natural to ask what types of models intrinsically lead to compressed spectra. Typically, this is the outcome in models where the gluino is lighter than other gauginos at the high scale, due to additional degrees of freedom carrying Standard Model charges that split the typical gaugino mass unification relations. This can be arranged in gravity mediation scenarios, or in gauge mediation where the doublet and triplet messengers have distinct couplings. Then running effects can lift the gluino mass while flowing to the IR, with the typical result that all gauginos are comparable in mass. So this is a fairly simple outcome that erodes limits using only a mild departure from parsimony. It will not enable SUSY to be hidden at the LHC forever, but at the moment still buys some erosion of the limit space.

\subsection{Stealth SUSY}

The reduction in MET due to compression was in some sense accidental; compressed SUSY events still involve heavy invisible particles carrying substantial energy, and only result in MET reduction if the heavy particles are produced back-to-back. It's therefore quite interesting to consider stealth supersymmetry, a systematically robust scenario in which the MET signal is reduced \cite{Fan:2011yu, Fan:2012jf}. To emphasize the distinction, keep in mind that

\begin{itemize}
\item Compressed SUSY is really reduction of $m_{eff}$, the reduction of MET is incidental or accidental. The LSP is heavy and energetic.

\item Stealth SUSY is genuinely the reduction of MET, not $m_{eff}$. The LSP is light, carries little energy.
\end{itemize}

The idea behind stealth SUSY is that there is a light, approximately supersymmetric multiplet in theory. Of course, we don't expect this multiplet to be a standard MSSM supermultiplet, since the non-observation of MSSM sparticles suggests a considerable splitting in the MSSM, but it's plausible to imagine additional degrees off freedom that feel SUSY breaking more weakly.

In the presence of such an approximately supersymmetric multiplet, decay chains proceed as normal, bleeding off all visible energy down to an $R$-odd stealth particle, which we'll call $\tilde S$. We imagine $\tilde S$ has a relatively small mass splitting with its $R$-even superpartner, $S$. Then we can have decays of the form $\tilde S \to S \tilde G$, where $\tilde G$ is a gravitino or other light $R$-odd particle that plays the role of the genuine LSP. The $R$-even particle $S$ must go back to visible particles, like $S \to jj$. If this is the case, then the true LSP $\tilde G$ carries off a small amount of MET simply due to kinematics; the decay that produces the LSP has small phase space for the invisible particle. Note that it's crucial that $S$ also decay back into visible particles on detector timescales, lest the missing energy signal result from the entirety of the decay products of $\tilde S$.

One expects the invisible momentum carried off in such process to be of order $p_{inv.} \sim \gamma \delta M \sim  \frac{m_{\tilde g}}{m_{\tilde S}} (m_{\tilde S} - m_S)$. Now limits from normal SUSY searches start to become effective around $m_{\tilde S} - m_S \gtrsim 20$ GeV. Thus to be effective in eroding limits, we need small splittings in the $S, \tilde S$ multiplet, on the order of $\lesssim 10$ GeV. So the $S, \tilde S$ are really coming from an approximately supersymmetric multiplet.

The generic challenge for model-building in this case is that there is a universal amount of SUSY breaking proportional to the gravitino mass, $m_{3/2}$. If $m_{3/2}$ is the primary scale of SUSY breaking, as in gravity- or anomaly-mediation, then it's hard to arrange an approximately supersymmetric multiplet unless these degrees of freedom are protected by conformal sequestering of some form. Alternately, we can imagine this scenario working in low-scale models where $m_{3/2}$ is small and the primary source of SUSY breaking in the MSSM comes from gauge loops.

\begin{figure}[htbp] 
   \centering
     \includegraphics[width=4in]{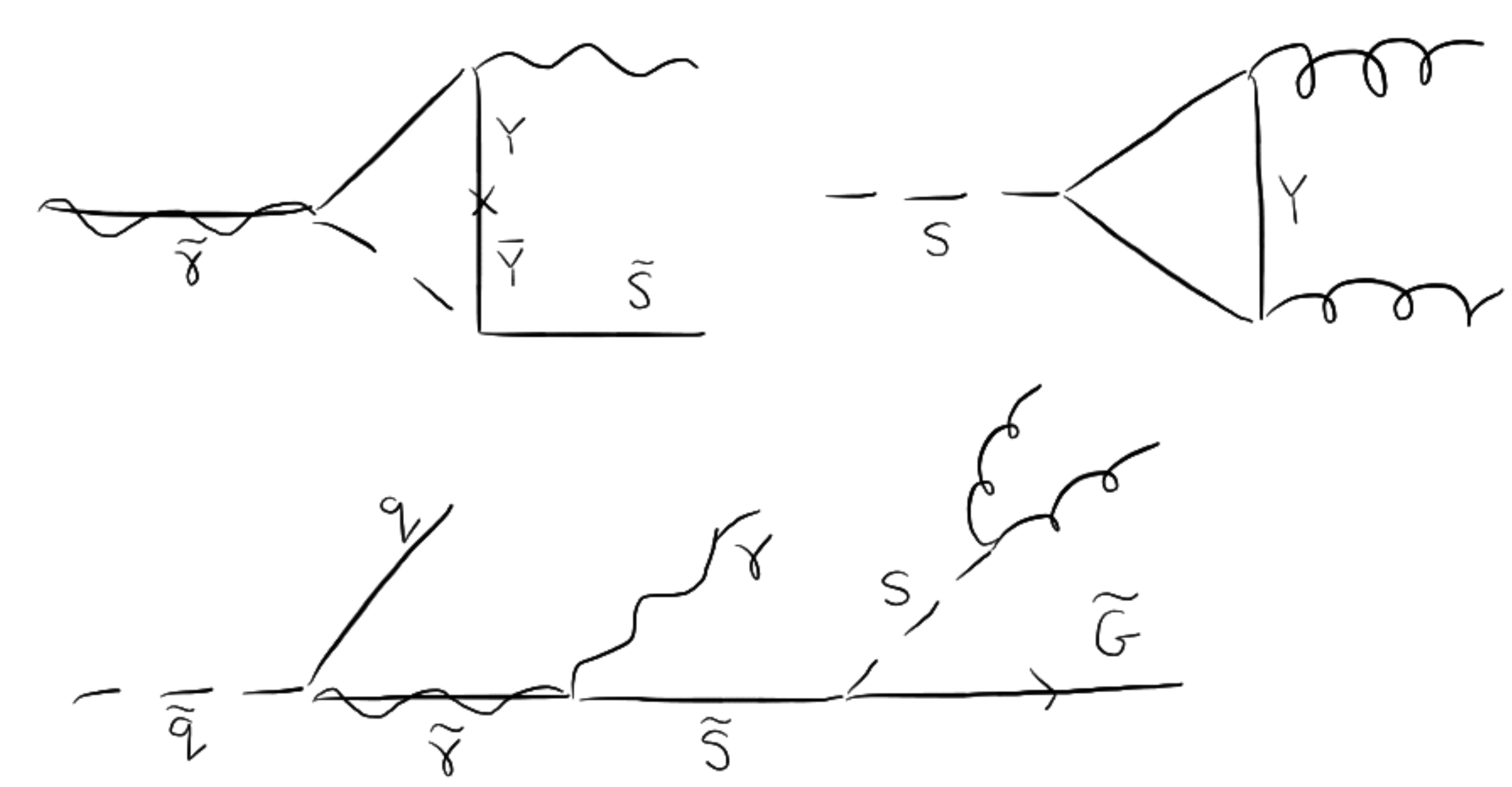} 
   \caption{Top row: Loop-induced couplings in a stealth toy model with new vector-like states. Bottom row: A sample stealth decay chain.}
   \label{fig:stealth}
\end{figure}

The other requirement for a viable model is an appropriate portal, since decay chains have to end in $\tilde S$ and the $S$ particle has to decay back to visible states. The simplest such portal is $S H_u H_d$, since $H_u H_d$ is the smallest gauge-invariant operator we can write down. We could also could add new vector-like states $Y, \bar Y$ charged under the SM and a $S Y \bar Y$ coupling. This can give a one-loop bino-photon-$\tilde S$ vertex allowing decays into $\tilde S$ radiating off a photon. Diagrams illustrating the effective one-loop couplings and example decay chains are shown in Fig.~\ref{fig:stealth}.

As in the case of compressed SUSY, experimentalists catch up quickly. In fact, CMS already places a limit on stealth scenarios where the decay into $S$ proceeds via a photon \cite{CMS:2012un}. Even if the missing energy signal is substantially eroded, high event activity and a distinctive final state such as a $\gamma$ are sufficient to discriminate from background. For the particular scenario they consider, the 7 TeV limit is 1.5 TeV on squarks whose decays go through a photon. More generally, the experimental handles are high jet multiplicity and large jet activity, since SUSY decay chains end up in light final states with several decay steps. If couplings are sufficiently small, there may also be displaced verticles, which provide an additional handle. Thus stealth SUSY, too, is not a panacea; in addition to eroding MET, one needs to hope that stealth decay chains do not throw off too many decisive tagging objects. So it's an interesting idea, and robust in its elimination of missing energy as a tagging object, but ultimately experiment is likely to catch up in a substantial sense.

\subsection{$R$-PV SUSY}

All of our discussion so far has assumed that SUSY events should involve MET due to a conserved $R$-parity. Recall that this assumption of $R$-parity was again the product of parsimony, since it naturally explained the absence of dimension-four proton decay as well as providing a viable dark matter candidate by ensuring the stability of the LSP. However, apart from any discomfort about parsimony, it's not entirely clear that these motivations are completely robust. Genuine supersymmetric WIMP dark matter is increasingly constrained, except along certain subspaces where cancellations in couplings erode direct search limits. Proton decay is a more powerful concern, but is not unavoidable; it's possible that $R$-parity is violated in specific ways that do not introduce fatal rates for proton decay. While $R$-parity provided a parsimonious avenue for dark matter and safe proton decay, it's not necessary for naturalness and its benefits can be realized in other ways without guaranteeing the stability of the LSP.

Recall that $R$-parity is a $Z_2$ subgroup of a continuous $U(1)_R$. Loosely speaking, $R$-parity is $(-1)^R$, where the theory contains $R$-even and $R$-odd particles charged as $+1, -1$, respectively. This is not a continuous symmetry, since an unbroken $U(1)_R$ forbids Majorana gaugino masses and gravitino masses. (As we've seen, there are scenarios where this may be made viable, but they require additional extensions of the MSSM). $R$-parity can be related to a twist of matter-parity, 
$(-1)^{2S} (-1)^{3B+L}$. If $R$-parity is preserved, it naturally forbids proton decay up to dimension 5 and renders the lightest R-parity odd state stable.

It's natural to ask what happens if we break $R$-parity. In the absence of $R$-parity, additional marginal and relevant terms allowed by gauge invariance are \cite{Barbier:2004ez}

\begin{equation}
W_{RPV} = \mu_i H_u L_i + \frac{1}{2} \lambda_{ijk} L_i L_j E_k^c + \lambda'_{ijk} L_i Q_j D_k^c+\frac{1}{2} \lambda_{ijk}'' U_i^c D_j^c D_k^c
\end{equation}

Note that the antisymmetric contraction forces antisymmetric flavor indices in the $LL$ and $DD$ parts in $\lambda, \lambda''$.  Thus the total parameter counting involves  3 dimensionful parameters $\mu_i$, 9 $\lambda$, 27 $\lambda'$, 9 $\lambda''$, giving us 48 new (possibly complex) parameters in total \cite{Barbier:2004ez}.

Of these, the couplings $\lambda''$ violate $B$-number, while the couplings $\lambda, \lambda'$ violate $L$-number. Note that both $B$ and $L$ number must be violated to induce proton decay, so that if for some reason only $B$- or $L$-violating operators are nonvanishing, then proton stability may be preserved.

Notice that we can rotate away the relevant terms via $H_d \propto \mu H_d + \mu_i L_i$, which regenerates remaining terms. So we should think of the relevant RPV operators as being physical, but not uniquely parameterizable. Historically, it was conventional to always rotate away the $\mu$, but in the context of predictive models it's often useful to obtain RPV coefficients in the full basis to preserve possible correlations between parameters.

Of course, just as in the MSSM there are soft parameters corresponding to all supersymmetric parameters, in the case of RPV we can also have soft parameters for each superpotential parameter. This gives rise to 51 new possibilities: 45 RPV $A$-terms, 3 $B$ terms, and 3 soft mass terms. Generically, these soft terms can cause many phenomenological problems, including the potential for inducing sneutrino vevs. There is some redundancy in the combined physical effects of the supersymmetric and soft RPV terms, but all together 48+51-3 = 96 complex free parameters introduced by RPV when supersymmetric and soft terms are considered together \cite{Barbier:2004ez}.

The parametric multiplicity aside, when RPV is turned on, the lightest R-odd particle can decay into $R$-even particles, i.e., SM states. As desired, this erases MET signals, but increases the event activity by creating high-multiplicity final states via decays involving the RPV couplings. There can also still be some MET in SUSY events if $W$ or $Z$ bosons are thrown off at intermediate steps in the decay chain, so small amounts of MET can still prove useful depending on the shape of the cascade. Finally, RPV decays often involve leptons as well, which makes for a distinctive final state. So even without substantial missing energy, the combination of high jet activity and multiplicity, potential (albeit small) MET from SM decays, and leptons from $L$-violating RPV couplings provide various experimental handles on RPV scenarios.

Needless to say, this scenario is strongly constrained in a generic realization:
\begin{itemize}
\item  Obviously having both $B$ and $L$ violation is prohibitive; the combination $\lambda'_{imk} \lambda_{11k}''^*$ ($i=1,2,3, m=1,2$) would lead to tree-level proton decay forbidden unless $\lesssim 10^{-26}$. 
\item In general, we can avoid deadly problems by conserving either $B$ or $L$ separately; this avoids proton decay. However, one needs a reason to turn on only certain RPV couplings without turning on others in order to ensure either $B$ or $L$ are conserved.
\item Even in this case, there are also constraints on pure $B$ or $L$ violation. For example, $\Delta B = 2$ processes like $n-\bar n$ oscillations, or $\mu \to e$ conversion for leptons place additional constraints on RPV couplings. The limits are most constraining on light generations.
\end{itemize}

Given these constraints, it would be nice to have a predictive framework in which an appropriate pattern of RPV couplings can be realized. One such framework is MFV SUSY \cite{Csaki:2011ge}. Here the idea is that there is no conserved $R$-parity, but RPV couplings follow the structure of minimal flavor violation, i.e., flavor violation governed solely by SM yukawas. Turning off the SM Yukawas, there is a $SU(3)^5$ global symmetry of the five species. MFV is a spurion analysis, assuming only source of symmetry breaking is SM Yukawas. 

The only renormalizable RPV operator allowed by MFV is $UDD$, proportional to $Y_u Y_d Y_d$. This gives a natural texture to the $UDD$ couplings and forbids $L$ violation. It also results in small RPV couplings for light generations, larger for heavy generations, which is favorable from the perspective of other indirect limits. It also leads to a predictive set of LHC signals, since processes involve heavy flavor.

Of course, at this level MFV SUSY is purely an ansatz. To realize it fully, we need a model. One option is to gauge a subset of the flavor symmetries of the MSSM, and spontaneously break them to give rise to SM yukawas \cite{Krnjaic:2012aj,Csaki:2013we}. In the process, this also gives rise to the RPV operators in a pattern approximately dictated by MFV SUSY. If realized in nature, it doesn't provide a total escape from limits. If RPV is concentrated in heavy flavor, this typically implies signals involving leptons and visible activity in decay chains, with better signals that can be searched for readily at the LHC. It's not a panacea, and limits on RPV processes are typically competitive with RPC processes save in a few rare cases. In general, one has to really work to benefit significantly.

\subsection{Summary}

If we want to save SUSY by breaking the signal, we should focus on altering typical expectations for either event activity or MET.  Compressed SUSY breaks the activity signal by leaving little space for visible decay products, but MET remains fairly sensitive -- especially in conjunction with ISR to eliminate any accidental MET reduction. On the other end, stealth SUSY and RPV break MET, but retain substantial event activity that can readily be probed at the LHC.

In general, whether reducing MET or event activity, these innovations gain at most a few hundred GeV in the current space of limits. Experimentalists are extremely clever and adaptable!

\section{Breaking the spectrum}

Alternately, we can ``break the spectrum'', by constructing non-minimal models that populate corners of parameter space under less stress from LHC limits. Having a satisfying model provides a reason to populate such under-constrained regions, and likewise can provide new observables or suggest new search strategies moving forward. There is still tremendous room to make progress in this direction.

\subsection{Natural SUSY}

Supersymmetric naturalness is not badly threatened by the LHC if only the particles most intimately involved in the naturalness of the Higgs are light. However, we would be more comfortable if this scenario were a {\it generic} prediction of some class of models.

It's certainly possible to construct theories where the third generation soft masses are preferentially distinguished from those of the first two generations. However, this alone is not terribly satisfying; we would like the lightness of third generation sparticles to somehow be tied more directly to the heaviness of third generation fermions and their role in electroweak naturalness.  An ad hoc light third generation of sparticles is a bit like an ad hoc $Z'$: there is no reason it can't be there, but it's fairly unsatisfying as a theory of nature. Theories featuring a light third generation (or at the very least, light stops) and heavy first- and second-generation particles have recently been dubbed ``natural SUSY'' models \cite{Papucci:2011wy}, perhaps a concession to the fact that universal models with conventional LHC signals are no longer natural. However,  models with the ``natural SUSY'' spectrum have a much longer history, dating back to the essentially identical ``more minimal'' models of the mid-1990's \cite{Dimopoulos:1995mi, Cohen:1996vb}, with many subsequent years of detailed model-building. These models were motivated more by flavor considerations than naturalness, owing to the fact that flavor violation in the third generation was (and remains) not as strongly constrained as flavor violation in the first and second generation.

There are two primary challenges to constructing a viable model of natural SUSY:
\begin{itemize}
\item Flavor. If SUSY breaking is to know about the third generation in a non-arbitrary way, it has some relation to flavor. This raises the prospect of flavor violation.
\item The Higgs mass. If SUSY is natural, and the stop masses are down around the periphery of LHC limits, then explaining the Higgs mass purely through radiative corrections from the stops is a bit challenging. One could always just add new degrees of freedom to raise the Higgs mass, but this is more satisfying if the new physics is somehow tied to the mechanism driving the spectrum.
\end{itemize}

Let's consider a selection of examples. This will be far from an exhaustive survey of natural SUSY model-building; notable models not discussed here include \cite{Csaki:2012fh, Larsen:2012rq, Cohen:2012rm, Randall:2012dm}.

\subsubsection{Single-sector SUSY breaking}

A most satisfying explanation involves flavor and the soft spectrum arising from the same source. In single-sector SUSY breaking, this is realized by the notion that supersymmetry is broken by some strong dynamics acting on degrees of freedom that also carry Standard Model charges. Light-fermion generations are identified with bound states of the strong dynamics, while heavy-fermion generations are elementary. The composite states feel supersymmetry breaking more directly than the elementary ones (often at one loop vs. two loops), correlating light fermions with heavy sfermions.

For example, if $H$, $Q_3$ are elementary chiral superfields, while $Q_{1,2}$ are composites $\sim (\psi \bar \psi)$, then certain Yukawa couplings and mixings arise as irrelevant operators generated at some flavor scale $M_F$:
\begin{equation}
W \supset H Q_3 \bar u_3 + \frac{1}{M_F} H Q_3 (\psi \bar \psi)_i + \frac{1}{M_F^2} H (\psi \bar \psi)_i (\psi \bar \psi)_j
\end{equation}
which leads to a Yukawa texture of the form
\begin{equation}
Y \sim \left( \begin{array}{ccc} \varepsilon^2 & \varepsilon^2 & \varepsilon \\
 \varepsilon^2 & \varepsilon^2 & \varepsilon \\
 \varepsilon & \varepsilon & 1
 \end{array} \right)
 \end{equation}
 where $\varepsilon \sim \Lambda / M_F$ is given by the ratio between the scale of strong coupling and the flavor scale. The irrelevant operators are generated by integrating out elementary degrees of freedom at the scale $M_F$ that may be charged under the SM but in general are neutral under the strong interactions. Presumably the composites $(\psi \bar \psi)$ are also connected more directly to SUSY breaking in the strong sector.
 
 To make this more substantive, it would be nice to have a sharp example in which these ideas are realized. The initial proposals \cite{ArkaniHamed:1997fq, Luty:1998vr} fully realized the idea through models with incalculable SUSY breaking. A calculable example arises in SUSY QCD \cite{Franco:2009wf}. Consider a SUSY QCD theory with $SU(N_c)$ gauge group and $N_f$ (anti)fundamental flavors $Q \; (\tilde Q)$, with superpotential $W = m Q \tilde Q$ that preserves a diagonal $SU(N_f)$ flavor symmetry. In the range $N_c < N_f < \frac{3}{2} N_c$ the theory becomes strongly coupled at a scale $\Lambda$, below which it can be described via Seiberg duality in terms of an IR-free dual theory consisting of and $SU(N_f - N_c)$ magnetic gauge group, $N_f$ (anti)fundamental magnetic flavors $q \;(\tilde q)$, and a meson $M \sim (Q \tilde Q)$ transforms as a bifundamental under the diagonal flavor symmetry. The magnetic superpotential is
 \begin{equation}
 W = h q M \tilde q - \mu^2 M
 \end{equation}
 where $\mu^2 = m \Lambda$. The trilinear $q M \tilde q$ has rank $N_f - N_c$, while $M$ has rank $N_f$, so not all $F$-terms of $M$ may be set to zero by giving vevs to magnetic quark bilinears. Thus SUSY is spontaneously broken in this theory by the so-called rank condition \cite{Intriligator:2006dd}. The theory also possesses supersymmetric vacua arising from irrelevant operators, so that SUSY-breaking is metastable. However, these SUSY vacua are parametrically distant in field space and lie far from the enhanced symmetry point at the origin of field space.
 
Decomposing the fields into
\begin{equation}
q \sim \left( \begin{array}{c} \sigma \\ \rho \end{array} \right) \hspace{1cm} \tilde q^T \sim \left( \begin{array}{c} \tilde \sigma \\ \tilde \rho \end{array} \right) \hspace{1cm} M \sim \left( \begin{array}{cc} Y & Z \\ \tilde Z^T & \Phi \end{array} \right)
\end{equation}
the fields $\sigma, \tilde \sigma$ acquire a vev in the metastable vacua by canceling off the maximal number of $F$-terms of $M$. The fields $\rho, Z$ acquire tree-level masses and tree-level SUSY-breaking splittings between the scalar and fermion components. The field $\Phi$ is massless at tree level, but the scalars acquire a soft mass at one loop of order 
\begin{equation}
\tilde m_\Phi^2 \sim \frac{h^2}{16 \pi^2} \mu^2
\end{equation}
while the fermions are massless. 

Then it's fairly clear to see how this theory can be leveraged into realizing the single-sector idea. If  part of the $SU(N_c)_f \subset SU(N_f)$ flavor symmetry left unbroken in the metastable vacuum is weakly gauged by the Standard Model, then $\Phi$ contains various SM representations with massless fermions and massive scalars at one loop; these can be identified with first- and second-generation fields. Moreover, the $\rho, Z$ fields are also charged under the SM and so serve as messengers of gauge mediation, giving two-loop soft masses to the scalars of elementary (third-generation) fields. Then we have
\begin{equation}
\tilde m_{1,2}^2 \sim \tilde m_\Phi^2 \sim \frac{h^2}{16 \pi^2} \mu^2 \gg \tilde m_3^2 \sim \left( \frac{\alpha}{4 \pi} \right)^2 \mu^2
\end{equation}

Where are the bodies buried? $M$ typically contains more fields beyond those of the first two generations, which must be eliminated from the low-energy spectrum. This can be achieved by adding additional elementary spectator fields who couple to the undesired components of $Q \tilde Q \sim M$ at high energies, leading to mass terms at the scale $\Lambda$. However, this is in general fairly ad hoc. The theory also entails a substantial amount of matter charged under the SM, so Landau poles in the SM gauge couplings are problematic. The theories presented above also have a tradeoff with respect to flavor. They possess a $U(2)$ sflavor symmetry since the first two generations are composites of the same meson species, which prevents prohibitive FCNC. On the other hand, this only leads to a crude $U(2)$ theory of flavor, with additional structure required to explain the flavor structure in the first and second generation fermions. One can construct more elaborate models where the first two generations arise from mesons with different classical dimension, but these models are tuned with respect to flavor \cite{Craig:2009hf}. Finally, the theory does not intrinsically explain the Higgs mass, and so, e.g., the NMSSM needs to be added as an external module.

Models have been constructed with somewhat more exotic gauge groups such as $Sp(N)$, where the origin of Standard Model chiral generations is more natural and low-scale Landau poles can be avoided \cite{Behbahani:2010wh}. There are also interesting realizations involving warped extra dimensions, as in \cite{Gabella:2007cp}.

\subsubsection{Split families}

Here the idea is that the first two generations are charged under a different set of Standard Model gauge groups at high energies \cite{Craig:2011yk, Craig:2012hc}. (See also \cite{Auzzi:2011eu}. Note that these models are related to earlier models with additional $U(1)$ symmetries \cite{Dvali:1996rj}.)  Flavor and naturalness are guaranteed by gauge invariance. 

\begin{figure}[htbp] 
   \centering
   \includegraphics[width=4in]{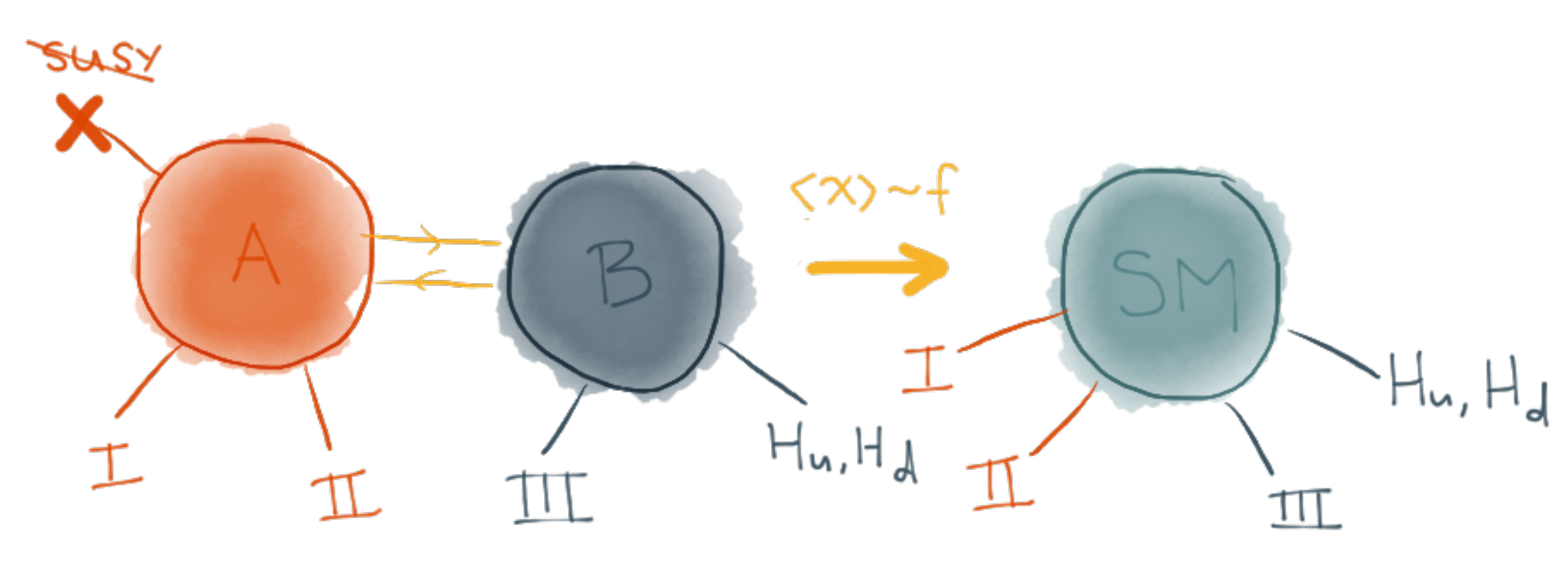} 
   \caption{The pattern of gauge groups and family assignments in the simplest split family model. On the left is the UV configuration, which higgses to the IR configuration on the right.}
   \label{fig:splitfam}
\end{figure}

Imagine that above some scale $f$, the SM gauge group is extended into a double copy, $[SU(3) \times SU(2) \times U(1)]_A \times [SU(3) \times SU(2) \times U(1)]_B$, with the first two generations transforming under the $A$ copy and the third generation and Higgs doublets $H_u, H_d$ transforming under the $B$ copy. These choices are all intrinsically anomaly-free. There are also some bifundamental fields $\chi, \tilde \chi$ charged under both groups. At the scale $f$, the $\chi, \tilde \chi$ acquire vevs that Higgs  $[SU(3) \times SU(2) \times U(1)]_A \times [SU(3) \times SU(2) \times U(1)]_B \to SU(3) \times SU(2) \times U(1)$, yielding the MSSM at low energies.  This configuration is illustrated in Fig.~\ref{fig:splitfam}

Gauge invariance forbids certain Yukawa couplings from arising as marginal operators. Instead, we must have
\begin{equation}
W \supset H Q_3 \bar u_3 + \frac{1}{M} \chi H Q_3 \bar u_i + \frac{1}{M^2} \chi^2 H Q_i \bar u_j
\end{equation}
which leads to the same flavor texture discussed earlier, but with $\varepsilon \sim f / M$. 

What about the soft spectrum? If we imagine SUSY breaking is mediated via gauge mediation through messengers charged under group $A$, then the first and second-generation soft masses are the usual GMSB ones, while the third-generation and Higgs masses are screened by an additional loop factor, and they are only the usual 2-loop GMSB diagrams below the scale $f$. This leads to 
\begin{equation}
\tilde m_i^2 \sim \left( \frac{\alpha_i}{4 \pi} \right)^2 \left( \frac{F}{M} \right)^2 \gg \tilde m_3^2 \sim \left( \frac{\alpha_i}{4 \pi} \right)^2 \left( \frac{f}{M} \right)^2 \left( \frac{F}{M} \right)^2
\end{equation}
Since the first- and second-generation soft masses are gauge-mediated, they enjoy an automatic $U(2)$ sflavor symmetry that protects against problematic flavor violation.

At this stage, the scales $f$ and $M$ could be arbitrarily large, with $f \sim M/10$ explaining the suppression of third-generation soft masses and the relative largeness of the top Yukawa. However, the theory acquires an additional novel feature if these scales are low, with $f \lesssim 10$ TeV. In that case, the Higgsing down to the SM gauge group occurs at low scales. This yields additional contributions to the MSSM $D$-terms, and hence an enhancement to the Higgs quartic precisely of the type discussed earlier. This provides an intrinsic, automatic way of explaining the Higgs mass in these models provided the scale of Higgsing is not too high to decouple the new contributions to the quartic.

If the Higgsing scale is low, it provides an additional novel feature to improve the naturalness of the theory. Since only the gauginos of the $A$ group acquire one-loop Majorana masses above the scale $f$, the renormalization of third-generation soft masses due to the gluino is cut off by the scale $f$. Thus radiative corrections are naturally of the size
\begin{equation}
\delta m_{\tilde t}^2 = \frac{2 g_s^2}{3 \pi^2} m_{\tilde g}^2 \ln \left(f / m_{\tilde g} \right)
\end{equation}
where $f \sim 10$ TeV keeps the logarithm small. Hence there is less radiative correlation between the soft parameters in this model as well.

This class of models is appealing in the sense that it provides a natural SUSY spectrum; a lower cutoff for radiative corrections; and an intrinsic explanation for the observed Higgs mass. On the other hand, as with the single-sector model presented in the previous section, the simplest two-site theories require additional structure to explain the flavor pattern of the first and second generation fermions. And while unification is possible in these models, it typically lowers the unification scale and requires additional structure to explain the absence of proton decay \cite{Craig:2012hc}.

\subsubsection{Flavor mediation}

Imagine that supersymmetry breaking is mediated through a gauged flavor symmetry. Such gauge symmetries must be spontaneously broken to yield the observed pattern of fermion masses and mixing, and in general -- if the gauged flavor symmetry also mediates supersymmetry breaking -- we expect the spontaneous breaking to leave its imprint on the soft mass spectrum \cite{Craig:2012yd}. In fact, this imprint works precisely in the direction we desire, giving rise to an inverse hierarchy of soft masses. It can, in general, also preserve desirable flavor properties \cite{Craig:2012di}.

To understand the physics, let's first consider the effects of higgsing on gauge mediation. For simplicity, consider a $U(1)$ gauge symmetry spontaneously broken at the scale $f$ along a $D$-flat direction, with messengers $\Phi_{\pm}$ coupled to a SUSY breaking spurion $X$ (with $\langle X \rangle = M + F \theta^2$) via $W \supset X \Phi_+ \Phi_-$. For fields charged under this $U(1)$, the scalar soft masses are parametrically
\begin{equation}
\tilde m^2 \sim \begin{array}{l} \left( \frac{M}{f} \right)^2 \left( \frac{\alpha_i}{4 \pi} \right)^2 \frac{F^2}{M^2} \hspace{1cm} M \ll f \\ 
 \left( \frac{\alpha_i}{4 \pi} \right)^2 \frac{F^2}{M^2} \hspace{2.1cm} f \ll M
 \end{array}
 \end{equation}
 which is as one expects; if the mass scale of the gauge bosons is much larger than the messenger scale, the contributions to gauge-mediated soft masses are suppressed accordingly, while if the mass scale is much smaller, it reverts to the usual GMSB expressions. 
 
\begin{figure}[htbp] 
   \centering
   \includegraphics[width=4in]{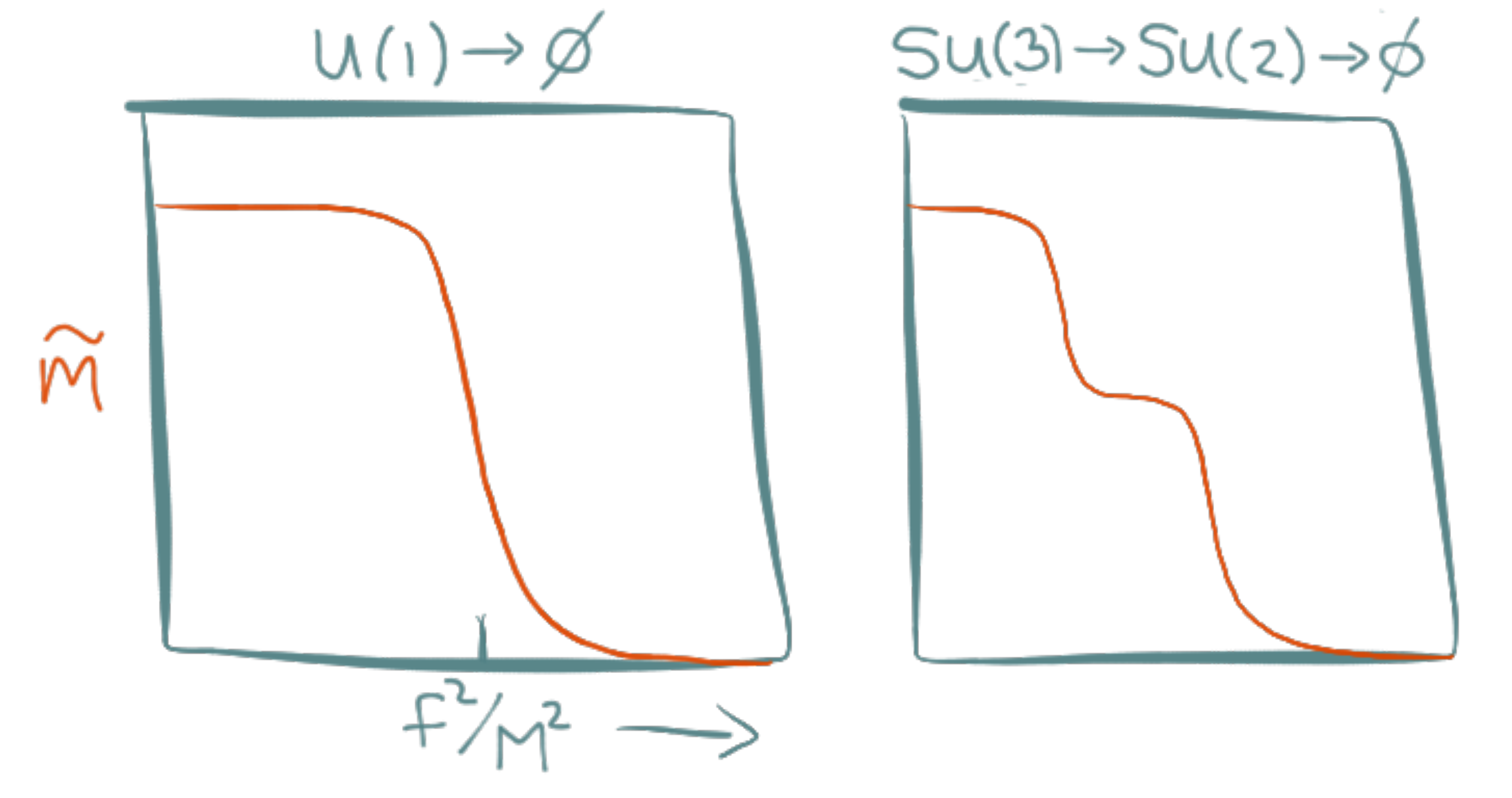} 
   \caption{Soft masses in Higgsed gauge mediation as a function of the ratio $f^2/M^2$. Left: Soft masses in a simple $U(1)$ example. Right: Soft masses in an $SU(3)$ example with the breaking pattern $SU(3) \to SU(2) \to \varnothing$. Adapted from a figure appearing in  \cite{Craig:2012di}. }
   \label{fig:flavormed}
\end{figure}

 In the more general case of non-abelian gauge symmetries, the soft masses instead factorize into a sum of separate contributions for each gauge boson mass eigenstate, with corresponding $M/f$ suppression for each mass eigenvalue. This is illustrated schematically in Fig.~\ref{fig:flavormed}, which illustrates the soft mass suppression as a factor of increasing $f/M$ for a $U(1)$ group and a $SU(3)$ group sequentially broken to $SU(3) \to SU(2) \to \varnothing$.
  
Now one can imagine how this might be put to work to generate an inverse sflavor hierarchy. If a non-abelian group is broken sequentially, the smallest soft mass terms are those arising from mediation of gauge bosons with the largest masses. So if the third generation scalars receive their soft masses from gauge bosons with larger masses, the third-generation scalars will be lighter than those of the first and second generation.

The simplest model realizing this in detail consists of a gauged $SU(3)_F$ flavor symmetry, which is the largest anomaly-free flavor symmetry that can be gauged in the SM without adding additional multiplets to cancel mixed anomalies. Under this flavor symmetry, $Q, u, d, L, e$ all transform as $\sim 3$. To cancel just the $SU(3)_F$ anomalies, we also require a right-handed neutrino multiplet $N_c \sim \bar 3$ and a pair of symmetric tensors $S_u, S_d \sim \bar 6$. Note also that this gauged flavor symmetry is completely compatible with gauge coupling unification, as it treats all SM species democratically and so commutes with e.g. $SU(5)$.

We imagine that $S_u$ and $S_d$ spontaneously break the flavor symmetry along $D$-flat directions and generate SM Yukawas via interactions of the form
\begin{equation}
W = \frac{1}{M_u} S_u H_u Q \bar u + \frac{1}{M_d} S_d H_d Q \bar d + \dots
\end{equation}
which gives rise to SM yukawas via the vevs (up to flavor rotations)
\begin{equation}
\langle S_u \rangle = \left( \begin{array}{ccc} v_{u_1} & 0 & 0 \\ 0 & v_{u_2} & 0 \\ 0 & 0 & v_{u_3} \end{array} \right) \hspace{1cm} 
\langle S_d \rangle = V_{CKM} \left( \begin{array}{ccc} v_{d_1} & 0 & 0 \\ 0 & v_{d_2} & 0 \\ 0 & 0 & v_{d_3} \end{array} \right) V_{CKM}^T
\end{equation}
where we have $v_{u_3} \gg v_{u_2} \gg v_{u_1}$ corresponding to the hierarchy $m_t \gg m_c \gg m_u$. In the limit $v_{u_i} \gg v_{d_i}$, as indicated by the mass hierarchy between the top and bottom quark, we have schematically the breaking pattern
\begin{eqnarray}
v_{u_3} &:& \hspace{5mm} SU(3)_F \to SU(2)_F \\
v_{u_2} &:& \hspace{5mm} SU(2)_F \to \varnothing
\end{eqnarray}
where the last breaking $SU(2)_F \to \varnothing$ occurs because $SU(2)$ is rank-1. Given this pattern of symmetry breaking, we arrive at a soft spectrum with
\begin{eqnarray}
\tilde m_3^2 &\sim& \left( \frac{\alpha_F}{4 \pi} \right)^2 \frac{F^2}{M^2} \frac{M^2}{v_{u_3}^2} \\
\tilde m_{1,2}^2 &\sim& \left( \frac{\alpha_F}{4 \pi} \right)^2 \frac{F^2}{M^2} \frac{M^2}{v_{u_2}^2} \gg \tilde m_3^2
\end{eqnarray}
Note that this naturally gives rise to a $U(2)$ sflavor symmetry since $SU(2)_F$ is rank-one, and so its breaking only generates one mass scale for the first two generations of scalars. There is also some off-diagonal mixing due to gauge bosons in e.g. $SU(3)_F / SU(2)_F$, but these off-diagonal soft terms are of the next-to-MFV form. 

This model naturally gives rise to
\begin{enumerate}
\item an inverse sflavor hierarchy
\item a direct connection between flavor and sflavor that is not ad hoc
\item a natural $U(2)$ sflavor symmetry
\end{enumerate}
and is entirely compatible with conventional gauge coupling unification since the $SU(3)_F$ treats all MSSM multiplets equally. Of course, there are some wrinkles; this model also
\begin{enumerate}
\item Does not explain gaugino masses for MSSM gauginos, which should be generated by additional standard gauge mediation or Dirac gaugino masses.
\item Does not explain the Higgs mass or, in general, the necessary parameters of the Higgs sector
\item Entails additional complications to guarantee that the Higgsing of the flavor symmetry is $D$-flat.
\end{enumerate}
Amusingly, although introduced as a model for natural supersymmetry, gauged flavor mediation also seamlessly gives rise to models of mini-split supersymmetry \cite{Kahn:2013pfa} (discussed briefly below).

\subsection{Supersoft SUSY}

So far we've focused on models that preferentially separate the mass scale of third-generation scalars from those of the first and second generation, allowing stops to remain light without running afoul of limits on squark-gluino associated production. Of course, another option is to keep all squarks around the same mass scale but decouple the gluino. This also alleviates direct gluino limits and the stringent limits from squark-gluino associated production \cite{Kribs:2012gx}. Of course, as we learned at the beginning of these lectures, such a separation between gluino and squark masses is typically unnatural due to the correlative effects of RG evolution:
\begin{equation}
\delta m_{\tilde q}^2 \sim \frac{2 g_s^2}{3 \pi^2} m_{\tilde g}^2 \ln \left(\Lambda / m_{\tilde g} \right) \rightarrow m_{\tilde q} \gtrsim m_{\tilde g}/2
\end{equation}

However, these RG effects again arise from parsimony, in this case the assumption that the gluino mass is Majorana. If instead the gluino mass is Dirac, then the radiative corrections can be truncated. A simple way to realize this truncation is if the vector sector of the MSSM is extended to $\mathcal{N}=2$ SUSY, in which case the theory becomes ``super-soft'' \cite{Fox:2002bu}. Of course, we can't make the entirety of the MSSM $\mathcal{N}=2$ due to the need for chiral matter, but there's no problem with extending the gauge sector. This requires adding an adjoint chiral multiplet $A_i$ to each vector multiplet $V_i$ of the MSSM.

\begin{figure}[htbp] 
   \centering
   \includegraphics[width=2in]{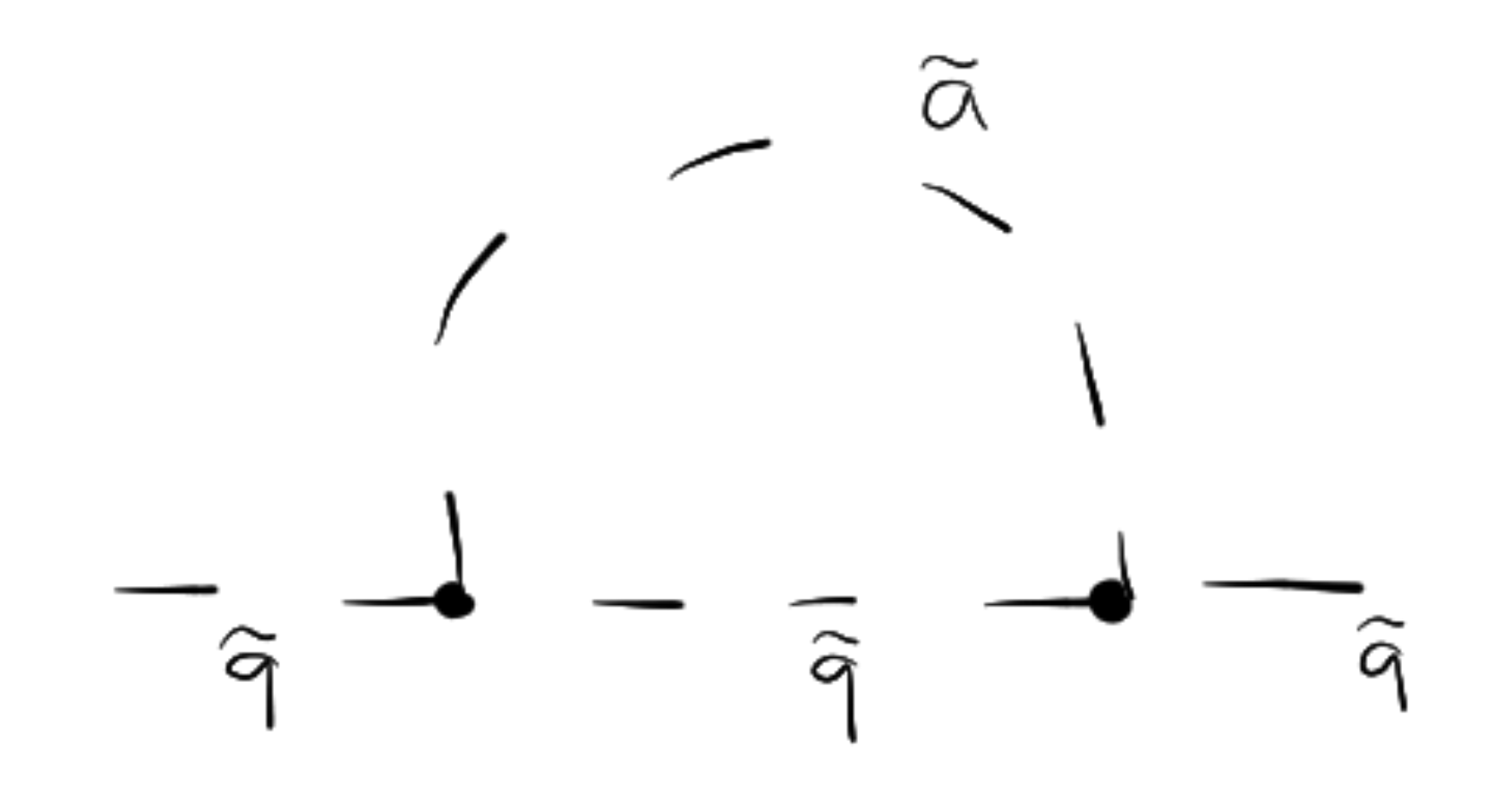} 
   \caption{The crucial added contribution to squark soft masses in super-soft SUSY breaking that renders the logarithm finite.}
   \label{fig:adjoint}
\end{figure}

In such a theory we can break supersymmetry not with $F$-terms, but rather with a $D$-term expectation value of a hidden sector $U(1)$. Then gaugino masses arise from superpotential terms of the form
\begin{equation} \label{eq:soft1}
W \supset \frac{W_\alpha' W_j^\alpha A_j}{M}
\end{equation}
where $W_\alpha'$ gets a $D$-term expectation value, yielding
\begin{equation}\label{eq:dirac}
\mathcal{L} \supset \frac{D}{M} \lambda \tilde a
\end{equation}
Note that we cannot write down large scalar masses for MSSM matter fields, since the leading scalar soft mass operator allowed by the symmetries is 
\begin{equation}
K \supset \frac{(W'^{\alpha} W_\alpha')^\dag W'^\beta W_\beta'}{M^6} Q^\dag Q
\end{equation}
Rather, the leading contribution to scalar soft masses comes from gaugino masses. However, there is now an additional diagram that renders such soft masses finite, rather than log-sensitive to the cutoff, shown in Fig.~\ref{fig:adjoint}. This additional diagram cancels the logarithmic sensitivity to the cutoff and replaces it with the scalar soft mass of $a$, such that
\begin{equation}
\tilde m_i^2 \sim \frac{\alpha_i}{\pi} m_D^2 \log \left( m_a^2 / m_D^2 \right)
\end{equation}
where $m_D$ is the Dirac mass arising from (\ref{eq:dirac}).

The natural question is then what we expect for the value of $m_a^2$. If no other SUSY-breaking operators are present for the gauge sector, then $m_a^2 = 2 m_D^2$ and the logarithm is small. However, there is one more operator allowed by the symmetries that gives rise to additional contributions to $m_a$:
\begin{equation} \label{eq:soft2}
W \supset \frac{W_\alpha' W'^\alpha}{M^2} A_j^2
\end{equation}
This operator is entirely unrelated to the primary super-soft operator generating gaugino masses. In fact, in a UV complete model, the operators (\ref{eq:soft1}) and (\ref{eq:soft2}) typically arise at the same loop order, but the former is dimension-1 and the latter is dimension-2. This implies that the mass scale of (\ref{eq:soft2}) is actually a loop factor larger than that of  (\ref{eq:soft1}). This brings the logarithm back to the same order as one expects from, e.g., low-scale gauge mediation with Majorana gaugino masses.

The other challenge is that the usual $D$-term contributions to the Higgs quartic vanish in this theory, so that quartic terms must be re-introduced and it becomes challenging to explain the observed Higgs mass.

\subsection{Colorless SUSY}

There is an exceptionally intriguing class of models that have retrieved little attention in recent years but become increasingly appealing in light of LHC superpartner limits: ``folded'' supersymmetry \cite{Burdman:2006tz}, in which the top partners at low energies are neutral under QCD. This leads to an effectively colorless theory of supersymmetry where the new degrees of freedom that control the dominant radiative contributions to the Higgs mass have no strong production modes. Absent a large QCD production rate, the limits on colorless top partners are substantially eroded, and naturalness is preserved. Such theories leverage the observation that the cancellation of the one-loop quadratic divergence coming from the top quark (in an effective theory with a cutoff) has no intrinsic reliance on the QCD representation of the superpartner. 

Of course, supersymmetry commutes with gauge symmetries, so that the full protection of supersymmetry does tie the gauge representation of particles and sparticles. This immediately suggests that colorless models of supersymmetry are somewhat contrived. Indeed, as we will see, the explicit colorless constructions on the market at present are somewhat baroque. Nonetheless, explicit examples exist, demonstrating the viability of natural colorless theories; the lack of elegant constructions is more a shortcoming of theorists than of the essential idea. To the extent that such theories provide a qualitatively new approach to naturalness, and point to entirely new search strategies at the LHC, in my mind this remains one of the most promising directions for future development. 

How does it work? In an effective theory with some cutoff $\Lambda$, we can think of the stop as cutting off the quadratic divergence in the Higgs mass $\propto 6 y_t^2 \Lambda^2 / 16 \pi^2$ coming from the top quark. (Needless to say, in a full theory one should not speak in terms of quadratic divergences, which are scheme-dependent, but in an effective theory with explicit cutoff this suffices to capture the correct physics.) This is guaranteed if the stops $\tilde t_L$ and $\tilde t_R$ couple with interactions of the form

\begin{equation} \label{eq:foldedL}
\mathcal{L} \supset \lambda_t^2 |H_u \cdot \tilde t_L|^2 + \lambda_t^2 |H_u|^2 |\tilde t_R|^2
\end{equation}

These tree-level interactions, plus a color factor of 3 coming from the dimension of the $\tilde t_L$ and $\tilde t_R$ $SU(3)_c$ representations, guarantee cancellation of the top quadratic divergence. There is no $SU(3)_c$ contraction with the Higgs field; the important feature is merely the counting factor of 3. But at one loop, this counting factor could come from something other than the dimension of $SU(3)_c$ representation. The challenge is to construct an effective theory where the couplings in (\ref{eq:foldedL}) are guaranteed at the cutoff and where the factor of 3 arises appropriately. 

\begin{figure}[htbp] 
   \centering
   \includegraphics[width=4in]{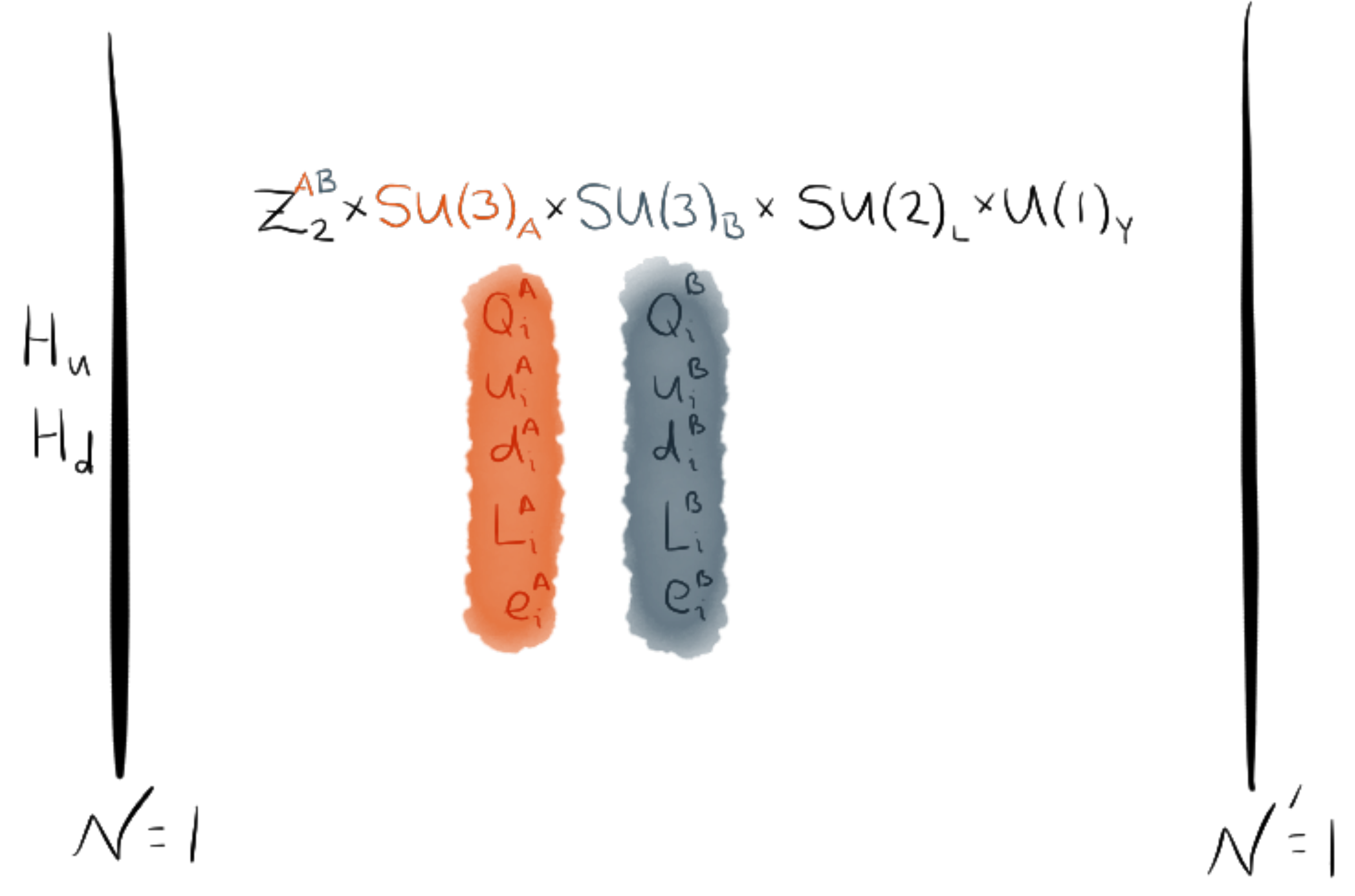} 
   \caption{Cartoon of a 5D realization of folded supersymmetry, consisting of a flat extra dimension with appropriate bulk gauge group and matter fields, Higgs multiplets on the boundary, and Scherk-Schwarz SUSY-breaking boundary conditions.}
   \label{fig:folded}
\end{figure}

An existence proof for such a theory is provided by folded supersymmetry. Consider a supersymmetric theory with a flat 5th dimension, illustrated in Fig.~\ref{fig:folded}. The bulk gauge symmetry consists of $SU(3)_A \times SU(3)_B \times SU(2)_L \times U(1)_Y$ augmented by an additional $Z_2^{AB}$ parity exchanging the $SU(3)_{A,B}$ sectors. The two $SU(3)$ factors are not Higgsed down to the diagonal; rather, $SU(3)_A$ is associated with $SU(3)_c$ at low energies, while $SU(3)_B$ is an additional gauge group that will confine at an appropriate scale. The bulk matter consists of two full copies of 5D SM matter fields, one set charged under $SU(3)_A$ and the other under $SU(3)_B$, with identical charges under the $SU(2)_L \times U(1)_Y$ factor. The $Z_2^{AB}$ parity guarantees that e.g. the superpotential couplings of $A$ multiplets and $B$ multiplets are identical. Finally, the Higgs doublets $H_u, H_d$ are 4D multiplets living on one of the branes.

\begin{figure}[htbp] 
   \centering
   \includegraphics[width=4in]{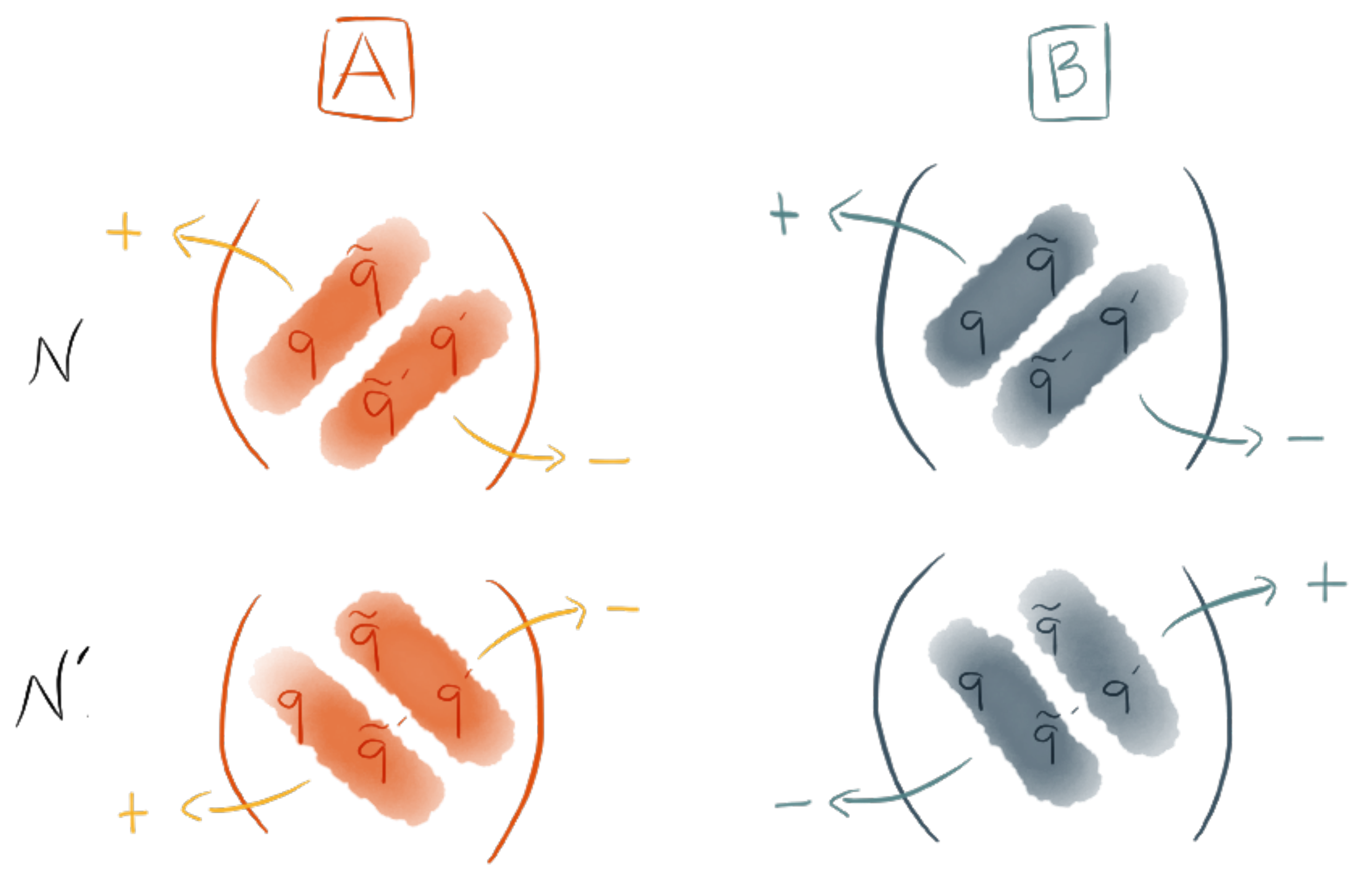} 
   \caption{Cartoon of the Scherk-Schwarz boundary conditions required to project out a $q_A$ fermionic zero mode and $\tilde q_B$ sfermionic zero mode. The surviving zero modes are those that transform as $++$ under the combined $\mathcal{N}=1$ and $\mathcal{N}'=1$ boundary conditions.}
   \label{fig:scherk}
\end{figure}

Supersymmetry is broken by Scherk-Schwarz boundary conditions. This corresponds to preserving different $\mathcal{N}=1$ supersymmetries on the two branes, so that the zero-mode spectrum is nonsupersymmetric. In a 5D theory with $\mathcal{N}=1$ supersymmetry, each bulk matter multiplet consists of fermions and scalars filling out two 4D $\mathcal{N}=1$ chiral multiplets, with the $\mathcal{N}=1$ on each brane pairing the fermions and scalars into chiral multiplets and preserving the zero mode from one chiral multiplet or another. If the two branes preserve different $\mathcal{N}=1$, corresponding to different pairings into chiral multiplets, then only a single zero mode (scalar or fermion) remains massless at tree level in the 4D effective theory. The trick lies in the assignment of parities under the Scherk-Schwarz projection, as illustrated in Fig.~\ref{fig:scherk}.

 For e.g., the three generations of quark doublet supermultiplets $Q_i^A$ we choose one $\mathcal{N} = 1$ to preserve the zero mode of one chiral multiplet consisting of scalar $\tilde q_i^A$ and fermion $q_i^A$, while we choose the other $\mathcal{N} = 1$ (call it $\mathcal{N}'$) to preserve the zero mode of a chiral multiplet consisting of the scalar $\tilde q_i'^A$ and fermion $q_i^A$. Hence the only surviving massless zero mode is the fermion $q_i^A$. We repeat the exercise for the $B$ copies, but we switch the Scherk-Schwarz assignments such that the surviving massless zero mode is the scalar $\tilde q_i^B$. Thus the zero mode spectrum for the $Q_i^{A,B}$ consists of a massless fermion $q_i^A$ and massless (at tree level) sfermion $\tilde q_i^B$, etc.

The $Z_2^{AB}$ guarantees the Yukawa couplings are of the form
\begin{equation}
W \supset \lambda_t H_u Q^A_3 u^A_3 + \lambda_t H_u Q^B_3 u^B_3 + \dots
\end{equation}
and so the Lagrangian terms for the surviving zero modes are 
\begin{equation}
\mathcal{L} \supset \lambda_t H_u q_3^A u_3^A + \lambda_t^2 |H_u \cdot \tilde q_3^B|^2 + \lambda_t^2 |H_u|^2 |\tilde u_3^B|^2
\end{equation}
(with a slight abuse of notation; $q/u$ denote fermions). As you can see, the couplings required for protecting the Higgs mass from the top quark at one loop are present, and the multiplicity factor is guaranteed by the dimension of the scalar representations under $SU(3)_B$ but {\it not} under QCD, i.e., $SU(3)_A$.  We have used the $Z_2$ and pattern of Scherk-Schwarz breaking to essentially trade out the $\tilde q_A$ zero modes for $\tilde q_B$ zero modes with the same tree-level couplings.

This orbifold projection leaves no gaugino zero modes under the gauge factors, so in particular the particles that control the stop masses and the worst two-loop divergences (the gluinos) are absent. Thus the theory must still have a low cutoff. In particular, the soft mass for the squark zero modes is cut off by the size of the extra dimension,
\begin{equation}
\tilde m_Q^2 \sim \left( \frac{\alpha_3}{4 \pi} \right)^2 \frac{1}{R}
\end{equation}
so naturalness requires $1/R \sim 5-10$ TeV. There is still a one-loop divergent contribution to the Higgs mass from the $W$ loop, but for this size of extra dimension this does not introduce a substantial fine-tuning.

Given a flat fifth dimension, the theory is not UV complete due to non-renormalizability of the gauge interactions; it also has a fundamental cutoff $\Lambda$ at which it must be embedded into a higher-dimensional theory. This scale cannot be decoupled from $R$, since the top Yukawa is volume-suppressed with $y_t \sim 1/(\Lambda R)$, which constrains $\Lambda \lesssim 20$ TeV. 

The physics of the Higgs sector is that of the MSSM, since $H_u$ and $H_d$ originate from complete 4D chiral multiplets on one brane. Thus the Higgs mass still requires enhancement to match the observed value, and in this case large $A$-terms are not generated. 

So this theory has a low cutoff, since the protection offered by SUSY is preserved only at one loop. However, the necessary relations between couplings are preserved by the UV completion into an extra dimension. This provides an existence proof for models where the immediate physics controlling the Higgs mass is neutral under QCD, but other examples are likely to exist. It would be particularly interesting to construct models where the shell game of swapping $SU(3)$ zero modes is unnecessary; it would also be attractive to construct models that work strictly in four dimensions. In my mind, this is one of the final frontiers for promising model-building in light of current limits.

\subsection{Focus point SUSY} 

The idea of focus point SUSY \cite{Feng:1999zg}  is qualitatively quite different from the cases discussed above. Rather than preserve naturalness by finding reasons for sparticles to satisfy their generic naturalness bounds consistent with current limits, focus point SUSY attempts to eliminate the radiatively sensitivity of the weak scale to the soft masses -- thereby allowing large soft masses but with small radiative effects due to cancellations along the RG trajectory. The construction relies on the observation that certain subspaces of UV soft parameters can lead to $m_{H_u}^2(m_W) \sim 0$ irrespective of the overall scale in the UV parameters.\footnote{For an excellent extended discussion of the virtues of focus point SUSY in light of LHC data, see \cite{Feng:2013pwa}.}

The challenge is that one requires UV physics to fix the boundary conditions that lead to this insensitivity. In this respect focus point SUSY itself an insightful observation in need of explicit models to enforce the necessary boundary conditions. Examples of such models may be found in e.g. \cite{Choi:2005hd, Kitano:2005wc, Yanagida:2013ah}.

To see the focus point effect, consider the RGEs for the soft terms most directly connected to naturalness of the weak scale -- $m_{H_u}^2, m_{U_3}^2,$ and $m_{Q_3}^2$, as well as the $A$-term $A_t$ (since we will be interested in viable predictions for the Higgs mass). Then assuming $\tilde m^2, A^2 \gg m_{\lambda}^2$, the RGEs are linear with the form \cite{Feng:2012jfa}

\begin{equation}
\frac{d}{d \ln Q} \left[ \begin{array}{c}
m_{H_u}^2 \\ m_{U_3}^2 \\ m_{Q_3}^2 \\ A_t^2
\end{array} \right]
= \frac{y_t^2}{8 \pi} \left[ \begin{array}{cccc}
3 & 3 & 3 & 3 \\
2 & 2 & 2 & 2 \\
1 & 1 & 1 & 1 \\
0 & 0 & 0 & 12
\end{array} \right]
\left[ \begin{array}{c}
m_{H_u}^2 \\ m_{U_3}^2 \\ m_{Q_3}^2 \\ A_t^2
\end{array} \right] \ .
\label{eq:fullrge}
\end{equation} 

If the input parameters at the high scale (typically taken to be $M_{GUT}$ in the case of focus point SUSY) enjoy particular relations, then $m_{H_u}^2$ at the weak scale can be made to vanish. For the soft terms of interest, the necessary relations are satisfied by a two-parameter family of input values that lead to $m_{H_u}^2(m_W) = 0$ \cite{Feng:2012jfa}:

\begin{equation}
\left[ \begin{array}{c} 
m_{H_u}^2(M_{GUT}) \\ m_{U_3}^2(M_{GUT}) \\ m_{Q_3}^2(M_{GUT}) \\ A_t^2(M_{GUT})
\end{array} \right] =
m_0^2  \left[ \begin{array}{c} 
1 \\ 1+x - 3y \\ 1-x \\ 9y
\end{array} \right] 
\to
\left[ \begin{array}{c} 
m_{H_u}^2(m_W) \\ m_{U_3}^2(m_W) \\ m_{Q_3}^2(m_W) \\ A_t^2(m_W)
\end{array} \right] =
m_0^2  \left[ \begin{array}{c} 
0 \\ \frac{1}{3} + x - 3y \\ \frac{2}{3} - x \\ y
\end{array} \right] .
\end{equation}

Since at this order $m_{H_u}^2$ is independent of the scale of the UV input parameters (provided the theory is constrained to lie along the two-parameter family of focus point relations), the usual fine-tuning measure $\Delta$ is insensitive to the overall scale of the soft masses. In principle, this allows for large $A$-terms to explain the Higgs mass and sparticle masses heavy enough to satisfy bounds, with corresponding values for $\Delta$ that are much smaller than one would expect from naive dimensional analysis. Technically speaking, the explanation of the weak scale is still somewhat {\it post hoc}, since there is no intrinsic connection between $m_W$ and $m_{H_u}^2(m_W) = 0$, but perhaps this is asking too much.

In any event, one requires a detailed theory to genuinely argue that the fine-tuning is reduced. The key assumption above is that one should compute tuning only by varying the overall scale of soft parameters and the two-parameter family of soft parameter relations. This is a reasonable assumption in a theory where symmetries restrict the input soft parameters to the appropriate subspace. If there is no such restriction, then there is a tuning associated with fixing the input parameters to their focus point relations.

If we shift any of the parameters off their focus point ansatz by an amount $\delta m^2$, then the shift in $m_{H_u}^2(m_W)$ is also of order $\delta m^2$. In general one might expect certain parts of the focus point ansatz to be singled out by symmetries or model-building finesse -- for example, the universal point $x = y = 0$ (though this has problems with the Higgs mass). But particularly when $A$-terms are turned on, it's hard to imagine guaranteeing the necessary relations by symmetries. Without such guarantees, there is a tuning percentage on the order of $\delta m^2 / m_0^2$ associated with the non-genericness of the UV boundary conditions. So this is an outstanding target for additional model-building aimed at singling out the space of UV parameters that lie on the focus point manifold. It also provides an attractive illustration of the fact that naive expectations for tuning can be derailed by subtle effects.

\subsection{Unnatural (mini-split) SUSY}

So far I have focused almost exclusively on models aimed at preserving the naturalness of the weak scale through supersymmetric physics -- that is to say, I have considered what happens if we discard parsimony as a principle and retain naturalness. It is, however, possible that supersymmetry exists in nature but does not provide a completely natural explanation of the weak scale, with some degree of fine-tuning keeping the mass of the Higgs light relative to the mass of other sparticles. In this case the actual structure of SUSY models can be quite simple, preserving parsimony at the price of naturalness. Perhaps anthropic selection takes care of the additional fine-tuning, or perhaps a moderate amount of fine-tuning is simply acceptable for the weak scale as long as SUSY takes care of ``most'' of the tuning up to the Planck scale. Signals may persist if the scalars are not too heavy, or if fermions such as the gauginos and higgsinos remain light for various reasons and can be directly produced at the LHC or play a role in dark matter. This scenario is intriguing to the extent that it provides qualitatively new signals to explore directly at the LHC and indirectly in measurements of flavor violation, EDMs, and dark matter.

Such ideas about split supersymmetry predate LHC limits \cite{ArkaniHamed:2004fb,Giudice:2004tc}, but have enjoyed a recent revival in light of the Higgs mass measurement \cite{Arvanitaki:2012ps, ArkaniHamed:2012gw}. As we have seen, the observed Higgs mass places an upper bound on sparticle masses assuming the quartic is given by the MSSM $D$-terms; this bound tantalizingly allows for scalars that lie one or two loop factors above the weak scale. Such scalars are quite a bit lighter than considered in early models of split SUSY, and invite model-building aimed at explaining possible hierarchies between the scalars and weak-scale gauginos. Many problems of weak-scale SUSY model-building are alleviated, but some finesse is still required to avoid problematic flavor violation and EDMs.

On one hand, this is a bit unsatisfying when seen through the lens of our initial objectives. It's hard to imagine that there's a ``natural-ishness'' principle at work, favoring supersymmetry at a scale that solves most of the weak hierarchy problem but not all of it. On the other hand, all of our discussion of naturalness has assumed that the biggest naturalness problem -- the cosmological constant problem -- somehow factorizes. Anthropic selection provides by far the cleanest explanation for the observed size of the cosmological constant. If anthropics is at play there, who's to say how much, or little, of a role anthropics plays in tidying up the much smaller hierarchy between the weak scale and the Planck scale. Since it's so hard to construct viable models of weak-scale SUSY -- even before LHC limits are taken into account -- perhaps nature simply wants SUSY down to the point where model-building becomes uncomfortable, and no further. 

In any event, at present there are many excellent advocates for mini-split supersymmetry, and so I will not discuss them further here, save to say that studying potentially novel signatures of higher-scale SUSY deserves more attention.

\section{Looking ahead}

\begin{quote}
{\it ``Measure what can be measured, and make measurable what cannot be measured.''}\\
-Galileo Galilei
\end{quote}

So here we are. With the LHC shutdown, we have a year or two to think about how best to devote our energies to the search for supersymmetry -- and naturalness in general -- at 13 or 14 TeV. Already ATLAS has published preliminary projections of its sensitivity in various key SUSY channels at the 14 TeV LHC \cite{ATL-PHYS-PUB-2013-002}:

\begin{itemize}

\item The exclusion reach for squarks and gluinos asymptotes to $m_{\tilde q} \sim 2.7$ TeV and $m_{\tilde g} \sim 2.3$ TeV with 300 \ifb, reaching  $m_{\tilde q} \sim$ 3.1 TeV and  $m_{\tilde g} \sim$ 2.7 TeV with 3000 \ifb.

\item The reach for stops will not extend much beyond 1 TeV. In particular, the discovery reach in $\tilde t \to t \chi_1^0$ with 300 \ifb extends out to 800 GeV for an LSP lighter than 300 GeV, while the exclusion reach in the same channel with 3000 \ifb extends out to just shy of 1.1 TeV for an LSP lighter than 500 GeV. I find this surprising; with optimization of search strategies, I expect that sensitivity will extend much beyond 1.1 TeV, perhaps closer to 1.5 TeV or beyond. In any event, we are sure to see careful projections in the near future. 

\item The discovery reach for winos is claimed to extend at or beyond 800 GeV with 3000 \ifb provided the LSP is lighter than 300 GeV. This is also somewhat surprising, since it represents a considerable improvement over the 300 \ifb reach.

\end{itemize}

I would take these sensitivity estimates as approximate for the time being, with the understanding that more detailed projections will be available soon.\footnote{My expectation is that the stop reach will increase and the wino reach may decrease relative to current projections, but we'll see.} Either way, ATLAS and CMS are in a position to decisively settle the question of whether there are superpartners beneath a TeV, except in only a few pathological or unlucky scenarios. We won't even need to wait until 300\ifb or 3000 \ifb are on tape; within the first year, if there are no hints of kinematics-limited new physics, we'll more or less have moral certainty that colored sparticles do not exist or lie in the multi-TeV range.

So what are the most promising avenues for further study? In my mind, three things immediately spring to mind:

\begin{itemize}
\item Improve searches without MET, especially purely hadronic searches.
\item Focus on looking for electroweak physics at the electroweak scale.
\item Pursue radically natural ideas like colorless supersymmetry.
\end{itemize}

On the search strategy end, we should focus as much as possible on improving searches that do not rely on substantial amounts of MET. There are already useful proposals along these lines, and sensitivity will benefit further with additional study.
To the extent that searches without MET often must rely on identifying signals through high hadronic activity or multiplicity, there is great utility in improving our understanding of QCD backgrounds and further developing search strategies to differentiate SM jets from BSM jets.

While the reach for colored sparticles is high and will only grow higher at 13-14 TeV, the situation is considerably different for electroweak physics. Our sensitivity to new electroweak degrees of freedom is barely better than LEP, and in the case of light charginos the LHC says nothing that we did not already know in 2001. Considering that the one particle discovered so far at the LHC was an electroweak degree of freedom, there's much to be said for focusing efforts on
\begin{enumerate}
\item Improving LHC sensitivity to new electroweak physics through improved kinematic variables and focused search strategies.
\item Constructing models of new physics where the intrinsically light degrees of freedom carry only electroweak quantum numbers.
\end{enumerate}

This ties into what I see as the ``final frontier'' in electroweak naturalness: theories where the naturalness of the weak scale is still guaranteed by light degrees of freedom, but where these light degrees of freedom do not carry QCD quantum numbers. Folded supersymmetry provides a proof of principle for how such a theory might function, but at the moment is a solitary example of what must be a broad equivalence class of theories. Only after we've extensively studied such theories and developed searches for the relevant partner particles can we decisively evaluate the state of electroweak naturalness at the LHC. \\

I began with a pithy summary of the state of supersymmetry after Run I of the LHC, claiming that
\begin{quote}
Candor compels me to declare that at this time there is no supersymmetry as our forebears understood the term, and as they meant it to be understood by us.
\end{quote}
I hope I have convinced you that this is an optimistic statement, rather than a pessimistic one. There is immense opportunity to build new models involving supersymmetry and develop new search strategies to discover them. And the wonderful news is that there will be immediate and decisive feedback from Run II. In a matter of only a few years we will know whether supersymmetry plays any role in physics at the TeV scale -- not only the supersymmetry of our forebears, but also the supersymmetry of our own era.

\section*{Acknowledgments}
I am grateful to the organizers of the ``Beyond the Standard Model after the first run of the LHC'' workshop at the Galileo Galilei Institute for giving me the opportunity to deliver these lectures during the training week, and grateful to the attendees of the training week for sitting through them. I particularly thank Emilian Dudas, Jamison Galloway, Stuart Raby,  Francesco Riva, Andrea Romanino, and Marco Serone for helpful feedback during the preparation and presentation of the lectures, and Jamison Galloway, Daniel Green, Simon Knapen, and Matthew McCullough for reading and commenting upon a draft of these lecture notes. I gratefully acknowledge the hospitality of the Kavli Institute for Theoretical Physics and the Galileo Galilei Institute where parts of these lecture notes were frantically prepared.  This work was supported in part by the DOE under grant 
DE-FG02-96ER40959, the NSF under grant PHY-0907744, and the Institute for Advanced Study.

\bibliography{lecbib}
\bibliographystyle{utphys}

\end{document}